\documentclass[aps,superscriptaddress,amsmath,amssymb,floatfix,twocolumn,showpacs,amsfonts,longbibliography]{revtex4-1}
\usepackage{times}
\usepackage[varg]{txfonts}
\usepackage{textcomp}
\usepackage{graphicx}
\usepackage{subfigure}
\usepackage{tabu}
\usepackage{color}
\usepackage[colorlinks=true,citecolor=blue,urlcolor=blue,linkcolor=blue,hyperindex]{hyperref}
\usepackage{braket}
\usepackage{overpic}
\usepackage{amssymb}
\usepackage{ulem}
\usepackage{multirow}
\usepackage{verbatim}

\allowdisplaybreaks

\begin{document}

\title{$K$ -edge and $L_{3}$ -edge RIXS study of columnar and staggered quantum dimer phases of the square lattice Heisenberg model}
\author{Meiyu He}
\affiliation{State Key Laboratory of Optoelectronic Materials and Technologies, School of Physics, Sun Yat-Sen University, Guangzhou 510275, China}

\author{Trinanjan Datta}
\email[Corresponding author:]{tdatta@augusta.edu}
\affiliation{Department of Chemistry and Physics, Augusta University, 1120 15$^{th}$ Street, Augusta, Georgia 30912, USA}
\affiliation{State Key Laboratory of Optoelectronic Materials and Technologies, School of Physics, Sun Yat-Sen University, Guangzhou 510275, China}

\author{Dao-Xin Yao}
\email[Corresponding author:]{yaodaox@mail.sysu.edu.cn}
\affiliation{State Key Laboratory of Optoelectronic Materials and Technologies, School of Physics, Sun Yat-Sen University, Guangzhou 510275, China}

\date{\today}

\begin{abstract}
We compute the $K$ and $L_{3}$ -edge resonant inelastic x-ray scattering (RIXS) spectrum of the columnar and the staggered quantum dimer states accessible to the square lattice Heisenberg magnet. Utilizing a bond-operator representation mean-field theory we investigate the RIXS features of the one- and two-triplon excitation spectrum supported by the quantum dimer model in the background of condensed singlet excitation. We find that the two-triplon excitation boundary lies within an energy range of $2.7J_{2}~(1.7 J_{2})$ to $9J_2~(7.5J_{2})$ for the columnar (staggered) phase, where $J_2$ is the interbond coupling strength. We estimated the two-triplon gap to be $ 2.7J_2$ ($ 1.7J_2$)  for the columnar (staggered) dimer phase. The highest intensity of the $K$ -edge RIXS spectrum is localized approximately around the $(\pi/2,\pi/2)$ point for both the columnar and the staggered phases. At the $L_{3}$ -edge we study the one- and two- triplon signal considering experimental scattering geometry, polarization restriction, and experimental resolution effects. Our calculations find a contribution to the two-triplon RIXS signal that originates from the local hard-core dimer constraint. This leads to a finite non-zero signal at the (0,0) momentum transfer which can offer an explanation for the existing ladder RIXS experiments and also predicts a non-zero signal for the two-dimensional quantum dimer system. We find that the $L_3$ -edge RIXS response of the one- and two-triplon signal could exist in antiphase rung modulation for zero and $\pi$ as found in inelastic neutron scattering. Since the disordered phase has the potential to harbor a variety of quantum paramagnetic states, our RIXS calculations provide useful signatures to identify the true nature of the ordering pattern.
\begin{description}
\item[PACS number(s)] 78.70.Ck, 75.25.-j, 75.10.Jm
\end{description}
\end{abstract}

\maketitle
\section{Introduction}
The success of $K$ and $L$ -edge resonant inelastic x-ray scattering spectroscopy in recent years has provided deep insight into the physical properties of correlated electronic and magnetic systems ~\cite{RevModPhys.83.705,DEAN20153}. Improved resolution and instrumentation techniques have allowed detection of novel physical phenomena \cite{doi:10.1063/1.2372731}. Computational \cite{PhysRevLett.110.265502,Jia_2012,PhysRevLett.106.157205,PhysRevB.82.064513,PhysRevB.83.245133,PhysRevB.85.064423,PhysRevB.85.064423} and theoretical \cite{PhysRevLett.96.107005,EPL.73.121,PhysRevLett.101.106406,PhysRevB.75.214414,PhysRevB.77.134428,PhysRevB.89.165103,PhysRevLett.109.117401,PhysRevB.86.125103} studies have predicted physical behavior that provide a conceptual understanding of the microscopic pathways which give rise to collective or fractionalized excitations ~\cite{Schlappa2018}. Indirect and direct RIXS study of the antiferromagnetic  N\'{e}el ordered state have been completed ~\cite{EPL.73.121,PhysRevB.77.134428,PhysRevB.81.085124,PhysRevB.81.085124,PhysRevLett.105.157006,LuoPhysRevB.89.165103,PhysRevB.96.144436}. Presently, Cu or O $K$-edge RIXS can be used to detect double spin flip bimagnon excitations~\cite{EPL.73.121,PhysRevB.77.134428}. Additionally, Cu $L$-edge RIXS has adequate resolution to detect magnon excitations \cite{PhysRevB.81.174533,PhysRevB.85.214527,PhysRevLett.105.157006}. This fact establishes RIXS on an equal footing with inelastic neutron scattering (INS)~\cite{PhysRevLett.98.027403}, which has reported measurements of the triplon excitation. RIXS can also detect selection rule allowed triplon excitations with $\Delta S = 0,1,2$, where $\Delta S$ refers to spin angular momentum change.

Dimer systems have been investigated experimentally using a host of techniques $-$ INS \cite{PhysRevLett.81.1702,PhysRevB.62.8903,PhysRevLett.98.017202,PhysRevLett.98.027403,PhysRevLett.100.205701,PhysRevB.80.094411,PhysRevLett.105.097202,PhysRevB.83.140413,PhysRevLett.108.167201,PhysRevB.88.014504}, Raman spectroscopy \cite{Schmidt_2001,PhysRevB.72.094419,PhysRevLett.90.167201}, neutron time-of-flight \cite{PhysRevB.88.094411}, optical conductivity ~\cite{PhysRevLett.87.127002} and RIXS \cite{PhysRevLett.103.047401,PhysRevB.85.224436,kumarPhysRevB.99.205130}. A quantum dimer model (QDM) can be used to model a short-range resonating-valence-bond system~\cite{PhysRevLett.61.2376}. QDMs can emerge in a variety of lattices. For example, the 3D material TlCuCl$_{3}$~\cite{PhysRevB.92.214401,PhysRevLett.100.205701}, the quasi-2D bilayer spin dimer compound BaCuSi$_{2}$O$_{6}$~\cite{PhysRevLett.93.087203}, quasi-1D spin ladder system ${\mathrm{Sr}}_{14}{\mathrm{Cu}}_{24}{\mathrm{O}}_{41}$~\cite{PhysRevLett.81.1702}, ${\mathrm{La}}_{6}{\mathrm{Ca}}_{8}{\mathrm{Cu}}_{24}{\mathrm{O}}_{41}$~\cite{PhysRevB.62.8903}, ${\mathrm{La}}_{4}{\mathrm{Ca}}_{10}{\mathrm{Cu}}_{24}{\mathrm{O}}_{41}$~\cite{PhysRevLett.98.027403}, ${({\text{C}}_{5}{\text{D}}_{12}\text{N})}_{2}{\text{CuBr}}_{4}$~\cite{PhysRevB.80.094411}, and an ultracold atomic system~\cite{PhysRevA.99.043623}.

Quantum fluctuations, hard-core constraints, and lattice geometry play important roles in dimer physics ~\cite{PhysRevA.99.043623}. Till date, dimer RIXS investigations have focused primarily on the two-leg spin ladder materials ~\cite{PhysRevLett.103.047401,PhysRevB.85.224436,kumarPhysRevB.99.205130}. The logical next step in this process of investigation would be to attempt an understanding of the two-dimensional (2D) dimer system probed by RIXS. Currently, there is no theoretical or experimental study on this topic. The dimerized Heisenberg model with competing intra- and inter- bond strength can be tuned from the N\'{e}el ordered to the quantum disordered dimer state ~\cite{PhysRevLett.61.2484,PhysRevB.49.11919}. There is a debate over the nature of the transition - weakly first order or deconfined quantum critical \cite{Senthil1490,PhysRevLett.101.050405,PhysRevLett.98.227202,PhysRevB.80.174403,PhysRevB.90.245143,PhysRevB.94.115120,PhysRevB.75.235122}. We study the RIXS behavior of the dimer model in a regime which is far from the critical transition point.

When the singlet bonds on the dimer lattice are broken due to an external or internal perturbation, it leads to mobile $S=1$ spin triplet excitations called the triplon. The triplon excitation is a quasiparticle with an energy gap. It is three-fold degenerate in the absence of a magnetic field. The $L_3$-edge RIXS experiment for the spin ladder system, ${\mathrm{Sr}}_{14}{\mathrm{Cu}}_{24}{\mathrm{O}}_{41}$, has observed a two-triplon excitation signal for a large range of momentum, including zero momentum transfer ~\cite{PhysRevLett.103.047401}. Theoretical calculation by Igarashi \cite{PhysRevB.85.224436} has revealed one-triplon excitation contribution to the RIXS intensity at $q_{a}=-\pi$. Additionally, a systematic computational and theoretical study of the Cu $L$-edge RIXS features of the spin ladder model at various coupling ratio as well as doping condition has been pursued~\cite{kumarPhysRevB.99.205130}.

\begin{figure}[t]  
\centering
{\subfigure[~Columnar dimer phase]{\label{fig:C-dimer}
\includegraphics[scale=0.12]{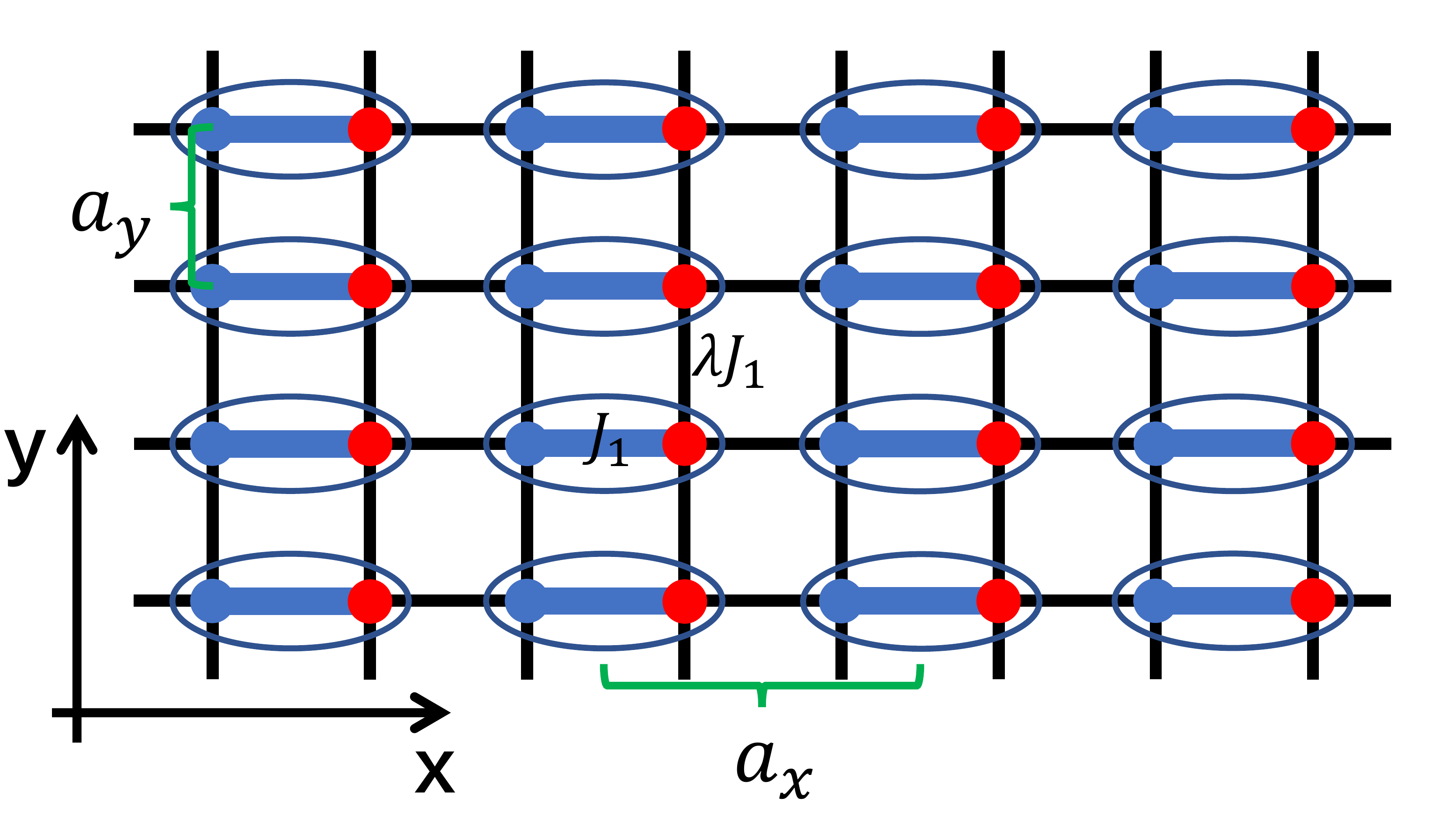}}}
{\subfigure[~Staggered dimer phase]{\label{fig:S-dimer}
\includegraphics[scale=0.12]{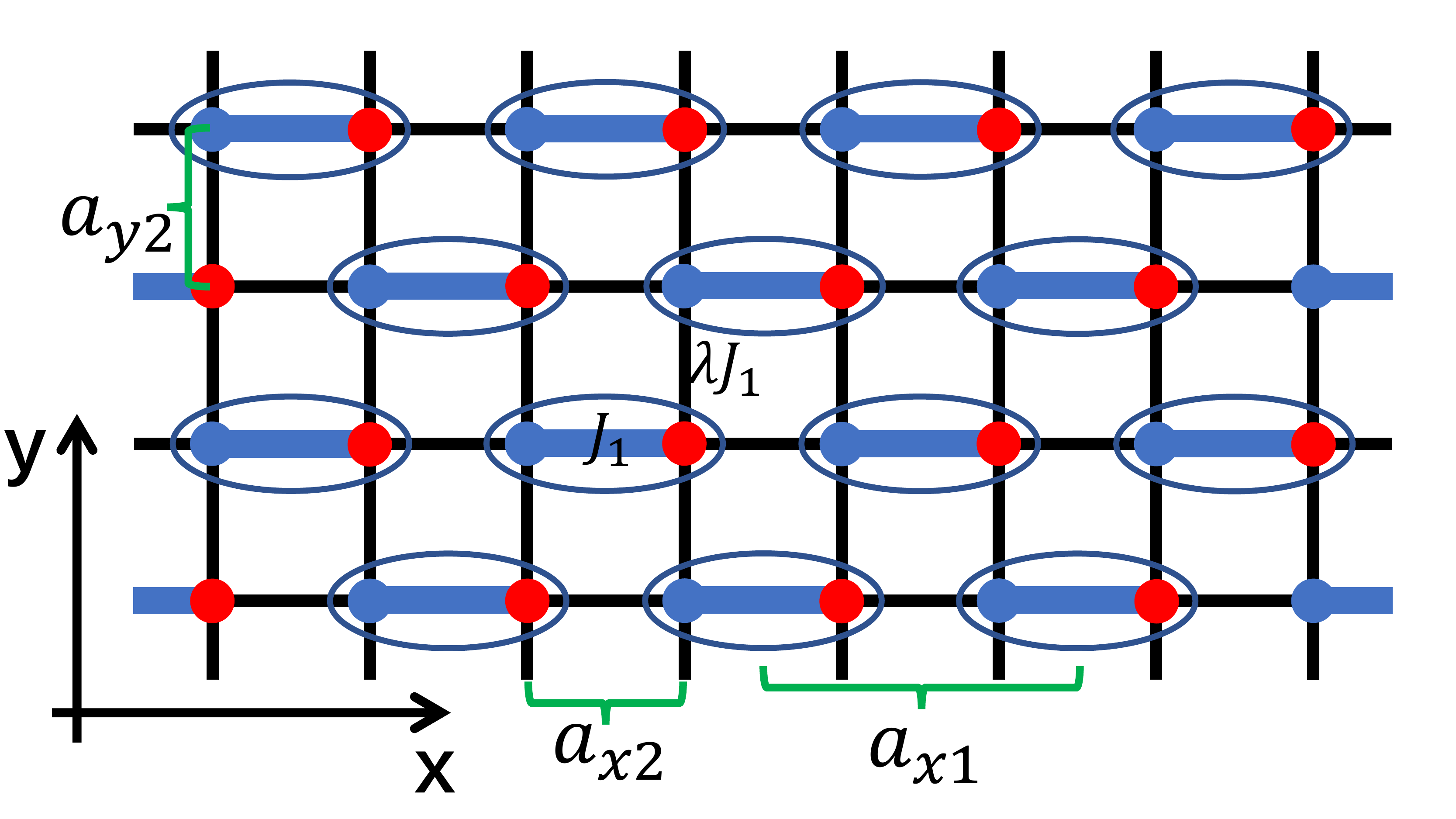}}}
\caption{Dimer ordering of the square lattice Heisenberg antiferromagnet. Two possible choices of the ordering pattern are illustrated. The intrabond exchange strength is given by $J_{1}$ and the interbond strength ratio $\lambda=J_{2}/J_{1}$. For the columnar dimer model (CDM) the critical interbond coupling strength can take values of $g_c=1.90951(1)$  \cite{PhysRevLett.121.117202}, $g_c=1.90948(4)$ \cite{doi:10.1063/1.3518900} and  $g_c=1.90947(3)$ \cite{PhysRevE.88.061301}. As for the staggered dimer model (SDM), $g_c=2.51943(1)$  \cite{PhysRevLett.121.117202} and $g_c=2.51941(2)$ \cite{PhysRevE.88.061301}. The lattice constants are defined as $a_{x}=2a$, $a_{y}=a$, $a_{x_{1}}=2a$, $a_{x_{2}}=a$, $a_{y_{2}}=a$.} 
\label{fig:dimer}
\end{figure}

A RIXS study on the 2D spin dimer model is valuable and necessary since it will allow the condensed matter and material science community to compare and contrast the features between magnon and triplon excitation. Till date, theory, computation, or experimental studies have focused on two-leg spin ladders. However, extended to a 2D system, the square lattice dimer system is capable of supporting a greater variety of lattice symmetries than a two-leg spin ladder. For example, the columnar dimer and the staggered dimer phases~\cite{PhysRevB.65.014407,PhysRevB.79.014410,PhysRevLett.101.127202,PhysRevB.40.10801,PhysRevB.41.9323,PhysRevB.63.104420,PhysRevB.44.12050}, see Fig.~\ref{fig:dimer}. Thus, as a starting point of our investigation, we choose the columnar dimer and the staggered dimer phases. Consequently, it's beneficial to study the RIXS response for a 2D dimerized lattice even though it has not been naturally realized in a material, at present. 

In this article, we analyze the indirect $K$ -edge and direct $L_{3}$ -edge RIXS response of the one- and the two- triplon excitations within a non-interacting bond-operator representation theory. At the $K$ -edge, we computed the RIXS intensity for the columnar and the staggered phases based on the UCL approximation~\cite{EPL.73.121,PhysRevB.77.134428}. The two-triplon excitation is the sole contribution at the $K$-edge. The RIXS response from both the phases follow the two-triplon density of states (DOS) very closely with the energy range more or less dictated by the DOS continuum. We find that the two-triplon excitation boundaries lie within an overall energy range of 1.7$J_2$ to 9$J_2$. Our calculation is general enough to be applicable to any system that can support dimers. Based on the previously stated examples, the cuprates are the most likely candidates, at present. The most intense RIXS signal lies primarily along the lower boundary which tracks the two-triplon dispersion continuum. The $K$ - edge two-triplon signal disappears at the $\Gamma$ point similar to the $K$ -edge bimagnon response~\cite{LuoPhysRevB.92.035109}. The peak intensity at the $K$ -edge occurs around the $(\pi/2,\pi/2)$ point for both the columnar and the staggered phases.

For the $L_{3}$-edge, we derived the scattering operator for a local dimer using the dipole transition approximation scheme. Our calculations considered the effects of the incident x-ray photon polarization ($\pi$ and $\sigma$) and optimal scattering geometry. Akin to the two-leg spin ladder system, the $L_3$ - edge has non-zero one- and two-triplon excitation RIXS contribution. The $L_3$-edge RIXS intensity has a sine or a cosine squared momentum dependence arising from the presence of the dimerized structure. Such an intensity modulation for the one- and two- triplon excitation spectra in RIXS is consistent with INS~\cite{PhysRevLett.98.027403}. Thus, RIXS could be used to select the different kinds of triplon excitation signal within the square lattice magnets. In the limit of strong rung interaction our RIXS intensity shows features similar to the spin ladder RIXS response. We also find a non-zero intensity at the $\Gamma$ point consistent with several ladder experiments. For our 2D system, this signal originates from the fluctuation of the hard-core constraint. Based on our calculations on the square lattice, we can offer an explanation for the existence of the two-triplon signal at zero momentum transfer in a two-leg spin ladder system. We find that the physical origin of this phenomenon is independent of the quasi-1D or 2D nature of the underlying lattice.

This paper is organized as follows. In Sec.~\ref{sec:model} we introduce our model Hamiltonian and the bond-operator representation. In Sec.~\ref{sec:kedge} and ~\ref{sec:ledge} we state our $K$ and $L_3$ - edge RIXS formalism, respectively. In Sec.~\ref{sec:analysis} we present our results and discussion on $K$ (Sec.~\ref{sec:kedgedisc}) and $L_{3}$ (Sec.~\ref{sec:l3edgedisc}) -edge RIXS spectra. In Sec.~\ref{sec:conclusion} we state our conclusions. In Appendix ~\ref{appendix A} we present the derivation of the $L_{3}$-edge RIXS scattering operator for the dimer system considering both angular and polarization effects. In Appendix ~\ref{sec:sladder} we describe the  $L_3$-edge RIXS features of the two-leg spin ladder model obtained from the limiting case of the columnar dimer lattice.
\section{Model}\label{sec:model}      
We consider a spin $S\!=\!\frac{1}{2}$ quantum magnet in a spontaneously dimerized quantum-disordered phase \cite{PhysRevB.41.9323}. The Hamiltonian for our study can be written as
  \begin{equation}
   H_{D}=J_{1}\sum_{<ij>\in D} {\bf S}_{i}\cdot {\bf S}_{j}+\lambda J_{1} \sum _{<ij>\notin D} {\bf S}_{i}  \cdot {\bf S}_{j} \label{Hamiltonian},
  \end{equation}
where $\left\langle ij \right\rangle$ denotes nearest-neighbor (NN) sites and $D$ refers to a specific choice of the dimer pattern, columnar or staggered. $J_1$ represents the intrabond exchange strength and $J_{2}=\lambda J_{1}$ is the interbond exchange strength. 

The competing intra- and inter- bond strength determines the ordered and the disordered states. In the limit $\lambda=1$, see Fig.~\ref{fig:dimer} for definition, the Hamiltonian is an isotropic Heisenberg model with a N\'{e}el ordered state. In the opposite $\lambda \rightarrow 0$ limit the ground state is the singlet product state formed out of two spins connected to the valence bonds. As the anisotropy coupling is tuned from zero to one, a quantum phase transition from the disordered state to the N\'{e}el ordered state can happen. We are interested in the disordered state in the regime $0< \lambda < \lambda_{c}$ where the valence bond solid breaks the lattice symmetry but retains the underlying spin-rotational symmetry~\cite{PhysRevB.41.9323}. Our parameter choice $g=1/\lambda =3$ can support either a columnar or a staggered phase \cite{PhysRevB.41.9323}. We study both phases in this paper. In this case the system should yield a singlet background and the triplets could disperse through the dimer sites against the singlet background. As mentioned in the introduction, the broken singlet bonds disperse as mobile $S=1$ spin triplet excitations called the triplon. 

The bond-operator technique developed by Sachdev and Bhatt offers a theoretical formalism to describe the triplon excitation~\cite{PhysRevB.41.9323}. Within this approach the singlets and triplets are treated as hard-core bosons on the dimer lattice. Under this formulation singlet bosons can be considered as a condensed mode in the dimerized system. To implement the bond-operator formalism \cite{PhysRevB.41.9323}, we define
  \begin{eqnarray}
  S_{1\alpha}&=&\frac{1}{2} (s^{\dagger}t_{\alpha}+t_{\alpha}^{\dagger}s-i\epsilon_{\alpha \beta \gamma} t_{\beta}^{\dagger}t_{\gamma}),  \\
  S_{2\alpha}&=&\frac{1}{2} (-s^{\dagger}t_{\alpha}-t_{\alpha}^{\dagger}s-i\epsilon_{\alpha \beta \gamma} t_{\beta}^{\dagger}t_{\gamma}),
  \end{eqnarray}
where $s$ ($s^{\dagger}$) is the annihilation (creation) operator for the singlet, $t_{\alpha}$ ($t^{\dagger}_{\alpha}$) is the  annihilation (creation) operator for the triplet, and $\alpha$ is equal to the components $x$, $y$, $z$. Under the singlet background condition, the singlet mode will condense and the triplets disperse against this background. We implement a mean-field theory treating the singlet operator as a $c$-number. Thus, $\langle s^{\dagger} \rangle = \langle s \rangle = \bar{s}$, which implies taking an average singlet amplitude on each dimer bond while the triplet operators are retained in the Hamiltonian. Meanwhile, the hard-core constraint $s^{\dagger}_{i}s_{i}+\sum_{\alpha}t_{i,\alpha}^{\dagger}t_{i,\alpha}=1$ should be applied on each dimer site. This introduces a $-\mu (s^{\dagger}_{i}s_{i}+\sum_{\alpha}t^{\dagger}_{i,\alpha}t_{i,\alpha})$ term into the Hamiltonian, where $\mu$ is the chemical potential. After completing the above steps, we can rewrite Eq. \eqref{Hamiltonian} at the quadratic level of triplet operators in the momentum space as
\begin{equation}
H_{2}=\sum_{\bf{k},\alpha} A_{\bf{k}}t_{\bf{k} \alpha}^{\dagger}t_{\bf{k}\alpha}+\frac{1}{2}\sum_{\bf{k},\alpha}B_{\bf{k}}(t_{\bf{k} \alpha}^{\dagger}t_{-\bf{k} \alpha}^{\dagger}+t_{\bf{k} \alpha}t_{-\bf{k} \alpha}),
  \end{equation}
where $A_{\bf k}$ and $B_{\bf k}$ depend on the spin coupling and the structure of the dimer sites. For the columnar dimer phase
   \begin{eqnarray}
   A_{\bf{k}}&=&\frac{J_{1}}{4}-\mu+\lambda J_{1}\bar{s}^{2}\left( -\frac{1}{2}\cos(k_{x} a_{x} )+\cos(k_{y} a_{y})\right), \label{eq:cak} \\
   B_{\bf{k}}&=&\lambda \bar{s}^{2} J_{1}\left( -\frac{1}{2}\cos(k_{x}  a_{x}) +\cos(k_{y}  a_{y})\right),\label{eq:cbk} 
   \end{eqnarray}
and for the staggered dimer phase
  \begin{eqnarray}
   &A_{\bf{k}}=\frac{J_{1}}{4}-\mu -\frac{1}{2}\lambda J_{1}\bar{s}^{2}\cos(k_{x} a_{x_{1}} )\nonumber \\
&\quad -\lambda J_{1}\bar{s}^{2}\cos(k_{x} a_{x_{2}})\cos(k_{y} a_{y_{2}}), \label{eq:sak}                  \\
   B_{\bf{k}}&=\lambda J_{1}\bar{s}^{2}\left(-\frac{1}{2}\cos(k_{x} a_{x_{1}} )-\cos(k_{x}a_{x_{2}})\cos(k_{y} a_{y_{2}}) \right ). \label{eq:sbk}  
    \end{eqnarray}
The Bogoliubov transformation, ~$\gamma_{\bf{k}\alpha }=u_{\bf{k}}t_{\bf{k}\alpha}+v_{\bf{k}}t_{-\bf{k}\alpha}^{\dagger}$  and  $\gamma_{\bf{k}\alpha }^{\dagger}=u_{\bf{k}}t_{\bf{k}\alpha}^{\dagger}+v_{\bf{k}}t_{-\bf{k}\alpha}$~, were utilized to diagonalize the Hamiltonian to obtain
 \begin{eqnarray}
  &H=E_{GS}+\sum_{\bf{k} \alpha} \omega_{\bf{k}} \gamma_{\bf{k}\alpha }^{\dagger}\gamma_{\bf{k}\alpha}^{}, \\
   &\omega_{\bf{k}}=\sqrt[]{A_{\bf{k}}^{2}-B_{\bf{k}}^{2}}. \label{omega}
 \end{eqnarray}
In the above, $E_{GS}$ stands for the ground state energy, $\gamma_{\bf{k} \alpha}$ ($\gamma_{\bf{k}\alpha }^{\dagger}$) is the triplon annihilation (creation) operator, while $\omega_{\bf{k}}$ is the dispersion relation for the triplon excitation. For our system, $u_{\bf{k}}$ and $v_{\bf{k}}$ are determined by 
 \begin{eqnarray}
   u_{\bf{k}}&=&\sqrt{\frac{A_{\bf{k}}}{2\sqrt{A_{\bf{k}}^{2}-B_{\bf{k}}^{ 2}}}+\frac{1}{2}},\\
   v_{\bf{k}}&=&sgn(B_{\bf{k}})\sqrt{u_{\bf{k}}^{2}-1},
   \end{eqnarray}
where $sgn(x)$ is the sign function. For a given value of $J_{1}$ and $\lambda$, we obtain the value of the self-consistency parameters $\mu$ and $\bar{s}$ by solving the saddle point equations 
 \begin{eqnarray}
       \frac{\partial E_{GS}}{\partial \bar{s}}&=&0, \\ \label{phps}
       \frac{\partial E_{GS}}{\partial \mu}&=&0.     \label{phpu}
 \end{eqnarray} 
We are interested in the RIXS spectrum of the triplon excitation for the columnar and staggered dimer phases, for reasons mentioned earlier. The disordered system in our study is far from phase transition (critical) point, where the physics is dominated by the NN valence bonds. Furthermore, triplon-triplon interactions have negligible influence ~\cite{PhysRevB.89.104415}.

\section{K-edge RIXS}\label{sec:kedge}
\begin{figure}[t]  
\centering
\hspace*{-2.25mm}\includegraphics[scale=0.26]{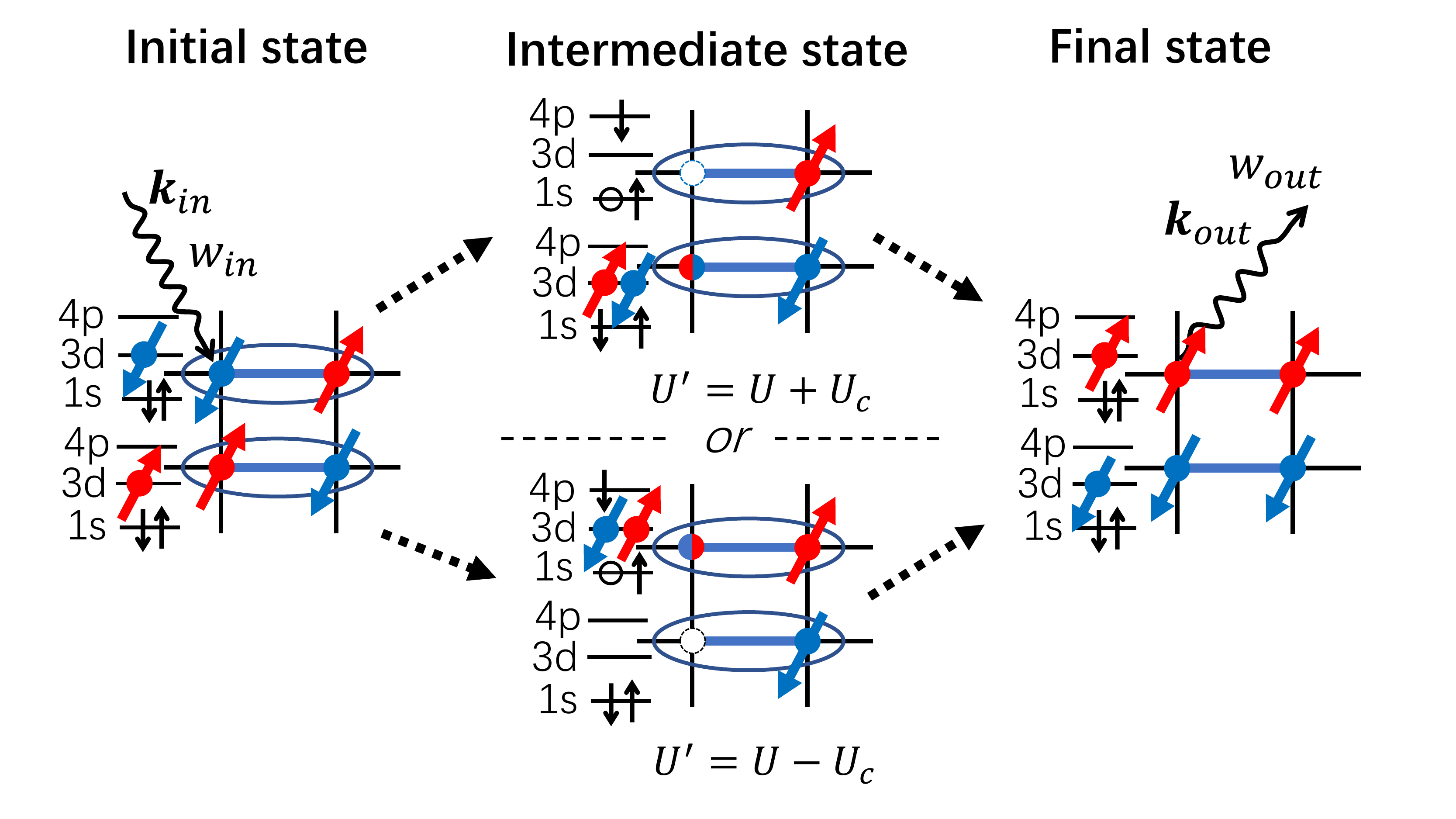}
\caption{Indirect $K$ -edge RIXS process for a dimerized system. The blue bond surrounded by the elliptical circle represents a singlet, while the blue bond without the elliptical circle implies a triplet excitation. The arrows indicate the spin state. A two-triplon excitation can occur via the superexchange process mediated by the core-hole potential $U_{c}$ as shown. Eq.~\eqref{eq:oqq2} is the expression for the $K$ -edge two-triplon scattering operator.}
\label{Kprocess}
\end{figure}

At the Cu $K$-edge, the transition of the excited electron can be described as $1s^{2}4p^{0} \rightarrow 1s^{1}4p^{1} \rightarrow1s^{2}4p^{0}$. Within the 4p-spectator approximation the dynamics of the promoted electron is neglected~\cite{PhysRevB.82.035113}. Also, the $3d$ electrons state is not directly affected. However, in the intermediate state the Coulomb interaction of the core-hole and the $3d$ electrons bring about changes to the exchange interaction. We use the ultrashort core-hole lifetime (UCL) expansion formalism in the indirect $K$-edge RIXS calculation. In this mechanism the core-hole potential $U_{c}$ in the intermediate state modifies the superexchange process \cite{EPL.73.121,PhysRevB.77.134428}. Since dimer systems are expected to appear in Cu- based Heisenberg magnets, we expect a similar mechanism to hold for our calculation. In Fig. \ref{Kprocess}, we show a couple of possible pathways to the triplet RIXS excitation which will disperse as the triplon through the lattice. 

We model the Hamiltonian in its intermediate state as
\begin{equation}
  H_{int}=H_{D}+\eta J_{1}\sum_{<ij>\in D} {\bf S}_{i}\cdot {\bf S}_{j}+\eta \lambda J_{1} \sum _{<ij>\notin D} {\bf S}_{i} \cdot {\bf S}_{j}, \label{H intermediate}
  \end{equation}  
where for simplicity of analysis we assumed that the core-hole mediated superexchange modification is the same along the intra- and the inter- bond interactions. Note, the intrabond interaction is along the  x-direction. But, the interbond coupling can be along the x- and the y- direction. In principle the overall scale factor $\eta$ is deduced from the ratio of $U$ and $U_c$. Even though the exchange bond strengths are different, we do not expect this to drastically modify the core-hole mediated prefactors to invalidate the results of the current RIXS analysis. The $K$ -edge RIXS operator is defined as
  \begin{equation}
\label{Orixs}
     \hat{O}_{\bf q}=\eta J_{1}\sum_{<ij>\in D} e^{i{\bf q}\cdot R_{i}}{\bf S}_{i}\cdot {\bf S}_{j}+\eta \lambda J_{1} \sum _{<ij>\notin D} e^{i{\bf q}\cdot R_{i}}{\bf S}_{i} \cdot {\bf S}_{j}.
  \end{equation}
Applying the bond-operator representation to Eq.~\eqref{Orixs} we obtain
\begin{eqnarray}
  \hat{O}_{\bf{q}}&=&\eta\sum_{{\bf k}\alpha}M_{{\bf k},{\bf q}} t_{{\bf k}+{\bf q}/2,\alpha}^{\dagger}t_{{\bf k}-{\bf q}/2,\alpha}  \nonumber 
\\&&\!+\! \eta \sum_{{\bf k}\alpha} N_{{\bf k},{\bf q}}\left(t_{{\bf k}+{\bf q}/2,\alpha}^{\dagger}t_{-{\bf k}+{\bf q}/2,\alpha}^{\dagger}\!+\!t_{{\bf k}-{\bf q}/2,\alpha}t_{-{\bf k}-{\bf q}/2,\alpha} \right), 
  \end{eqnarray}
where
 \begin{equation}
  M_{\bf{k},\bf{q}}=\frac{J_{1}}{2}\cos({q}_{x} \frac{{a}}{2})- 2\mu  + 2N_{\bf{k},\bf{q}}.
  \end{equation}
For the columnar dimer pattern
  \begin{eqnarray}
   N_{\bf{k},\bf{q}}&=&\frac {\lambda J_{1} \bar{s}^{2}}{4}( -\cos(k_{x}a_{x} +\frac{q_{x}}{2}a_{x}-{q}_{x}\frac{a}{2})  \nonumber \\
                    &\quad& -\cos(k_{x}a_{x} -\frac{q_{x}}{2}a_{x} +q_{x}\frac{a}{2})  \nonumber \\
                    &\quad&+2\cos(q_{x}\frac{a}{2})\cos(k_{y}a_{y} +\frac{q_{y}}{2}a)  \nonumber \\
                    &\quad&+2\cos(q_{x}\frac{a}{2})\cos(k_{y}a _{y}-\frac{q_{y}}{2}a),
  \end{eqnarray}
 and for the staggered dimer pattern
   \begin{eqnarray}
   N_{\bf{k},\bf{q}}&=&\frac {\lambda J_{1} \bar{s}^{2}}{4}( -\cos(k_{x}a_{x_{x}} +\frac{q_{x}}{2}a_{x_{1}}-{q}_{x}\frac{a}{2})  \nonumber \\
                    &\quad& -\cos(k_{x} a_{x_{1}} -\frac{q_{x}}{2} a_{x_{1}}+q_{x}\frac{a}{2})  \nonumber \\
                    &\quad&-2\cos(k_{x}a_{x_{2}} \!+\!\frac{q_{x}}{2} a_{ x_{2}}\!-\!q_{x}\frac{a}{2})\cos(k_{y}a_{ y_{2}} \!+\!\frac{q_{y}}{2} a_{y_{2}})  \nonumber \\
                    &&\!-\!2\cos(k_{x}a_{x_{2}} \!-\!\frac{q_{x}}{2} a_{ x_{2}}\!+\!q_{x}\frac{a}{2})\cos(k_{y} a_{ y_{2}} \!-\!\frac{q_{y}}{2} a_{y_{2}}).
  \end{eqnarray}
In the above equations, $q_{x}$ ($q_{y}$) is the projection of the momentum transfer vector along the $x$ ($y$) direction, respectively. Next, we apply the Bogoliubov transformation to the Fourier transformed operator. Subsequently, we decompose the scattering operator $\hat{O}_{\bf q}$ into its one- and two-triplon scattering contribution given by
  \begin{eqnarray}
\hat{O}_{\bf q}^{(1)}&=& \sum_{{\bf k},\alpha}\eta[M_{{\bf k},{\bf q}}( u_{{\bf k} +{\bf q}/2}u_{{\bf k}-{\bf q}/2}+v_{{\bf k} +{\bf q}/2}v_{{\bf k} -{\bf q}/2})  \nonumber \\
&\quad& \quad -2N_{{\bf k},{\bf q}}( u_{{\bf k}+{\bf q}/2}v_{{\bf k} -{\bf q}/2} +v_{{\bf k} +{\bf q}/2}u_{{\bf k} -{\bf q}/2})]\nonumber \\
&\quad&  \quad \gamma_{{\bf k} +{\bf q}/2,\alpha}^{\dagger}\gamma_{{\bf k}-{\bf q}/2,\alpha}^{},
  \end{eqnarray}
and
  \begin{eqnarray}
\label{eq:oqq2}
   \hat{O}_{{\bf q}}^{(2)}=&&\sum_{{\bf k},\alpha}\eta [-M_{{\bf k},{\bf q}}u_{{\bf k}+{\bf q}/2}v_{{\bf k}-{\bf q}/2}+ \nonumber     \\
   && \quad N_{{\bf k},{\bf q}}( u_{{\bf k} +{\bf q}/2}u_{{\bf k} -{\bf q}/2} +v_{{\bf k} +{\bf q}/2}v_{{\bf k}-{\bf q}/2})]\nonumber  \\
   && \quad (\gamma_{{\bf k}+{\bf q}/2,\alpha}^{\dagger}\gamma_{-{\bf k}+{\bf q}/2,\alpha}^{\dagger} +\gamma_{{\bf k}-{\bf q}/2,\alpha}^{}\gamma_{-{\bf k}-{\bf q}/2,\alpha}^{}).
  \end{eqnarray}

We calculated the cross section using the Kramers-Heisenberg formula \cite{RevModPhys.83.705}
   \begin{equation} 
      \frac{d^{2} \sigma}{d \Omega d \omega} \biggr\vert_{res} \propto I= \left\langle \sum_{f} \vert A_{fi} \vert^{2} \delta(\omega -\omega_{fi})\right\rangle. \label{Ik}
   \end{equation}
The scattering amplitude is given by
    \begin{equation}
     A_{fi}=\omega_{res} \sum_{n} \frac{\langle f \vert \hat{D}\vert n\rangle \langle n \vert \hat{D}\vert i\rangle}{\omega_{i}-E_{n}-i\Gamma},
    \end{equation}
where $\vert i \rangle$  ($\vert f \rangle$) denotes the initial (final) state. $\Gamma$ represents the core-hole lifetime broadening in the $K$-edge indirect RIXS process, $\hat{D}$ is the dipole transition operator, $\vert n \rangle$ denotes the intermediate states, while $E_{n}$ is the eigenvalue of the intermediate state. The core-hole lifetime broadening at the $K$ -edge is given by $\Gamma=750$ meV, assuming $J_{1}/\Gamma\approx 5$ \cite{PhysRevB.77.134428}. We assume the initial state $\vert i \rangle$ is the ground state of the system without any triplon excitation. Hence $ H_{D} \vert i\rangle =E_{i}\vert i\rangle =E_{GS}\vert i\rangle$. For simplicity we choose the ground state energy as the reference energy,  $E_{GS}=0$. The energy of the incident photons is $\omega_{i} $, while $\omega_{res}$ stands for the resonance energy. Using the UCL approximation $A_{fi}$ simplifies to \cite{EPL.73.121}  
  \begin{equation}
\label{afi}
  A_{fi}=\frac{\omega_{res}}{i\Gamma} \frac{1}{i\Gamma +\omega} \langle f\vert \hat{O}_{\bf{q}} \vert i \rangle.
  \end{equation}
Note, at zero temperature, $\hat{O}_{\bf q}^{(1)} \vert i \rangle=0$. The two-triplon excited state $\gamma_{{\bf k}+{\bf q}/2,\alpha}^{\dagger}\gamma_{-{\bf k}+{\bf q}/2,\alpha}^{\dagger}~\vert 0 \rangle$ is generated through  $\hat{O}_{\bf q}^{(2)}$. Hence, the intensity is given by
\begin{equation}
I({\bf q},\omega)=\sum_{{\bf k},\alpha} \vert A_{fi}\vert ^{2} \delta(\omega - \omega_{{\bf k}+{\bf q}/2}-\omega_{-\bf{k}+{\bf q}/2}) \label{IK},
\end{equation}
where only $\hat{O}_{\bf q}^{(2)}$ contributes. The incident photons at Cu $K$-edge carry energy up to 8979 eV. Furthermore, these photons at the $K$-edge carry quite large momentum. Thus, the first Brillouin zone can be comprehensively probed. The coupling constant for the nearest-neighbor superexchange interaction lies between 120~-~150 meV. We took the intrabond exchange $J_{1}=138$ meV in our calculations ~\cite{PhysRevLett.86.5377}. The interbond strength of $J_{2}=46$~meV is one-third of the intrabond strength since $\lambda=1/3$. The choice of the interbond strength is guided by the phase diagram of the dimerized square lattice Heisenberg magnet~\cite{PhysRevB.41.9323}. For simplicity, we set the overall energy scale factor $\omega_{res}$ and $\eta$ as unity in the $K$ -edge intensity calculation.

\section{$L_3$ -edge RIXS}\label{sec:ledge}
Magnetic excitation at $L$-edge RIXS process has been widely studied \cite{PhysRevLett.102.167401,PhysRevLett.103.117003}. We study the $L_{3}$ -edge RIXS spectrum of the 2D dimer system. For a pure magnetic excitation under the dipole approximation the scattering amplitude can be written in an effective form as \cite{PhysRevLett.103.117003}
   \begin{equation}
   A_{f i}=\frac{1}{\Delta} \langle f \vert  \sum_{i} e^{i\bf{q}\cdot \bf{R_{i}}}(1-r_{\bf{q}}+r_{\bf{q}}S_{i}^{x}) \vert i \rangle.
   \end{equation} 
Using a similar approach, we can derive the scattering operator for a local dimer within the first term of the UCL expansion, see Appendix \ref{appendix A} for derivation details. 

\begin{figure}[t]
\centering
\includegraphics[scale=0.24]{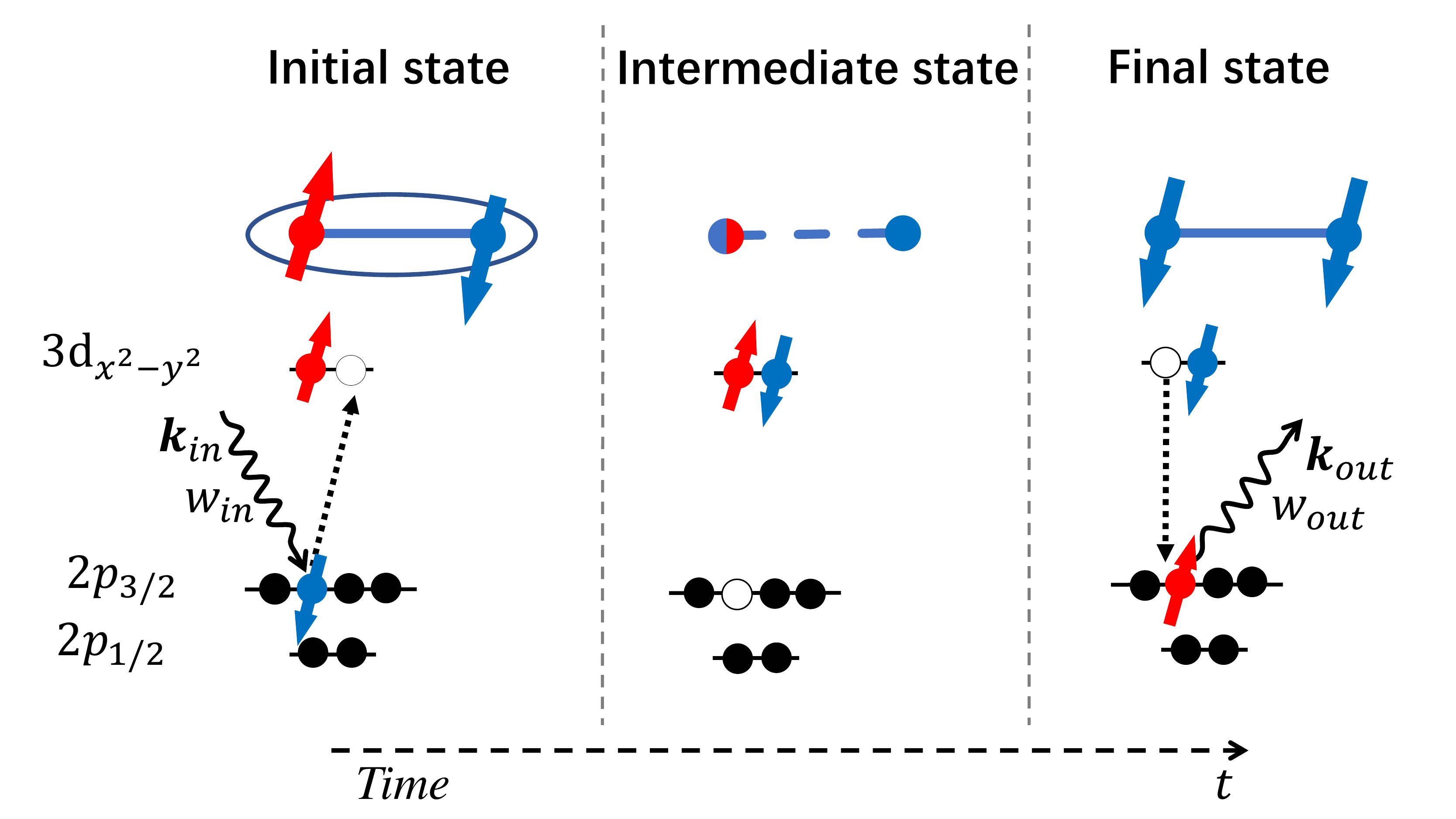}
\caption{The diagram for a $L_{3}$ -edge RIXS process. The circles stand for the electrons which are filled in different orbitals at Cu sites while the arrows indicate the spin state. The local triplet excitation could happen in the direct RIXS process.}
\label{fig3} 
\end{figure}

In Fig.~\ref{fig3}, we describe the $L_3$ -edge RIXS process for a dimer. Considering the scattering of the two local spins on one dimer in the direct RIXS process, we can write the scattering operator for the local dimer site in the hole representation as~\cite{1367-2630-13-4-043026} 
    \begin{equation}
    \hat{O}^{\epsilon_{i}}=\frac{1}{i \Gamma}\sum_{\epsilon_{f}}\sum_{i=1,2}T_{s}(\epsilon^{f},\epsilon^{i})\hat{O}_{s_{i}}+\frac{1}{i \Gamma}\sum_{\epsilon_{f}}T_{d}(\epsilon^{f},\epsilon^{i})\hat{O}_{d},
    \end{equation}
where
    \begin{eqnarray}
    \hat{O}_{s_{i}}&=&\sin{\theta_{s}}\cos{\phi_{s}}S_{i}^{x}- \sin\theta_{s}\sin\phi_{s}S_{i}^{y}+\cos{\theta_{s}}S_{i}^{z},  \label{Osi}\\
    \hat{O}_{d}    &=&\sum_{j=1,2}s_{m_{j}}^{\dagger}s_{m_{j}}+\sum_{\alpha ,j=1,2} t_{m_{j} \alpha}^{\dagger}t_{m_{j} \alpha}.
\label{Od}
    \end{eqnarray}
The RIXS operator $\hat{O}^{\epsilon_{i}}$ includes polarization dependence factor. $T_{\text{s}}(\epsilon^{f},\epsilon^{i})$ and $T_{\text{d}}(\epsilon^{f},\epsilon^{i})$ stands for the polarization factor for the experimental geometry, see Eqs.~\eqref{A6} - \eqref{A11} in Appendix \ref{appendix A}. $\theta_{s}$  is the polar angle of the spins and $\phi_{s}$ is azimuthal angle of the spins in spherical coordinates. 

The contribution arising from $\hat{O}_{s_{i}}$ corresponds to the spin contribution. For the $\hat{O}_{d}$ part there are a couple of possible scenarios. In the first scenario, on grounds of strict hard-core constraint implementation the entire equation just becomes unity, see Table \ref{Table 1}. This is the trivial perspective within our RIXS calculation. However, our calculation is being performed in a singlet background and fluctuations in the number of singlet or triplet terms could occur (quantum fluctuation or interaction effects). Thus, even though this term appears as a strict constraint like expression, the sum of the singlet and the triplet may not always be unity (constraint softening). Hence, the triplet part could and does have a finite RIXS intensity contribution, see Fig.~\ref{fig6} and Fig.~\ref{fig7}. This implies that $\hat{O}_{d}$ could detect the weakening of the local hard-core constraint.

We extend the local scattering process to collective excitation modes. At zero temperature we find three distinct contributions to the RIXS intensity spectrum $-$ a one-triplon contribution ($\hat{O}_{\bf q}^{(s1)}$) and a couple of two-triplon operators ($ \hat{O}_{\bf q}^{(s2)}$ and $ \hat{O}_{\bf q}^{(d2)}$). Here,  $\hat{O}_{\bf q}^{(s1)}$ and $ \hat{O}_{\bf q}^{(s2)}$ originate from $ \hat{O}_{s_{i}}$, see Eq.~\eqref{Osi}.~$\hat{O}_{\bf q}^{(d2)}$ originates from $\hat{O}_{d}$, see Eq.~\eqref{Od}.
\begin{figure}[t]  
\centering
{\subfigure[]{\label{fig:experiment}
\includegraphics[scale=0.2]{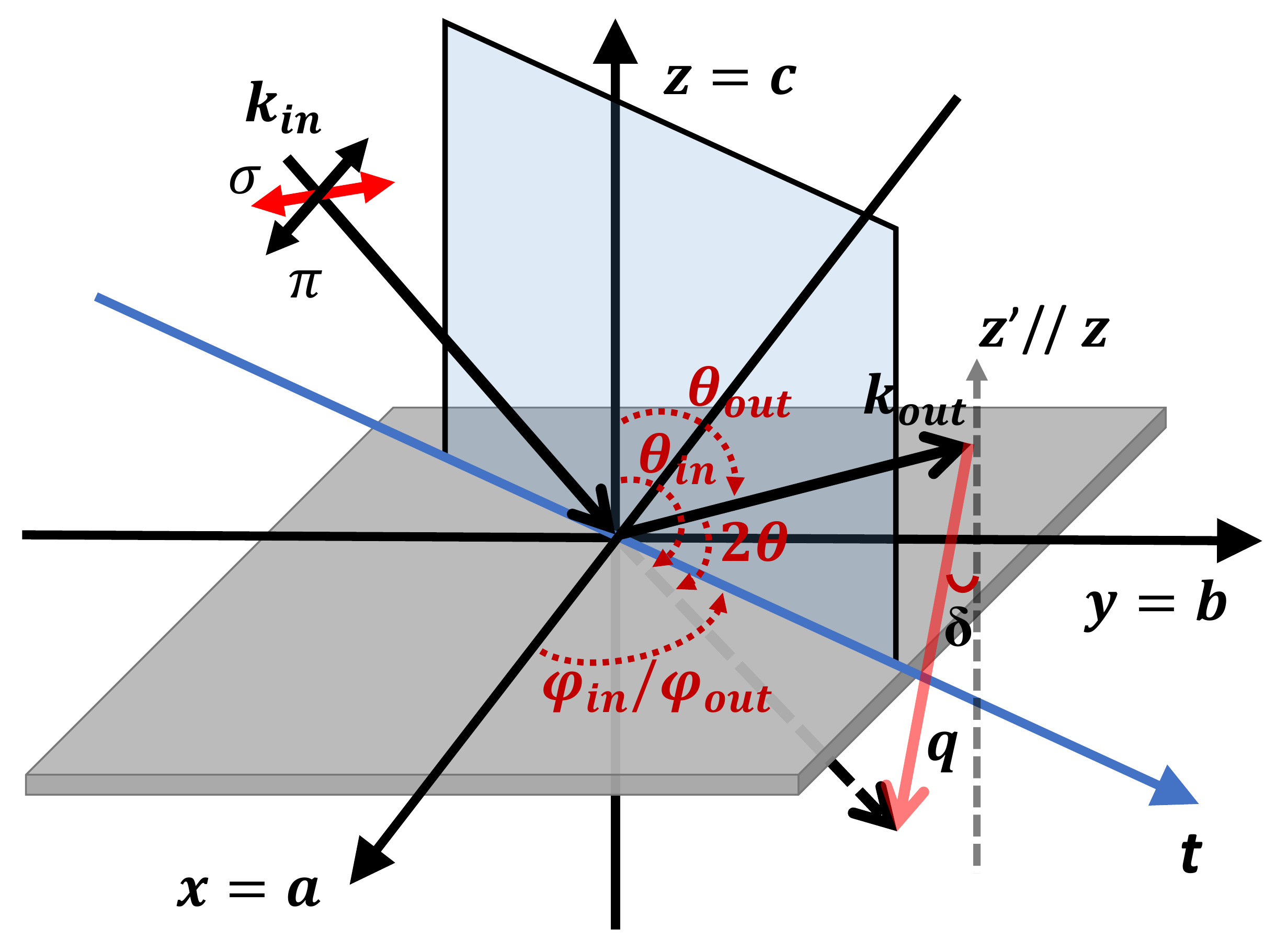}}}
{\subfigure[]{\label{fig:boundary}
\includegraphics[scale=0.14]{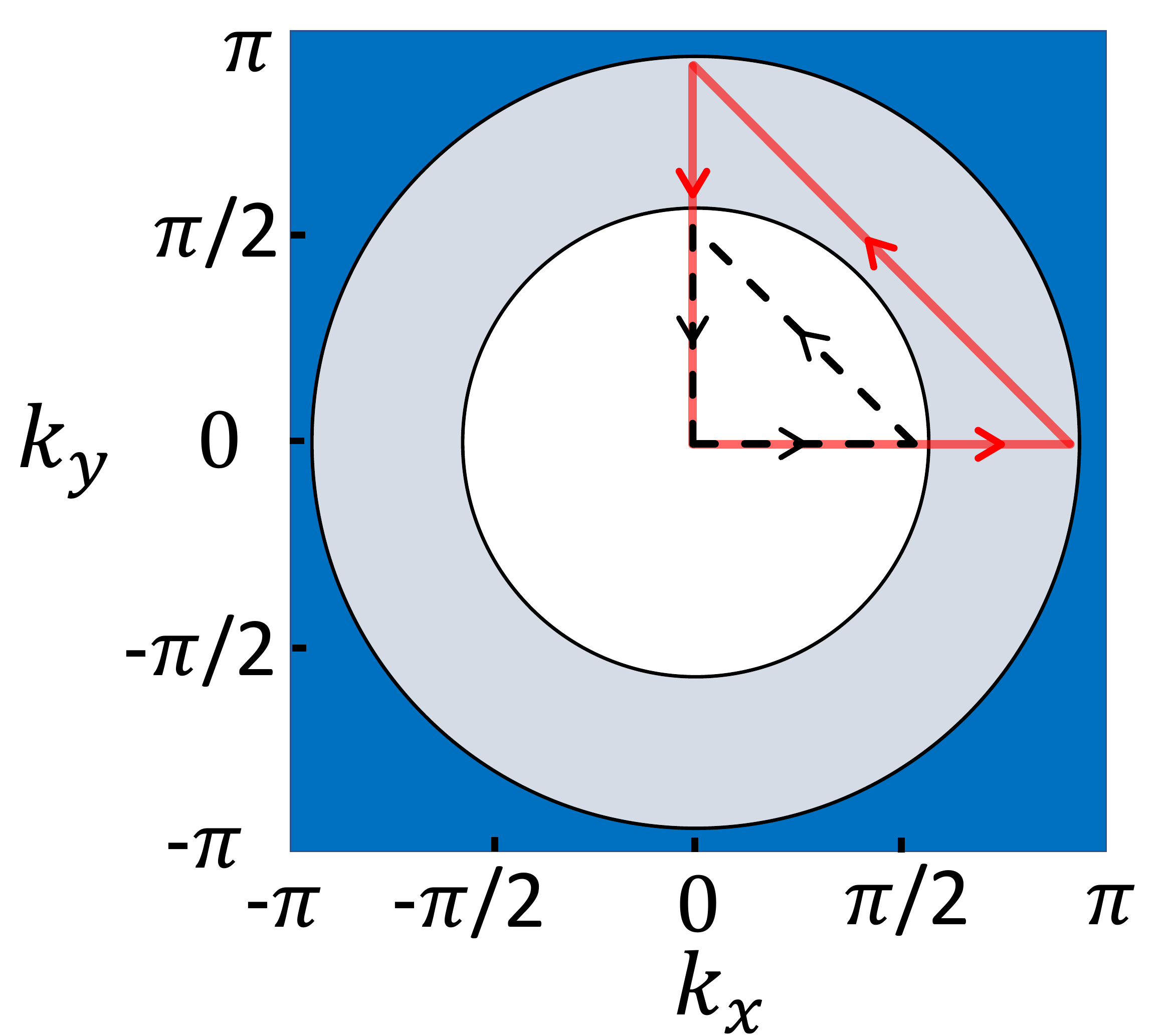}}}
\label{fig:Experiment&boundary}
\caption{(a) Experimental scattering geometry set up for $L$ -edge dimer RIXS. The sample is fixed on the gray plane while the scattering plane is shaded light blue. The dark blue line $t$ is the intersection line of the sample plane and the scattering plane. ${\bf k}_{in}$ (${\bf k}_{out}$) is the momentum of the incident (outgoing) photons and ${\bf q}$ is the momentum transfer to the material. $\theta_{in}$ ($\theta_{out}$) is the angle between the $z$ -axis and the vector ${\bf k}_{in}$ (${\bf k}_{out}$), while $2\theta$ is the scattering angle. $\phi_{in}$ and $\phi_{out}$ is the angle between the $t$ and the $x$ -axis. The scattering plane is perpendicular to the sample plane, which means $\phi_{in} = \phi_{out}$. $\delta$ is the angle between momentum transfer vector and $z$ -direction. (b) Detectable area and the momemtum loops at the $L_3$ -edge. The white area stands for the momentum zone that can be detected under  $2\theta =90^{\circ}$ geometry, while the white and gray area stands for the momentum range that can be detected under $2\theta=130^{\circ}$. The blue area is not accessible for the $L_{3}$ -edge x-ray experiments.}
\end{figure}
In the absence of a magnetic field the triplon excitation is triply degenerate, hence the intensity has spherical symmetry. For simplicity, we take the spin orientation as $(\theta_{s}, \phi_{s})=(\frac{\pi}{2},0)$. The resulting individual scattering operator components are given by
     \begin{eqnarray}
     \hat{O}_{\bf{q}}^{(s1)}&=& -2i\bar{s}\sin\left({q_{x}} \frac{a}{2} \right)  (u_{\bf{q}}\!-\!v_{\bf{q}})(\gamma_{{\bf q},x}^{\dagger}\!+\!\gamma_{-{\bf q},x}),   \label{Oqs1}      \\
\hat{O}_{\bf{q}}^{(s2)}&=& \sum_{{\bf{k}}}  2i\cos\left({q_{x}}\ \frac{a}{2}\right)(u_{{\bf{k}}+{\bf{q}}/2}v_{{\bf{k}}-{\bf q}/2}-u_{{\bf k}-{\bf q}/2}v_{{\bf k}+{\bf q}/2})\nonumber \\&& \quad \quad \gamma_{{\bf k}+{\bf q}/2,y}^{\dagger}\gamma_{-{\bf k }+{\bf q}/2,z}^{\dagger},  \label{Oqs2}  \\   
     \hat{O}_{\bf q}^{(d2)}&=&-2\sum_{{\bf k},\alpha}\cos\left(q_{x}\frac{a}{2}\right)u_{{\bf k}+{\bf q}/2}v_{{\bf k}-{\bf q}/2} \gamma_{{\bf k}+{\bf q}/2,\alpha}^{\dagger}\gamma_{-{\bf k}+{\bf q}/2,\alpha}^{\dagger}. \label{Oqd2} 
     \end{eqnarray}
Hence, the corresponding amplitude and intensity are given by
       \begin{equation}
         A_{f i}^{(\xi)}=\frac {1}{i \Gamma}\langle f \vert \hat{O}_{\bf{q}}^{(\xi)}\vert i \rangle, \label{Axi}
      \end{equation} 
          \begin{eqnarray}
       I_{\xi}({\bf q},\omega)&=&\sum_{f}\vert A_{fi}^{(\xi)}\vert ^{2} \delta(\omega-\omega_{f}).   \label{Iqxi}
       \end{eqnarray}
The label $\xi$ refers to $s1$, $s2$, and $d2$. For the one-triplon channel, $\omega_{f}=\omega_{{\bf q}}$, while for the two-triplon channel $\omega_{f}=\omega_{{\bf k}+{\bf q}/2}+\omega_{-{\bf k}+{\bf q}/2}$. We find that the two-triplon intensity will have contributions from the pure spin and the triplet parts of the RIXS operator. 

We can conclude from Eqs.~\eqref{Oqs1}~-~\eqref{Oqd2} that there is a $\sin^{2}(q_{x}a/2)$ momentum dependence for the one-triplon excitation and a $\cos^{2}(q_{x}a/2)$ momentum dependence for the two-triplon excitation. This complimentary intensity modulation feature has already been found in INS \cite{PhysRevLett.98.027403}. Thus, based on our calculations we find that the RIXS intensity depends not only on the polarization dependence, but also on the $\sin^{2}(q_{x}a/2)$ and $\cos^{2}(q_{x}a/2)$ factors. This characteristic feature for the triplon excitation originates from the broken lattice symmetry created by the dimer structure.

For a realistic comparison to RIXS experimental setup we need to consider the polarization dependence effect. $T(\epsilon^{f},\epsilon^{i})$ depends on the polarization of both the incident and the out-going photons. From the local $d\!-\!d$ excitation process we can conclude that polarization dependence has the rotational symmetry of the $c$-axis, that is $\phi_{in}$ and $\phi_{out}$ do not contribute to the polarization effect. The polarization factor is connected with the incident angle $\theta_{in}$ as well as the scattering angle $2\theta$. The total dimer RIXS intensity spectrum is given by
\begin{eqnarray}
     I({\bf q},\omega,\epsilon^{f},\epsilon^{i}) &=&\vert T_{s}(\epsilon^{f},\epsilon^{i})\vert ^{2}I_{s1}({\bf q},\omega)+\vert T_{s}(\epsilon^{f},\epsilon^{i})\vert ^{2}I_{s2}({\bf q},\omega) \nonumber \\
 &\quad&+\vert T_{d}(\epsilon^{f},\epsilon^{i})\vert ^{2}I_{d2}({\bf q},\omega)  \label{full intensity}.
     \end{eqnarray}
$T_{\text{s}}(\epsilon^{f},\epsilon^{i})$ and $T_{\text{d}}(\epsilon^{f},\epsilon^{i})$ reflect the polarization effect which modulates the $\hat O_{s}$ and $\hat O_{d}$ operator respectively, see Eqs. \eqref{A6}~-~\eqref{A11}. In  momentum space RIXS intensity depends on both the geometry and the polarization factor. Fig.~\ref{fig:experiment} shows the experimental geometry. Fig.~\ref{fig:boundary} shows the boundary that can be detected by Cu $L_{3}$ -edge RIXS, which is limited by the constraint Eqs.~\eqref{momentum transfer:q}~-~\eqref{angle delta} (for calculation detail refer to Appendix~\ref{appendix A}). We compute the $L_{3}$ -edge spectrum for both the columnar (see Fig. \ref{fig6}) and the staggered (see Fig. \ref{fig7}) dimer phase which utilizes similar momentum transfer limitation and polarization conditions. The experimental resolution was set to 30 meV and the core-hole lifetime broadening $\Gamma$ to 300 meV.  

 \begin{figure*}[t]  
\centering
\hspace*{-2.25mm}\includegraphics[scale=0.5]{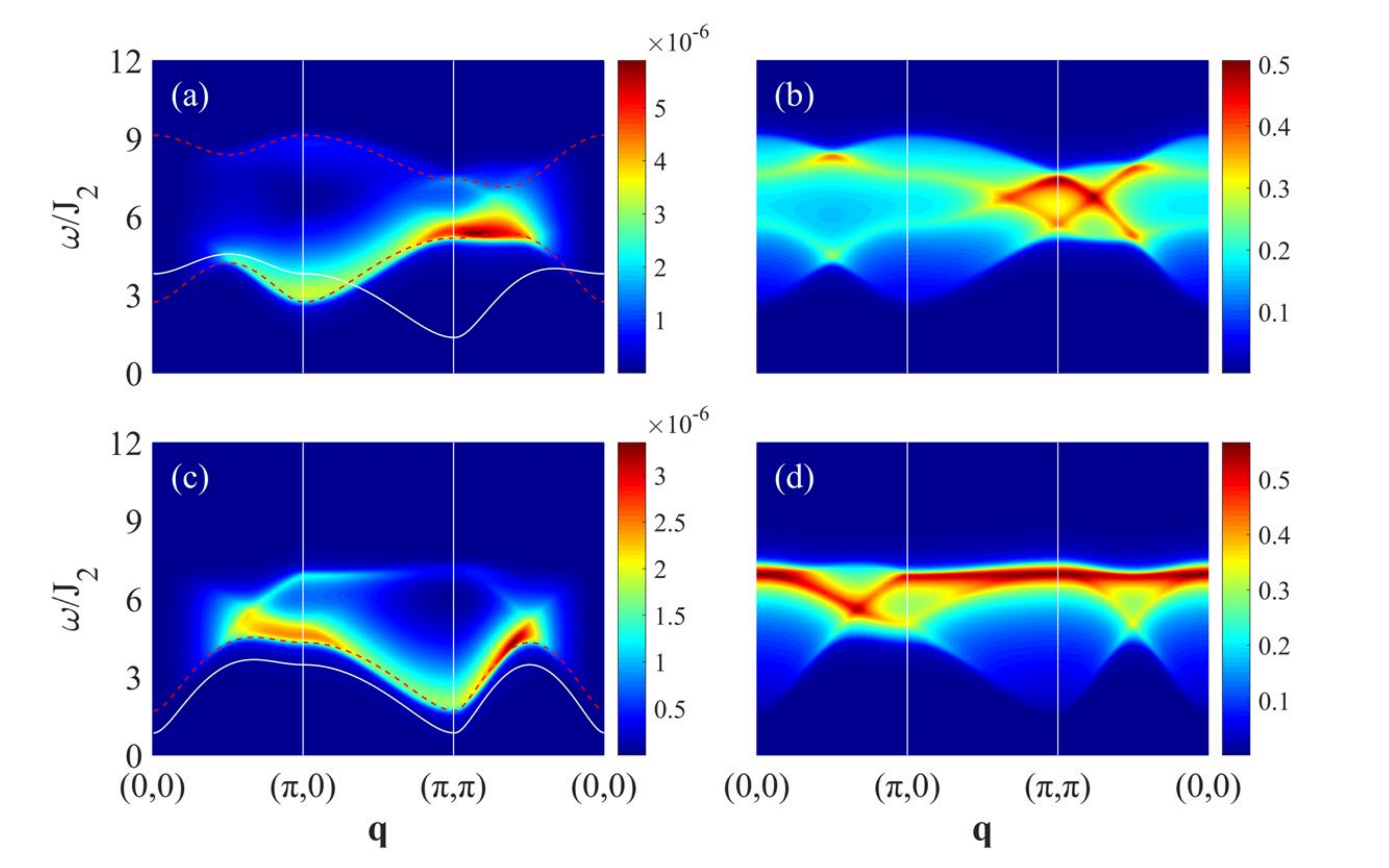}
\caption{$K$-edge RIXS response of two-triplon excitation in (a) columnar and (c) staggered dimer phase, respectively. Corresponding density of states (DOS) in (b) columnar and (d) staggered dimer phases, respectively. Solid white line represents one-triplon dispersion. Dashed red lines represent the upper and lower boundary of the two-triplon excitation. ${\bf q}$ represents momentum transfer. $\omega/J_2$ is scaled energy.}
\label{K intensity}
\end{figure*}
\section{Results and discussion} \label{sec:analysis}
\subsection{K-edge RIXS}\label{sec:kedgedisc}
In Fig.~\ref{K intensity}, we display the $K$-edge RIXS spectrum of the columnar and the staggered dimer ordering which were computed using Eq.~\eqref{IK}. The plots display the RIXS intensity, the DOS, and the boundaries for the one- and two-triplon dispersion. We {calculated the RIXS intensity along the momentum loop, $\Gamma (0,0) \! \rightarrow \! (\pi,0) \! \rightarrow \! (\pi,\pi)\! \rightarrow \!\Gamma (0,0)$. RIXS intensity is scaled by $J_{2}=\lambda J_{1}=1$.  The lower and upper boundaries of the two-triplon continuum is confined approximately between $2.7J_2 ~(1.7J_2)$ to $9J_2~(7.5J_2)$, which implies an energy range of 124 (78) meV - 414 (345) meV for the columnar (staggered) phase. At the $K$-edge polarization and geometry dependence can typically be factored out as an overall multiplicative constant. This dependence should not change the qualitative features of our conclusion. Additionally, if the x-ray edge is tuned to resonance based on x-ray absorption features, then the RIXS intensity factor can be enhanced \cite{NomuraPhysRevB.96.165128}.

The two-triplon dispersion boundaries were determined from the minimum and maximum energy values, as well as the corresponding wave vectors of the one-triplon dispersion, which is given by Eq.~\eqref{omega}. The lower boundary of the two-triplon continuum was tracked using $\omega_{2lower}= \omega (k_{x_0}, k_{y_0} )+\omega (-k_{x_0}+q_{x},-k_{y_0}+q_{y})$, where $\omega (k_{x_0},k_{y_0})$ is the energy gap. Similarly, the upper boundary was determined from the maximum value of $\omega (k_{x_0},k_{y_0})$. This method is valid when the one-triplon disperison has only a finite number of minimum or maximum values. For example, for the columnar dimer phase, when the energy gap of the one-triplon disperison is $\omega(0,\pi)$, the maximum energy of the one-triplon dispersion is given by $\omega(\pi/2,0)$. Thus, the lower boundary is $\omega_{2lower}=\omega (0, \pi )+\omega (q_{x},-\pi+q_{y})$ and the upper boundary is $\omega_{2upper}=\omega (\pi/2,0)+\omega (-\pi/2+q_{x},q_{y})$. For the staggered phase, the energy gap of the one-triplon dispersion is $\omega(0,0)$, which leads to $\omega_{2lower}(q_x,q_y)=\omega(0,0)+\omega(q_x,q_y)$ . Note, for the staggered phase there are a countless number of maximum values which make it difficult to track the upper boundary of the continuum in the staggered phase. Thus, we did not display the upper boundary. For all the other plots (including the $L_3$ -edge) we were able to successfully apply our technique.

We notice that the lower and upper boundaries of the two-triplon excitation RIXS signal closely follows the DOS, see the solid white lines and the dashed red lines in Fig.~\ref{K intensity}. However, the actual RIXS response is a convolution of the scattering amplitude and the energy conserving delta function. Thus, the intensity does not simply follow the DOS. Both for the columnar and the staggered phases the highest intensity is located around $(\pi/2,\pi/2)$. The next prominent intensity for the columnar phase appears at $(\pi,0)$, but for the staggered phase it is situated at $(\pi/2,0)$. Additionally, the excitation energy range seems to track the lower boundary of the DOS continuum. For both phases the intensity vanishes when ${\bf q}=0$, similar to the bimagnon excitation in the N\'{e}el state \cite{EPL.73.121,PhysRevB.77.134428}. We also note that for both spectra the non-zero intensity is confined between the momentum path $(\pi/2,0)$ to $(\pi/2,\pi/2)$. The single triplon dispersion is lower in energy for both phases for most parts along the chosen momentum path. But, in the columnar phase in vicinity of $(\pi,0)$ there are a range of momenta values where the two-triplon excitation is lower in energy. Since, the single triplon excitation is absent at the $K$ -edge due to conservation rules, we do not anticipate this to be an issue with experimental detection. For the staggered phase the lower boundary is consistently above the one-triplon line. 

\subsection{ $L_{3}$-edge RIXS}\label{sec:l3edgedisc}
 \begin{figure*}[t]  
\centering
\hspace*{-5mm}\includegraphics[scale=0.5]{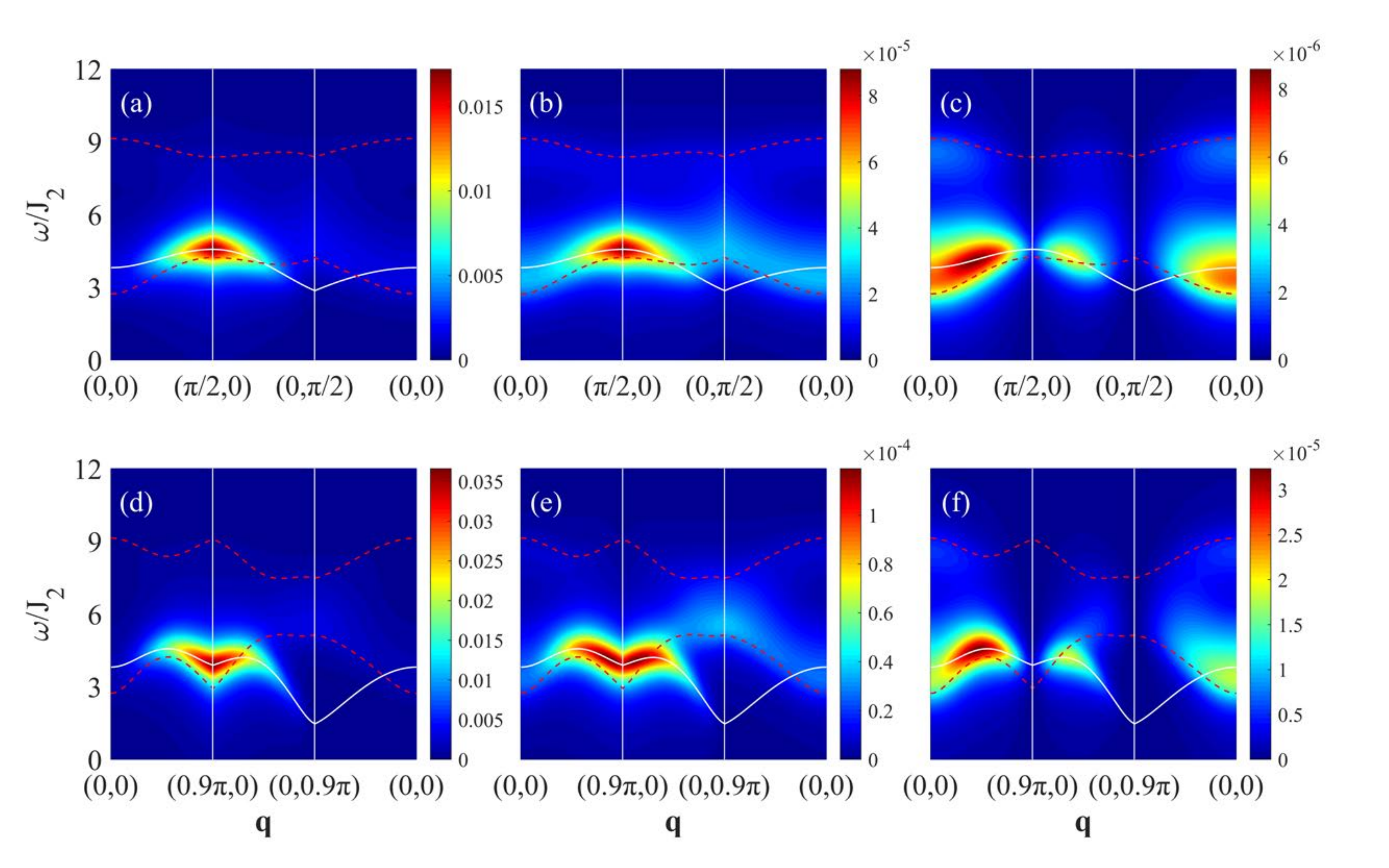}
\caption{ $L_{3}$ -edge RIXS spectrum $I(\bf{q},\omega)$ of columnar dimer pattern for Bragg scattering angle $2\theta=90^{\circ}$ [(a), (b), (c)] and $2\theta=130^{\circ}$ [(d), (e), (f)]. The three panels represent (a) and (d) without, (b) and (e) $\sigma$, and (c) and (f) $\pi$ beam polarization states. Solid white line represents one-triplon dispersion. Dashed red lines represent the upper and lower boundary of the two-triplon excitation. ${\bf q}$ represents momentum transfer. $\omega/J_2$ is scaled energy.} 
\label{fig6}
\end{figure*}
 \begin{figure*}[t]  
\centering
\hspace*{-5mm}\includegraphics[scale=0.5]{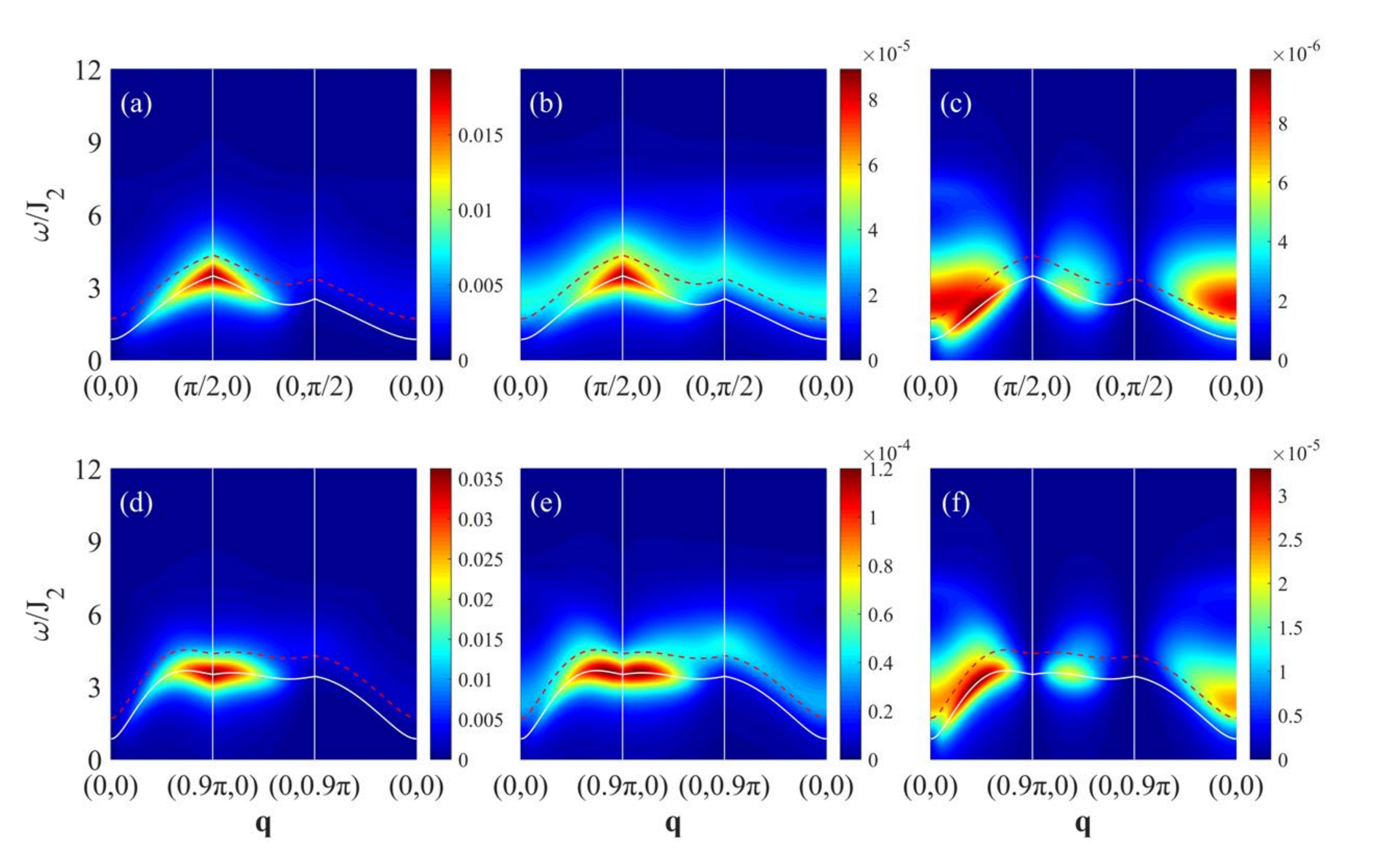}
\caption{ $L_{3}$ -edge RIXS spectrum $I(\bf{q},\omega)$ of staggered dimer pattern for Bragg scattering angle $2\theta=90^{\circ}$ [(a), (b), (c)] and $2\theta=130^{\circ}$ [(d), (e), (f)]. The three panels represent (a) and (d) without, (b) and (e) $\sigma$, and (c) and (f) $\pi$ beam polarization states. Solid white line represents one-triplon dispersion. Dashed red lines represent the lower boundary of the two-triplon excitation. ${\bf q}$ represents momentum transfer. $\omega/J_2$ is scaled energy.} 
\label{fig7}
\end{figure*}


In  Fig.~\ref{fig6}, based on Eqs.~\eqref{Iqxi}~-~\eqref{full intensity}, we display the results for the columnar dimer state along with the lower and upper boundary for the two-triplon dispersion computed using the same method described for the $K$ -edge plots. The momentum path choices for scattering angles $2\theta =90^{\circ}$ (dashed black path) and  $2\theta =130^{\circ}$ (solid red path) is guided by Fig.~\ref{fig:boundary}. However, limitations on the momentum transfer and the choice of lattice constant, prohibit the wave vector ${\bf k}=(0,\pi)$ from being accessed.  

We find that there will be a strong contribution from the one-triplon excitation (see the solid white line) for $q_{x} \neq 0$ along the dimer bond direction. Meanwhile, relatively weak but continuous signal for the two-triplon excitation exists at $q_{x} =0$, in addition to other momentum choices. We find that for some regions of the momentum path the lower boundary of the two-triplon spectrum lies underneath the one-triplon spectrum, see the solid white lines and the dashed red lines in Fig.~\ref{fig6}. This is in contrast to the bimagnon response where the lower boundary of the bimagnon spectrum coincides with the single magnon dispersion. Based on the contribution from $\hat{O}^{(d2)}_{{\bf q}}$, the two-triplon gap $\omega_{\Delta 2}$ could be inferred from the RIXS intensity at ${\bf q}=0$, where $\omega_{\Delta 2}=\omega(0, \pi)+\omega(0, -\pi)=2\omega_{\Delta 1}$. $\omega_{\Delta 1}=\omega(0, \pi)$ is the one-triplon gap. We estimated the two-triplon gap to be around $2.7J_2$ or 124 meV. The one-triplon gap is $1.35J_2$ or 62meV.}

In Figs.~\ref{fig6}(b)~-~\ref{fig6}(c) and \ref{fig6}(e)~-~\ref{fig6}(f) we consider x-ray polarization effects. Incorporating the polarization factor as per Eq. \eqref{full intensity} leads to a reduction in the intensity by $\vert \frac{1}{15 \Gamma } \vert ^{2}$ for both the $\sigma$ and the $\pi$ polarization, compared to Figs.~\ref{fig6}(a) and~\ref{fig6}(d) where polarization effects have not been considered. If we ignore incident angle modulation, we find $\vert T_{s}(\pi_{f},\sigma_{i}) \vert ^{2} \propto \vert \frac{1}{15} \vert ^{2}$,  $\vert T_{s}(\sigma_{f},\pi_{i}) \vert ^{2} \propto \vert \frac{1}{15} \vert^{2}$, and $\vert T_{d}(\pi_{f},\pi_{i}) \vert ^{2} \propto \vert \frac{2}{15} \vert^{2}$. While the one-triplon signal dominates the RIXS spectrum, a careful choice of the polarization and scattering geometry leads to an enhanced two-triplon signal. The $\sigma$ polarization result closely follows the unpolarized RIXS signal analysis. But, the $\pi$ polarization creates a redistribution of spectral weight with nodes of very weak RIXS intensity developed around the $(\pi/2,0), (0,\pi/2), (0.9\pi,0)$, and $(0,0.9\pi)$ points. The one- and two-triplon RIXS intensities are well separated for the $\sigma$ polarization, making it easy to be distinguished in an experimental setting. But, for the $\pi$ polarization, these two signals overlap. Thus, it is difficult to disentangle the signals in this situation.

We computed the staggered dimer state RIXS intensity using  Eqs.~\eqref{Iqxi}~-~\eqref{full intensity}, see Fig.~\ref{fig7}. Similar to the situation in the columnar dimer case, there are one-  and two-triplon excitation contributions to the RIXS spectrum. However, unlike the columnar dimer case, we find the lower boundary of the two-triplon dispersion is always on top of the one-triplon, see the solid white lines and the dashed red lines in Fig.~\ref{fig7}.  The reason for this trend is explained in the next paragraph. We estimated the two-triplon gap to be $1.7J_2$ (78 meV). Hence, the single triplon gap is $ 0.85J_2$ (39 meV). The polarization features in the staggered dimer case is the same as in the columnar dimer case. We find that the intensity of the $\sigma$ polarization incident light is stronger than the $\pi$ polarization incident light. This fact is evident when we compare subfigures (b) to (c) and (e) to (f) in both Figs.~\ref{fig6} and Fig.~\ref{fig7}.

The difference in the behavior of the one- and the two- triplon dispersion between the columnar and the staggered dimer phase can be understood as follows. The lowest energy of the one-triplon is located at ${\bf k}_c=(0,\pi)$ and ${\bf k}_s=(0,0)$ for the columnar and the staggered dimer phases, respectively. If we consider the lowest energy of two quasiparticles with total energy $\omega_{2lower}$ and total momentum ${\bf q}$ for the columnar dimer phase, then one of these quasiparticles with momentum ${\bf k}_c=(0,\pi)$ can be easily excited. Thus, the other quasiparticle, based on momentum conservation, should carry a momentum of ${\bf q}^{\prime} =(q_x,-\pi+q_y)$. Hence, for the columnar dimer phase, the lower boundary of the two-triplon dispersion is given by $\omega_{2lower}(q_x,q_y)= \omega (0, \pi )+\omega (q_{x},-\pi+q_{y}) \neq \omega (q_x,q_y)$. At ${\bf q}=0$, we find $\omega(0,0) \textgreater \omega_{\Delta 2}$. As a result, the one-triplon dispersion and the lower boundary of the two-triplon dispersion intersect each other in some parts of the momentum region. As for the staggered dimer phase, we can apply the same logic for the case of two excited triplons which carry momentum $(0,0)$ and $(q_x,q_y)$. Then we obtain $\omega_{2lower}(q_x,q_y)=\omega(0,0)+\omega(q_x,q_y)=\omega_{\Delta 1}+\omega(q_x,q_y)$, where $\omega_{\Delta 1}$ is the one-triplon energy gap. This implies the lower boundary of the two-triplon dispersion is shifted from the one triplon dispersion with an extra energy gap of $\omega_{\Delta 1}$. For reasonable energy resolution, it is possible to determine the one-triplon energy gap in the staggered dimer phase by separating the peaks of the one- and two- triplon signal.

Next, we compare our two-dimensional dimerized Heisenberg system results to those of the quasi-1D spin ladder materials \cite{PhysRevLett.103.047401,PhysRevB.85.224436}. Experimentally Braicovich and collaborators~\cite{PhysRevLett.103.047401} have observed a two-triplon excitation in the spin ladder cuprates Sr$_{14}$Cu$_{24}$O$_{41}$. Subsequently, Nagao and Igarashi ~\cite{PhysRevB.85.224436} predicted the existence of both the one- and the two-triplon excitation in the RIXS spectrum. It was suggested that the one-triplon excitation signal could be detected along the rung wave vector $q_{a} =-\pi$~\cite{PhysRevB.85.224436}, which is realizable by rotating the sample. Note, the rung direction of the spin ladder is equivalent to the $x$ -direction of our columnar and staggered dimer models. Thus, according to Eq.~\eqref{Oqs1} we obtain a $\sin^{2}(q_{x}a/2)$ momentum variation for the one-triplon excitation intensity. The origin of this factor is related to the dimerized structure. Using this modulation one can explain the disappearance of the one-triplon excitation at $q_{x}=0$, and the emergence at $q_{x}\neq 0$. In principle, our method can be applied to study the limiting case of a 2D square lattice dimer model. In Appendix \ref{sec:sladder}, we detail and discuss this application to the case of a two-leg spin ladder model. The results in Fig.~\ref{fig spin ladder} validate our approach to studying dimer physics within the context of RIXS.

Both in Raman~\cite{PhysRevB.72.094419} and RIXS~\cite{PhysRevLett.102.167401} experiments, finite two-triplon excitation intensity is observed in spin ladders at zero momentum transfer. According to Eqs. \eqref{Oqs1} and \eqref{Oqs2}, while ${\bf q}=0$, we find that $\hat{O}_{\bf q}^{(s)}$ does not have any contribution to the intensity because $\sin\left({q_{x}} \frac{a}{2} \right)$ in Eq. \eqref{Oqs1} and the coefficient $u_{{\bf{k}}+{\bf{q}}/2}v_{{\bf{k}}-{\bf q}/2}-u_{{\bf k}-{\bf q}/2}v_{{\bf k}+{\bf q}/2}$ in  Eq. \eqref{Oqs2} are zero. However, $\hat{O}_{\bf q}^{(d2)}$, see Eq. \eqref{Oqd2}, implies that the two-triplon excitation signal does not vanish. Based on the scattering operator derived in this paper, we suggest that the two-triplon excitation will contribute to the direct RIXS spectrum even at the zero momentum transfer point in the spin gap dimerized system. 

In the absence of a magnetic field three-fold degeneracy is a distinct feature of the triplon excitation mode. This feature is helpful in making some predictions regarding the local dimer spin orientation. According to Eq. \eqref{Osi}, if the spin components (thus the orientation) have a nonzero projection along the {\bf $\alpha=x, y, z$} direction, the contribution for the $L_3$ -edge RIXS intensity consists of both the one- and two-triplon excitation. The situation simplifies if we detect the intensity along $q_{x}=\pi$, when only the one-triplon excitation survives. Of course, the accessibility of these momentum transfer in a real experiment is constrained by the scattering rules and the lattice constant as discussed previously. If the three-fold degeneracy is lifted, for example, by an external magnetic field, then the SU(2) symmetry is broken and the behavior of the RIXS intensity will change. In this situation, we expect to have three different RIXS branches. By measuring the relative branching intensity ratios, it is possible to compute the local dimer spin orientation angles. Since orientation is information, such a RIXS measurement can potentially aid the quantum information science community to utilize RIXS to understand how to prepare, manipulate, and recover qubits. We also believe that such RIXS experiments can assist with the analysis of quantum entanglement properties, since RIXS can give us local dimer specific information.

\section{Conclusion} \label{sec:conclusion}
In this article, we studied the $K$ and $L_{3}$ -edge RIXS response within a bond-operator representation theory at the mean-field level with a background of condensed singlets. We calculated the RIXS intensities for both the columnar and the staggered dimerized systems. At the $K$-edge, the RIXS process is a core-hole mediated modification of the superexchange interaction within the UCL approximation. We find that this gives rise to the two-triplon excitation whose intensity vanishes at zero momentum transfer. The intensity tracks the lower boundary of the two-triplon dispersion continuum quite well. The peak position of the columnar and the staggered dimer states are around $(\pi/2,\pi/2)$.

In the direct RIXS process, we studied the intensity at the typical experimentally investigated $L_{3}$-edge. We derived the RIXS scattering operator expression for a local dimer using the hole representation within a first-order UCL expansion. We found that the local dimer scattering operator has contributions from spin scattering processes and an additional contribution that originates from the local dimer hard-core constraint. This additional term can offer insight into experimental data which finds non-zero RIXS intensity at the $L_3$ -edge~\cite{PhysRevLett.103.047401}.

We generalized the local RIXS scattering operator to the case of collective one- and two-triplon excitations for the dimer system, considering effects of x-ray polarization and experimental scattering geometry. Our $L_3$ edge result has an antiphase rung modulation behavior that is generalized from the ladder case. Our findings are consistent with current INS experiments on ladders which capture either the $\sin ^{2}(q_{x}a/2)$ or $\cos ^{2}(q_{x}a/2)$ momentum variation for the one- or the two- triplon excitation RIXS intensity, respectively. Note, for the square lattice there are choices of the wave vector where the one- and the two- triplon intensity can be mixed, but they clearly separate out as in the ladder case with the antiphase behavior at the zero and $\pi$ wave vector. The main peak of the two-triplon intensity follows the lower-boundary of the two-triplon dispersion continuum.

An important outcome of our RIXS calculation is that the two-triplon gap can be detected directly at zero momentum transfer using a square lattice Heisenberg compound. This is possible because we observe non-zero RIXS signal at zero momentum transfer. Current experimental estimate for the two-triplon gap is $100\pm30$ meV for Sr$_{14}$Cu$_{24}$O$_{41}$~\cite{PhysRevLett.103.047401}. For our parameter choice, $J_{1}=138$ meV and $J_{2}=\lambda J_{1}=46$ meV, the two-triplon gap is estimated to be around 124 (78) meV for the columnar (staggered) dimer phase. Since resolution issues at both the Cu $K$ and $L$ -edge are much less severe compared to other systems, we hope that experimentalists can take motivation from our calculation to measure the two-triplon gap using the dimer phases of a square lattice quantum Heisenberg magnet. For future consideration it would be interesting to investigate triplon-triplon interaction effects at the $K$ and $L$ -edge, a study of which is beyond the scope and interest of the present work~\cite{PhysRevB.80.174403}.

In summary, our dimer RIXS calculation provides experimentalists with physical signatures to identify the exact nature of the quantum dimer state that can be hosted in the disordered phase of the square lattice Heisenberg system. 
\begin{acknowledgments}
T.D. acknowledges invitation, hospitality, and kind support from Sun Yat-Sen University. T. D. acknowledges funding support from Augusta University Scholarly Activity Award and from Sun Yat-Sen University Grant No. OEMT--2017--KF--06. M.H. and D.X.Y. are support by NKRDPC Grants No. 2017YFA0206203, No. 2018YFA0306001, NSFC-11574404 and Leading Talent Program of Guangdong Special Projects. M.H. thanks Zijian Xiong and Zhihui Luo for the discussion.
\end{acknowledgments}
\appendix
\section{$L_3$-edge dimer RIXS scattering operator} \label{appendix A}
 We derive the scattering operator for the dimer pattern within the first term of the UCL expansion. For a local spin site, the scattering operator takes the form \cite{PhysRevX.6.021020}
 \begin{equation}
         O_{j,\epsilon}=\frac{1}{i \Gamma} D_{j,\epsilon^{f}}^{\dagger}\widetilde{P}_{j} \sum_{L_{3}}\vert L_{3}\rangle \langle L_{3} \vert \sum_{l=0}^{+\infty}\frac{\bar{H}^{l}}{(i \Gamma)^{l}}\widetilde{P}_{j}D_{j,\epsilon^{i}},
      \end{equation}
where $\widetilde{P}_{j}$ is the projection operator, which prohibits the double occupancy, therefore limits electron transition follows $3d^{9}2p^{6}\rightarrow 3d^{10}2p^{5}\rightarrow3d^{9}2p^{6}$ process. The first term ($l=0$) of the expansion is
      \begin{equation}
      O_{j,\epsilon}^{(0)}=\frac{1}{i \Gamma} D_{j,\epsilon^{f}}^{\dagger}\widetilde{P}_{j} \sum_{L_{3}}\vert L_{3}\rangle \langle L_{3} \vert \widetilde{P}_{j}D_{j,\epsilon^{i}}.
      \end{equation}
We concentrate on the pure magnetic excitation and do not consider the orbital excitation. Hence we introduce the local spin flip or non-flip concept to the dimer site. 
\begin{table}[]
\centering
\begin{tabular}{|l|l|l|l|l|}
\hline
           & $\vert s \rangle$ & $\vert t_{x}\rangle$ & $\vert t_{y} \rangle$ & $\vert t_{z} \rangle$ \\
\hline
\multirow{2}{*}{$\langle s \vert$ } &  $\frac{1}{2}(O_{\downarrow \downarrow}+O_{\uparrow \uparrow})$ & $\frac{1}{2}(O_{\downarrow \uparrow}+O_{\uparrow \downarrow})$ & $\frac{i}{2}(O_{\downarrow \uparrow}-O_{\uparrow \downarrow})$ & $\frac{1}{2}(O_{\downarrow \downarrow}- O_{\uparrow \uparrow})$\\
& $\qquad \times \hat{A}$   & $\qquad \times \hat{B}$  & $\qquad \times \hat{B}$ & $\qquad \times \hat{B}$   \\
\hline 
\multirow{2}{*}{$\langle t_{x} \vert$ } &  $\frac{1}{2}(O_{\downarrow \uparrow}+O_{\uparrow \downarrow})$ & $\frac{1}{2}(O_{\downarrow \downarrow}+O_{\uparrow \uparrow})$  & $-\frac{i}{2}(O_{\downarrow \downarrow}- O_{\uparrow \uparrow})$ & $-\frac{1}{2}(O_{\downarrow \uparrow}-O_{\uparrow \downarrow})$ \\
& $\qquad \times \hat{B}$ & $\qquad \times \hat{A}$ & $\qquad \times \hat{A}$ & $\qquad \times \hat{A}$ \\
\hline  
\multirow{2}{*}{$\langle t_{y} \vert$ } & $\frac{i}{2}(O_{\downarrow \uparrow}- O_{\uparrow \downarrow})$ & $\frac{i}{2}(O_{\downarrow \downarrow}- O_{\uparrow \uparrow})$ & $\frac{1}{2}(O_{\downarrow \downarrow}+ O_{\uparrow \uparrow})$ &  $-\frac{i}{2}(O_{\downarrow \uparrow}- O_{\uparrow \downarrow})$ \\ 
& $\qquad \times \hat{B}$ & $\qquad \times \hat{A}$ &  $\qquad \times \hat{A}$   & $\qquad \times \hat{A}$\\
\hline  
\multirow{2}{*}{$\langle t_{z} \vert$ } &  $\frac{1}{2}(O_{\downarrow \downarrow}- O_{\uparrow \uparrow})$ &  $\frac{1}{2}(O_{\downarrow \uparrow}- O_{\uparrow \downarrow})$ &   $\frac{i}{2}(O_{\downarrow \uparrow}+ O_{\uparrow \downarrow})$ &  $\frac{1}{2}(O_{\downarrow \downarrow}+ O_{\uparrow \uparrow})$ \\
& $\qquad \times \hat{B}$ & $\qquad\times \hat{A}$  & $\qquad \times \hat{A}$ & $\qquad \times \hat{A}$   \\
\hline 
\end{tabular}
\caption{Table for the scattering process for one dimer, based on the flip or non-flip process of the two spins. The elements in the table are calculated based on the 3d holes transition scenario. $\hat{A}=1^{\sigma_{R_{mi}}}$,$\hat{B}=(-1)^{\sigma_{R_{mi}}-1}$, and $\sigma_{R_{m1}}=1$,$\sigma_{R_{m2}}=2$. }
\label{Table 1}
\end{table}
Comparing Table \ref{Table 1} to the bond-operator representation, we can write
    \begin{equation}
    \hat{O}=\frac{1}{i \Gamma}\sum_{i=1,2}\hat{O}_{s_{i}}+\frac{1}{i \Gamma}\hat{O}_{d},
    \end{equation}
where
    \begin{eqnarray}
    \hat{O}_{s_{i}}&=&\frac{1}{2} \!\{ (O_{\downarrow \uparrow}\!+ \!O_{\uparrow \downarrow})\hat{S}_{i}^{x} \!+ \! i(O_{\downarrow \uparrow} \!-  \!O_{\uparrow \downarrow})\hat{S}_{i}^{y} \nonumber \\&&+ \! (O_{\downarrow \downarrow} \!- \! O_{\uparrow \uparrow})\hat{S}_{i}^{z} \} ,     \\
     \hat{O}_{d} &=&\frac{1}{2}(O_{\downarrow \downarrow}+ O_{\uparrow \uparrow})  [\sum_{j=1,2}s_{m_{j}}^{\dagger}s_{m_{j}}+\sum_{\alpha ,j=1,2} t_{m_{j} \alpha}^{\dagger}t_{m_{j} \alpha} ].
    \end{eqnarray}
Here $\hat{O}_{s}$ corresponds to the spin contribution and $\hat{O}_{d}$ corresponds to the diagonal terms' contribution in Table \ref{Table 1}. $O_{{\sigma}^{\prime} \sigma}$ is the amplitude for a local spin scattering process, from $\sigma$ to $\sigma^{\prime}$, where $\sigma$ and $\sigma^{\prime}$ could refer to the spin state $\uparrow$ and $\downarrow$. In the  experiment, there would be polarization dependence effect on the scattering intensity. From the local calculation, we work out the element of $\hat{O}$ under various polarization conditions. According to different polarization states of the photon, we have polarization dependence factors given by
\begin{eqnarray}
    & T_{\text{s}}(\sigma_{f},\sigma_{i})= T_{\text{s}}(\pi_{f},\pi_{i})=0,        \label{A6} \\    
    &T_{\text{d}}(\pi_{f},\sigma_{i})= T_{\text{d}}(\sigma_{f},\pi_{i})=0,  \label{A7}    \\
    & T_{\text{s}}(\pi_{f},\sigma_{i}) = \frac{1}{i}\frac{1}{15}  \sin (\theta_{in}+\frac{\pi}{2}-2\theta),  \label{A8}\\
    & T_{\text{s}}(\sigma_{f},\pi_{i}) = - \frac{1}{i}\frac{1}{15}  \sin (\theta_{in}+\frac{\pi}{2}),  \label{A9}\\
    & T_{\text{d}}(\sigma_{f},\sigma_{i})=  \frac{2}{15}  \label{A10},\\
    & T_{\text{d}}(\pi_{f},\pi_{i}) = \frac{2}{15} \sin(\theta_{in}-2\theta+\frac{\pi}{2})\sin(\theta_{in}+\frac{\pi}{2}).  \label{A11}
    \end{eqnarray}
$T_{\text{s}}(\epsilon^{f},\epsilon^{i})$ and $T_{\text{d}}(\epsilon^{f},\epsilon^{i})$ stands for the polarization effect, which modulate $\hat O_{s}$ and $\hat O_{d}$ accordingly. Finally, we obtain
    \begin{eqnarray}
\label{eq:osiapp}
     \hat{O}_{s_{i}}^{\sigma}&=&-\frac{1}{i}\frac{1}{15}\sin{(\theta_{in}-2\theta+\pi/2)}\{\sin{\theta_{s}}\cos{\phi_{s}}S_{i}^{x}  \nonumber \\
             &\quad&- \sin\theta_{s}\sin\phi_{s}S_{i}^{y}+\cos{\theta_{s}}S_{i}^{z} \},
    \end{eqnarray}
    \begin{eqnarray}
    \hat{O}_{d}^{\sigma} &=&\frac{2}{15}\sum_{m=1}^{N}\{\sum_{j=1,2}s_{m_{j}}^{\dagger}s_{m_{j}}+\sum_{\alpha ,j=1,2} t_{m_{j} \alpha}^{\dagger}t_{m_{j} \alpha}\},
    \end{eqnarray}
 for $\sigma$ incident polarization x-ray beam, and 
    \begin{eqnarray}
     \hat{O}_{s_{i}}^{\pi}&=&-\frac{1}{i}\frac{1}{15}\sin{(\theta_{in}+\pi/2)}\{\sin{\theta_{s}}\cos{\phi_{s}}S_{i}^{x}  \nonumber \\
             &\quad&- \sin\theta_{s}\sin\phi_{s}S_{i}^{y}+\cos{\theta_{s}}S_{i}^{z} \},
    \end{eqnarray}
    \begin{eqnarray}
    \hat{O}_{d}^{\pi}&=&\frac{2}{15}\sin{(\theta_{in}-2\theta+\pi/2)}\sin{(\theta_{in}+\pi/2)} \nonumber    \\
              &&\{\sum_{j=1,2}s_{m_{j}}^{\dagger}s_{m_{j}}+\sum_{\alpha ,j=1,2} t_{m_{j} \alpha}^{\dagger}t_{m_{j} \alpha}\}, 
    \end{eqnarray}   
 for $\pi$ incident polarization x-ray beam. For the collective excitation, we turn to the momentum space. Notice the scattering operator in the momentum space $O_{\bf{q},\epsilon} \propto \sum_{j}e^{i\bf{q}\cdot R_{j}}O_{j,\epsilon}$.For $\sigma$ polarization
    \begin{eqnarray}
    A_{fi}^{s,\sigma}&=&\frac {1}{i \Gamma}\langle f \vert \sum_{m=1}^{N}\sum_{i}\hat{O}_{s_{i}}^{\sigma}e^{i\bf{q}\cdot R_{mi}}\vert i \rangle, \\
    A_{fi}^{d,\sigma} &=&\frac{1}{i\Gamma} \langle f \vert\sum_{m=1}^{N}  \hat{O}_{d}^{\sigma} e^{i\bf{q}\cdot R_{m}}\vert i \rangle,    
    \end{eqnarray}
and for the $\pi$ polarization
\begin{eqnarray}
 A_{fi}^{s,\pi}&=&\frac {1}{i \Gamma}\langle f \vert \sum_{m=1}^{N}\sum_{i} \hat{O}_{s_{i}}^{\pi}e^{i\bf{q}\cdot R_{mi}}\vert i \rangle,      \\
    A_{fi}^{d,\pi}&=&\frac{1}{i\Gamma}\langle f \vert\sum_{m=1}^{N}\hat{O}_{d}^{\pi}e^{i\bf{q}\cdot R_{m}} \vert i \rangle.   
\end{eqnarray}   
From the above equations, we find the only difference between the $\sigma$ and $\pi$ polarization scattering intensity is the polarization dependence factor, which depend on the incident angle and scattering angle. The system parts are the same. Hence, we redefine
    \begin{eqnarray}
    \hat{O}_{s_{i}}&=&\sin{\theta_{s}}\cos{\phi_{s}}S_{i}^{x}- \sin\theta_{s}\sin\phi_{s}S_{i}^{y}+\cos{\theta_{s}}S_{i}^{z},  \label{Os}\\
    \hat{O}_{d}    &=&\sum_{j=1,2}s_{m_{j}}^{\dagger}s_{m_{j}}+\sum_{\alpha ,j=1,2} t_{m_{j} \alpha}^{\dagger}t_{m_{j} \alpha} 
    \end{eqnarray}
to obtain
    \begin{eqnarray}
     \hat{O}_{s_{i}}^{\epsilon_{i}}=\sum_{\epsilon_{f}}T_{s}(\epsilon^{f},\epsilon^{i})\hat{O}_{s_{i}},
    \end{eqnarray} 
    \begin{eqnarray}
     \hat{O}_{d}^{\epsilon_{i}}=\sum_{\epsilon_{f}}T_{d}(\epsilon^{f},\epsilon^{i})\hat{O}_{d}.
    \end{eqnarray}
Using the bond-operator for Eq.~\eqref{Os} and then applying the Fourier and Bogoliubov transformations, we find the operator contributes both to one-triplon and two-triplon terms as
       \begin{eqnarray}
      \hat{O}_{q}^{(s1)} &=&\sum_{\epsilon_{f}}-2i\bar{s}\sin(q_{x} \frac{a}{2} ) (u_{{\bf q}}-v_{{\bf q}}) [ \nonumber \\
        &&\sin{\theta_{s}}\cos{\phi_{s}}(\gamma_{{\bf q},x}^{\dagger}+\gamma_{-{\bf q},x}) \nonumber \\
        && -\sin\theta_{s}\sin\phi_{s}(\gamma_{{\bf q},y}^{\dagger}+\gamma_{-{\bf q},y})    \nonumber \\
        &&+\cos\theta_{s} (\gamma_{{\bf q},z}^{\dagger}+\gamma_{-{\bf q},z}],                
     \end{eqnarray}   
and
      \begin{eqnarray}
         \hat{O}_{fi}^{(s2)} &=& 2i\cos(q_{x} \frac{a}{2} )   \sum_{k} (u_{{\bf k}+{\bf q}/2}v_{{\bf k}-{\bf q}/2}-u_{{\bf k}-{\bf q}/2}v_{{\bf k}+{\bf q}/2}) \nonumber \\
   &&[\quad \sin{\theta_{s}}\cos{\phi_{s}}\gamma_{{\bf k}+{\bf q}/2,y}^{\dagger}\gamma_{-{\bf k}+{\bf q}/2,z}^{\dagger} \nonumber \\
     &&- \sin\theta_{s}\sin\phi_{s}\gamma_{{\bf k}+{\bf q}/2,z}^{\dagger}\gamma_{-{\bf k}+{\bf q}/2,x}^{\dagger} \nonumber \\
     &&+\cos\theta_{s}\gamma_{{\bf k}+{\bf q}/2,x}^{\dagger}\gamma_{-{\bf k}+{\bf q}/2,y}^{\dagger} \quad \quad \quad]. \label{Aoqs2}
      \end{eqnarray}
As for the contribution from diagonal terms, applying Fourier and Bogoliubov transformation on Eq.~\eqref{Od}, we obtain 
 \begin{eqnarray}
     \hat{O}_{\bf q}^{(d2)}
     \!&=& \! \!\sum_{{\bf k},\alpha}\! \!-\!2\cos(q_{x} \frac{a}{2} )u_{{\bf k}+{\bf q}/2}v_{{\bf k}-{\bf q}/2}\gamma_{{\bf k}+{\bf q}/2,\alpha}^{\dagger}\gamma_{-{\bf k}+{\bf q}/2,\alpha}^{\dagger}. \label{Aoqd2}
    \end{eqnarray}
Here we retain the two-triplon creation operator terms in Eqs~\eqref{Aoqs2} and ~\eqref{Aoqd2} since we are calculating the zero temperature intensity.

Finally, we consider the angular and polarization modulation to the RIXS spectrum on momentum space based on the experiment geometry. Following the method outlined in Ref.~\onlinecite{1367-2630-13-4-043026} we can analyze the relation between momentum transfer and the scattering angle $2\theta$ as well as $\delta$ as 
\begin{eqnarray}
\vert {\bf{q}}_{\parallel} \vert &=&2k_{0} \sin \theta  \vert \sin \delta \vert,  \label{momentum transfer:q}\\
        \theta_{in} &=& \delta + \frac{\pi}{2} + \theta,         \\
  \vert \delta \vert &=& \arcsin\left(\frac{\vert {\bf q}_{\parallel} \vert}{2  k_{0}}\right),
\end{eqnarray}
where $\vert {\bf{q}}_{\parallel} \vert$ is the projection length on the sample plane of the momentum transfer $ {\bf{q}}$ and $\delta$ is the angle between z-axis and ${\bf k}_{out}-{\bf k}_{in}$. For a fixed  value of $\vert \bf{q}_{\parallel} \vert$, we can find $\delta \!=\! \pm \vert \delta \vert $ since the polarization dependence will not change while rotating the sample  with c-axis. Theoretically both $\delta \!=\! \pm \vert \delta \vert $ could be used to detect the signal with momentum transfer ${\bf{q}}_{\parallel}$ corresponding to two different choice of the incident angle, $\theta_{in} \!=\! \pm \vert \delta \vert \!+\! \frac{\pi}{2} \!+\! \theta$. We chose $\delta \!=\! \!-\! \vert \delta \vert$, since ${\bf{q}}\!=\! -({\bf{k}}_{out}\!-\!{\bf{k}}_{in})$. Hence
 \begin{equation}
   \theta_{in} = - \arcsin\left(\frac{\vert {\bf q}_{\parallel} \vert}{2  k_{0}}\right) +\theta +\frac{\pi}{2}.
 \end{equation}
At Cu $L$ -edge the photons carry momentum which is much smaller compared to $K$ -edge. Thus, we should carefully consider the momentum range that can be detected under $L$ -edge.  The vector of the photons is shown in Eqs.~\eqref{kin} - \eqref{kout}
\begin{eqnarray}
{\bf{k}}_{in} &=& k_{0}(\sin\theta_{in} \cos\phi_{in},\sin\theta_{in}\sin\phi_{in},\cos\theta_{in})   \label{kin}, \\
{\bf{k}}_{out} &=& k_{0}(\sin(\theta_{in}-2\theta) \cos\phi_{in},\sin(\theta_{in}-2\theta)\sin\phi_{in},  \nonumber \label{kout}\\
                    &\quad&\cos(\theta_{in}-2\theta)). 
\end{eqnarray}
In experiments, the incident photons hit the sample surface from above and are collected above the sample plane (assuming the photons penetrate the sample with a very small probability). This process requires that the photons' momentum vector should obey some specified condition, that is, $k_{\text{in;z}} \leq 0$ and $k_{\text{out;z}} \geq 0$. $k_{\text{in;z}}$ ($k_{\text{out;z}}$) is the momentum component along $z$-direction for the incident (outgoing) photons. This leads to $ -\theta \leq \delta \leq \theta$, which restrict the maximum momentum transfer as
\begin{eqnarray}
\vert {\bf{q}}_{\parallel} \vert_{max} &=& 2k_{0} \sin^{2} \theta  \label{angle delta}. 
\end{eqnarray}
It is quite evident that the momentum transfer is related to the incident angle of the x-ray beam. Using Eqs.~\eqref{momentum transfer:q}-\eqref{angle delta} we calculate the momentum range that can be detected by RIXS in $L$ -edge case, see Fig.~\ref{fig:boundary}. In cuprates we set the lattice constant $a \!=\! 3.8~\overset{\circ}A$ ~\cite{PhysRevB.41.1926}, which implies $\frac{\pi}{a} \!= \!0.8267~\overset{\circ}A^{-1}$, while the x-ray photon at the $L_{3}$ -edge carry momentum $q_{in}\simeq 0.47~\hbar \overset{\circ}A^{-1}$ corresponding to $k_{in}\simeq 0.47~\overset{\circ}A^{-1}$~\cite{1367-2630-13-4-043026}.
\section{Spin ladder: a limiting case of the columnar dimer lattice}\label{sec:sladder}
 \begin{figure*}[t]  
\centering
\hspace*{-5mm}\includegraphics[scale=0.5]{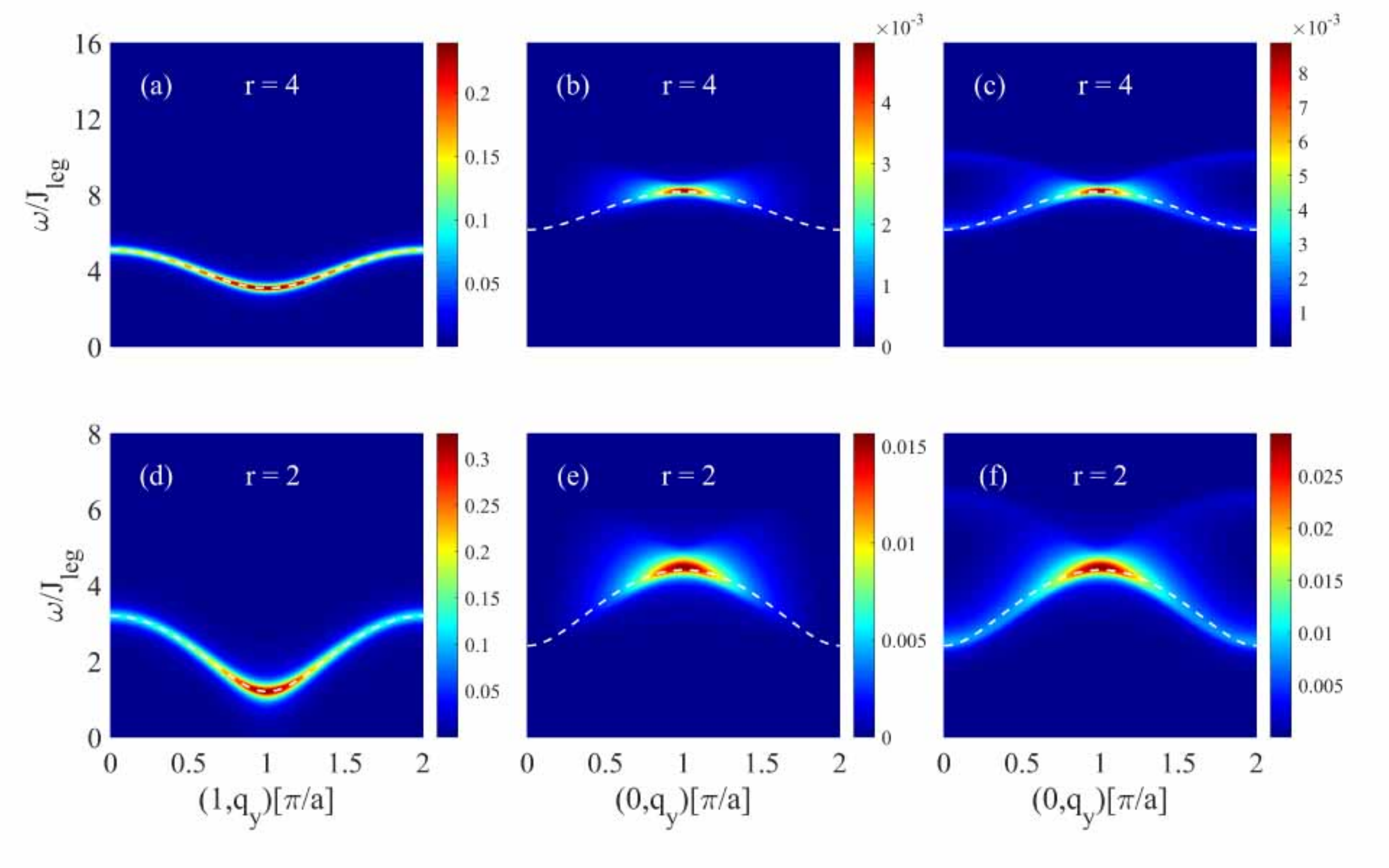}
\caption{$L_3$ -edge RIXS intensity for spin ladder model using our theory at various rung coupling strength $r= J_{rung}/J_{leg}$. The x-ray resolution was set to $J_{leg}/6$. Upper panel: (a), (b), and (c) are the intensities for $r=4$. Lower panel: (d), (e), and (f) are the intensities for $r=2$. (a) and (d) are the intensities of $I_{s1}({\bf q},\omega)$, while (b) and (e) are the intensities of $I_{s2}({\bf q},\omega)$, (c) and (f) are the intensities of $I_{2}^{total}({\bf q},\omega)=I_{s2}({\bf q},\omega)+I_{d2}({\bf q},\omega)$, compare with \cite{kumarPhysRevB.99.205130}. Dashed white line represents the one-triplon dispersion in (a), (d) while the lower boundary of two-triplon dispersion in (b), (c), (d), (f).  $\omega/J_{leg}$ is scaled energy.} 
\label{fig spin ladder}
\end{figure*}

The columnar dimer lattice can be thought of as a collection of decoupled spin ladders. To reach the limit of a single decoupled ladder we write Eq.~\ref{Hamiltonian} in a slightly modified format as
\begin{equation}
   H_{D}=J_{rung}\sum_{<ij>\in D} {\bf S}_{i}\cdot {\bf S}_{j}+\lambda^{\prime} J_{rung} \sum _{<ij>\notin D} {\bf S}_{i}  \cdot {\bf S}_{j}+J_{leg} \sum _{<ij>\notin D} {\bf S}_{i}  \cdot {\bf S}_{j} \label{hamladd},
\end{equation}
where $J_{rung}$ is the coupling constant for the two spins on the rungs, $J_{leg}$ is the coupling constant between the spins on the legs, and $\lambda^{\prime} J_{rung}$ is the inter-ladder coupling. If we set $\lambda^{\prime}=0$, we obtain our single spin ladder Hamiltonian $H_{L}$ as~\cite{PhysRevB.49.8901}
\begin{equation}
\label{eq:slad}
      H_{L}/J_{leg}= \sum_{m,j=1,2}S_{m,j}\cdot S_{m+1,j} + r\sum_{m}S_{m,1} \cdot S_{m,2},
      \end{equation}
where $r= J_{rung}/J_{leg}$ and the site indices sum over a single ladder configuration. The ladder is assumed to be oriented along the $y$-direction to be consistent with our previous analysis on the columnar dimer phase. For simplicity we ignore any ring-exchange interaction. Applying the bond-operator method to Eq.~\eqref{eq:slad} we then have
      \begin{eqnarray}
       A_{k} &=&\frac{J_{rung}}{4}- \mu + J_{leg} \bar{s}^{2} \cos(k_{y}a_{y}), \\
       B_{k} &=& J_{leg} \bar{s}^{2} \cos(k_{y}a_{y}).
      \end{eqnarray}
 One should keep in mind that $k_{y}$ here is along the direction of the leg. Note, as expected these equations are a special case of Eqs.~\eqref{eq:cak} and ~\eqref{eq:cbk} with the $k_{x}$ term set to zero. We utilize the above expressions for $A_{k}$ and $B_{k}$ to compute $u_{k}$ and $v_{k}$ which has the usual standard form. Repeating the procedure for the $L_{3}$-edge RIXS intensity for the spin ladder model we obtain the Fourier transformed Bogoliubov operators as
\begin{eqnarray}
     \hat{O}_{\bf{q},\alpha}^{(s1)}= -2i\bar{s}\sin\left(\frac{q_{x}a}{2}\right) (u_{q_{y}}\!-\!v_{q_{y}})(\gamma_{q_{y},\alpha}^{\dagger}\!+\!\gamma_{- q_{y},\alpha)}\label{eq:lads1},         
 \end{eqnarray}
\begin{eqnarray}
       \hat{O}_{\bf{q}, \alpha}^{(s2)}&=& 2i\sum_{k_{y}}  \cos\left(\frac{q_{x} a}{2}\right)\vert\epsilon_{\alpha \beta \rho}\vert (u_{k_{y}+q_{y}/2}v_{k_{y}-q_{y}/2}-u_{ k_{y}- q_{y}/2}v_{k_{y}+q_{y}/2})\nonumber \\&& \quad \quad \gamma_{k_{y}+ q_{y}/2,\beta}^{\dagger}\gamma_{-k_{y}+q_{y}/2,\rho}^{\dagger}\label{eq:lads2}, 
\end{eqnarray}
\begin{eqnarray}
 \hat{O}_{\bf q}^{(d2)} \!=\! \!-\!2 \! \sum_{ k_{y},\alpha}\! \cos\left(\frac{q_{x}a}{2}\right)u_{k_{y}\!+\! q_{y}\!/\!2}v_{k_{y}\!-\! q_{y}\!/\!2}\gamma_{k_{y}\!+\! q_{y}\!/\!2,\alpha}^{\dagger}\gamma_{-k_{y}\!+\! q_{y}/2,\alpha}^{\dagger}\label{eq:ladd2}.
    \end{eqnarray}
Here the $q_{x}$ wave vector is along the rung direction and the triplon mode components are given by $\alpha,\beta,\rho=x,y,z$. The repeated indices $\rho$ and $\beta$ are set to the triplon modes rather than Einstein summed. Using Eqs.~\eqref{eq:lads1}~-~\eqref{eq:ladd2} and Eqs.~\eqref{Axi}~-~\eqref{Iqxi}, we calculate the RIXS intensity for the spin ladder.

In  Fig.~\ref{fig spin ladder} we compare and contrast the RIXS signal for the one- and two-triplon response. We calculated the RIXS intensity for $q_{x} = 0,\pi/a$ and $r$= 2, 4. Taking advantage of the lattice modulation behavior we can use $q_{x}=\pi/a$ and 0 to separate the one- and two-triplon RIXS intensity. The upper panel is for strong rung coupling $r=4$ and the lower panel is for the weaker limit of $r=2$. In Figs.~\ref{fig spin ladder}(a) and (d) we show the results of the one-triplon intensity $I_{s1}({\bf q},\omega)$, in Figs.~\ref{fig spin ladder}(b) and (e) we plot the two-triplon intensity $I_{s2}({\bf q},\omega)$, and in the last column Figs.~\ref{fig spin ladder}(c) and (f) we show the total two-triplon contribution $I_{2}^{total}({\bf q},\omega)=I_{s2}({\bf q},\omega)+I_{d2}({\bf q},\omega)$. Polarization and experimental geometry consideration are ignored for the present discussion for simplicity.
We observe from the figures that the weaker rung coupling limit produces a more intense RIXS signal.Our results for $I_{s1}({\bf q},\omega)$ as well as $I_{s2}({\bf q},\omega)$ agree well with the findings of Ref.~\onlinecite{kumarPhysRevB.99.205130} in the undoped case for the strong-rung limit in the non-spin conserving channel. Our calculation focuses on the first-order term in the UCL expansion which dominates the RIXS spectral weight, when allowed. Hence we compare our result to their undoped strong-rung limit case in the NSC channel. 

Based on our bond-operator representation formulation calculation we find an additional finite signal at zero momentum transfer. This finding is in qualitative agreement with cuprates $L_3$ edge experimental result on a real spin ladder material \cite{PhysRevLett.103.047401}, but different from past theoretical and computational analysis ~\cite{PhysRevB.85.224436,kumarPhysRevB.99.205130} which do not find any non-zero signal. Hence, we infer that the scattering process $\hat{O}_{d}$ is important for an appropriate description of the two-triplon RIXS signal in a dimer system. 
\bibliographystyle{apsrev4-1}
\newpage
\bibliography{triplon}

\begin{thebibliography}{76}%
\makeatletter
\providecommand \@ifxundefined [1]{%
 \@ifx{#1\undefined}
}%
\providecommand \@ifnum [1]{%
 \ifnum #1\expandafter \@firstoftwo
 \else \expandafter \@secondoftwo
 \fi
}%
\providecommand \@ifx [1]{%
 \ifx #1\expandafter \@firstoftwo
 \else \expandafter \@secondoftwo
 \fi
}%
\providecommand \natexlab [1]{#1}%
\providecommand \enquote  [1]{``#1''}%
\providecommand \bibnamefont  [1]{#1}%
\providecommand \bibfnamefont [1]{#1}%
\providecommand \citenamefont [1]{#1}%
\providecommand \href@noop [0]{\@secondoftwo}%
\providecommand \href [0]{\begingroup \@sanitize@url \@href}%
\providecommand \@href[1]{\@@startlink{#1}\@@href}%
\providecommand \@@href[1]{\endgroup#1\@@endlink}%
\providecommand \@sanitize@url [0]{\catcode `\\12\catcode `\$12\catcode
  `\&12\catcode `\#12\catcode `\^12\catcode `\_12\catcode `\%12\relax}%
\providecommand \@@startlink[1]{}%
\providecommand \@@endlink[0]{}%
\providecommand \url  [0]{\begingroup\@sanitize@url \@url }%
\providecommand \@url [1]{\endgroup\@href {#1}{\urlprefix }}%
\providecommand \urlprefix  [0]{URL }%
\providecommand \Eprint [0]{\href }%
\providecommand \doibase [0]{http://dx.doi.org/}%
\providecommand \selectlanguage [0]{\@gobble}%
\providecommand \bibinfo  [0]{\@secondoftwo}%
\providecommand \bibfield  [0]{\@secondoftwo}%
\providecommand \translation [1]{[#1]}%
\providecommand \BibitemOpen [0]{}%
\providecommand \bibitemStop [0]{}%
\providecommand \bibitemNoStop [0]{.\EOS\space}%
\providecommand \EOS [0]{\spacefactor3000\relax}%
\providecommand \BibitemShut  [1]{\csname bibitem#1\endcsname}%
\let\auto@bib@innerbib\@empty
\bibitem [{\citenamefont {Ament}\ \emph {et~al.}(2011)\citenamefont {Ament},
  \citenamefont {van Veenendaal}, \citenamefont {Devereaux}, \citenamefont
  {Hill},\ and\ \citenamefont {van~den Brink}}]{RevModPhys.83.705}%
  \BibitemOpen
  \bibfield  {author} {\bibinfo {author} {\bibfnamefont {L.~J.~P.}\
  \bibnamefont {Ament}}, \bibinfo {author} {\bibfnamefont {M.}~\bibnamefont
  {van Veenendaal}}, \bibinfo {author} {\bibfnamefont {T.~P.}\ \bibnamefont
  {Devereaux}}, \bibinfo {author} {\bibfnamefont {J.~P.}\ \bibnamefont {Hill}},
  \ and\ \bibinfo {author} {\bibfnamefont {J.}~\bibnamefont {van~den Brink}},\
  }\href {\doibase 10.1103/RevModPhys.83.705} {\bibfield  {journal} {\bibinfo
  {journal} {Rev. Mod. Phys.}\ }\textbf {\bibinfo {volume} {83}},\ \bibinfo
  {pages} {705} (\bibinfo {year} {2011})}\BibitemShut {NoStop}%
\bibitem [{\citenamefont {Dean}(2015)}]{DEAN20153}%
  \BibitemOpen
  \bibfield  {author} {\bibinfo {author} {\bibfnamefont {M.}~\bibnamefont
  {Dean}},\ }\href {\doibase http://dx.doi.org/10.1016/j.jmmm.2014.03.057}
  {\bibfield  {journal} {\bibinfo  {journal} {Journal of Magnetism and Magnetic
  Materials}\ }\textbf {\bibinfo {volume} {376}},\ \bibinfo {pages} {3 }
  (\bibinfo {year} {2015})}\BibitemShut {NoStop}%
\bibitem [{\citenamefont {Ghiringhelli}\ \emph {et~al.}(2006)\citenamefont
  {Ghiringhelli}, \citenamefont {Piazzalunga}, \citenamefont {Dallera},
  \citenamefont {Trezzi}, \citenamefont {Braicovich}, \citenamefont {Schmitt},
  \citenamefont {Strocov}, \citenamefont {Betemps}, \citenamefont {Patthey},
  \citenamefont {Wang},\ and\ \citenamefont {Grioni}}]{doi:10.1063/1.2372731}%
  \BibitemOpen
  \bibfield  {author} {\bibinfo {author} {\bibfnamefont {G.}~\bibnamefont
  {Ghiringhelli}}, \bibinfo {author} {\bibfnamefont {A.}~\bibnamefont
  {Piazzalunga}}, \bibinfo {author} {\bibfnamefont {C.}~\bibnamefont
  {Dallera}}, \bibinfo {author} {\bibfnamefont {G.}~\bibnamefont {Trezzi}},
  \bibinfo {author} {\bibfnamefont {L.}~\bibnamefont {Braicovich}}, \bibinfo
  {author} {\bibfnamefont {T.}~\bibnamefont {Schmitt}}, \bibinfo {author}
  {\bibfnamefont {V.~N.}\ \bibnamefont {Strocov}}, \bibinfo {author}
  {\bibfnamefont {R.}~\bibnamefont {Betemps}}, \bibinfo {author} {\bibfnamefont
  {L.}~\bibnamefont {Patthey}}, \bibinfo {author} {\bibfnamefont
  {X.}~\bibnamefont {Wang}}, \ and\ \bibinfo {author} {\bibfnamefont
  {M.}~\bibnamefont {Grioni}},\ }\href {\doibase 10.1063/1.2372731} {\bibfield
  {journal} {\bibinfo  {journal} {Review of Scientific Instruments}\ }\textbf
  {\bibinfo {volume} {77}},\ \bibinfo {pages} {113108} (\bibinfo {year}
  {2006})}\BibitemShut {NoStop}%
\bibitem [{\citenamefont {Lee}\ \emph {et~al.}(2013)\citenamefont {Lee},
  \citenamefont {Johnston}, \citenamefont {Moritz}, \citenamefont {Lee},
  \citenamefont {Yi}, \citenamefont {Zhou}, \citenamefont {Schmitt},
  \citenamefont {Patthey}, \citenamefont {Strocov}, \citenamefont {Kudo},
  \citenamefont {Koike}, \citenamefont {van~den Brink}, \citenamefont
  {Devereaux},\ and\ \citenamefont {Shen}}]{PhysRevLett.110.265502}%
  \BibitemOpen
  \bibfield  {author} {\bibinfo {author} {\bibfnamefont {W.~S.}\ \bibnamefont
  {Lee}}, \bibinfo {author} {\bibfnamefont {S.}~\bibnamefont {Johnston}},
  \bibinfo {author} {\bibfnamefont {B.}~\bibnamefont {Moritz}}, \bibinfo
  {author} {\bibfnamefont {J.}~\bibnamefont {Lee}}, \bibinfo {author}
  {\bibfnamefont {M.}~\bibnamefont {Yi}}, \bibinfo {author} {\bibfnamefont
  {K.~J.}\ \bibnamefont {Zhou}}, \bibinfo {author} {\bibfnamefont
  {T.}~\bibnamefont {Schmitt}}, \bibinfo {author} {\bibfnamefont
  {L.}~\bibnamefont {Patthey}}, \bibinfo {author} {\bibfnamefont
  {V.}~\bibnamefont {Strocov}}, \bibinfo {author} {\bibfnamefont
  {K.}~\bibnamefont {Kudo}}, \bibinfo {author} {\bibfnamefont {Y.}~\bibnamefont
  {Koike}}, \bibinfo {author} {\bibfnamefont {J.}~\bibnamefont {van~den
  Brink}}, \bibinfo {author} {\bibfnamefont {T.~P.}\ \bibnamefont {Devereaux}},
  \ and\ \bibinfo {author} {\bibfnamefont {Z.~X.}\ \bibnamefont {Shen}},\
  }\href {\doibase 10.1103/PhysRevLett.110.265502} {\bibfield  {journal}
  {\bibinfo  {journal} {Phys. Rev. Lett.}\ }\textbf {\bibinfo {volume} {110}},\
  \bibinfo {pages} {265502} (\bibinfo {year} {2013})}\BibitemShut {NoStop}%
\bibitem [{\citenamefont {Jia}\ \emph {et~al.}(2012)\citenamefont {Jia},
  \citenamefont {Chen}, \citenamefont {Sorini}, \citenamefont {Moritz},\ and\
  \citenamefont {Devereaux}}]{Jia_2012}%
  \BibitemOpen
  \bibfield  {author} {\bibinfo {author} {\bibfnamefont {C.~J.}\ \bibnamefont
  {Jia}}, \bibinfo {author} {\bibfnamefont {C.-C.}\ \bibnamefont {Chen}},
  \bibinfo {author} {\bibfnamefont {A.~P.}\ \bibnamefont {Sorini}}, \bibinfo
  {author} {\bibfnamefont {B.}~\bibnamefont {Moritz}}, \ and\ \bibinfo {author}
  {\bibfnamefont {T.~P.}\ \bibnamefont {Devereaux}},\ }\href {\doibase
  10.1088/1367-2630/14/11/113038} {\bibfield  {journal} {\bibinfo  {journal}
  {New Journal of Physics}\ }\textbf {\bibinfo {volume} {14}},\ \bibinfo
  {pages} {113038} (\bibinfo {year} {2012})}\BibitemShut {NoStop}%
\bibitem [{\citenamefont {Klauser}\ \emph {et~al.}(2011)\citenamefont
  {Klauser}, \citenamefont {Mossel}, \citenamefont {Caux},\ and\ \citenamefont
  {van~den Brink}}]{PhysRevLett.106.157205}%
  \BibitemOpen
  \bibfield  {author} {\bibinfo {author} {\bibfnamefont {A.}~\bibnamefont
  {Klauser}}, \bibinfo {author} {\bibfnamefont {J.}~\bibnamefont {Mossel}},
  \bibinfo {author} {\bibfnamefont {J.-S.}\ \bibnamefont {Caux}}, \ and\
  \bibinfo {author} {\bibfnamefont {J.}~\bibnamefont {van~den Brink}},\ }\href
  {\doibase 10.1103/PhysRevLett.106.157205} {\bibfield  {journal} {\bibinfo
  {journal} {Phys. Rev. Lett.}\ }\textbf {\bibinfo {volume} {106}},\ \bibinfo
  {pages} {157205} (\bibinfo {year} {2011})}\BibitemShut {NoStop}%
\bibitem [{\citenamefont {Johnston}\ \emph {et~al.}(2010)\citenamefont
  {Johnston}, \citenamefont {Vernay}, \citenamefont {Moritz}, \citenamefont
  {Shen}, \citenamefont {Nagaosa}, \citenamefont {Zaanen},\ and\ \citenamefont
  {Devereaux}}]{PhysRevB.82.064513}%
  \BibitemOpen
  \bibfield  {author} {\bibinfo {author} {\bibfnamefont {S.}~\bibnamefont
  {Johnston}}, \bibinfo {author} {\bibfnamefont {F.}~\bibnamefont {Vernay}},
  \bibinfo {author} {\bibfnamefont {B.}~\bibnamefont {Moritz}}, \bibinfo
  {author} {\bibfnamefont {Z.-X.}\ \bibnamefont {Shen}}, \bibinfo {author}
  {\bibfnamefont {N.}~\bibnamefont {Nagaosa}}, \bibinfo {author} {\bibfnamefont
  {J.}~\bibnamefont {Zaanen}}, \ and\ \bibinfo {author} {\bibfnamefont {T.~P.}\
  \bibnamefont {Devereaux}},\ }\href {\doibase 10.1103/PhysRevB.82.064513}
  {\bibfield  {journal} {\bibinfo  {journal} {Phys. Rev. B}\ }\textbf {\bibinfo
  {volume} {82}},\ \bibinfo {pages} {064513} (\bibinfo {year}
  {2010})}\BibitemShut {NoStop}%
\bibitem [{\citenamefont {Forte}\ \emph {et~al.}(2011)\citenamefont {Forte},
  \citenamefont {Cuoco}, \citenamefont {Noce},\ and\ \citenamefont {van~den
  Brink}}]{PhysRevB.83.245133}%
  \BibitemOpen
  \bibfield  {author} {\bibinfo {author} {\bibfnamefont {F.}~\bibnamefont
  {Forte}}, \bibinfo {author} {\bibfnamefont {M.}~\bibnamefont {Cuoco}},
  \bibinfo {author} {\bibfnamefont {C.}~\bibnamefont {Noce}}, \ and\ \bibinfo
  {author} {\bibfnamefont {J.}~\bibnamefont {van~den Brink}},\ }\href {\doibase
  10.1103/PhysRevB.83.245133} {\bibfield  {journal} {\bibinfo  {journal} {Phys.
  Rev. B}\ }\textbf {\bibinfo {volume} {83}},\ \bibinfo {pages} {245133}
  (\bibinfo {year} {2011})}\BibitemShut {NoStop}%
\bibitem [{\citenamefont {Kourtis}\ \emph {et~al.}(2012)\citenamefont
  {Kourtis}, \citenamefont {van~den Brink},\ and\ \citenamefont
  {Daghofer}}]{PhysRevB.85.064423}%
  \BibitemOpen
  \bibfield  {author} {\bibinfo {author} {\bibfnamefont {S.}~\bibnamefont
  {Kourtis}}, \bibinfo {author} {\bibfnamefont {J.}~\bibnamefont {van~den
  Brink}}, \ and\ \bibinfo {author} {\bibfnamefont {M.}~\bibnamefont
  {Daghofer}},\ }\href {\doibase 10.1103/PhysRevB.85.064423} {\bibfield
  {journal} {\bibinfo  {journal} {Phys. Rev. B}\ }\textbf {\bibinfo {volume}
  {85}},\ \bibinfo {pages} {064423} (\bibinfo {year} {2012})}\BibitemShut
  {NoStop}%
\bibitem [{\citenamefont {Markiewicz}\ and\ \citenamefont
  {Bansil}(2006)}]{PhysRevLett.96.107005}%
  \BibitemOpen
  \bibfield  {author} {\bibinfo {author} {\bibfnamefont {R.~S.}\ \bibnamefont
  {Markiewicz}}\ and\ \bibinfo {author} {\bibfnamefont {A.}~\bibnamefont
  {Bansil}},\ }\href {\doibase 10.1103/PhysRevLett.96.107005} {\bibfield
  {journal} {\bibinfo  {journal} {Phys. Rev. Lett.}\ }\textbf {\bibinfo
  {volume} {96}},\ \bibinfo {pages} {107005} (\bibinfo {year}
  {2006})}\BibitemShut {NoStop}%
\bibitem [{\citenamefont {van~den Brink}\ and\ \citenamefont {van
  Veenendaal}(2006)}]{EPL.73.121}%
  \BibitemOpen
  \bibfield  {author} {\bibinfo {author} {\bibfnamefont {J.}~\bibnamefont
  {van~den Brink}}\ and\ \bibinfo {author} {\bibfnamefont {M.}~\bibnamefont
  {van Veenendaal}},\ }\href {http://stacks.iop.org/0295-5075/73/i=1/a=121}
  {\bibfield  {journal} {\bibinfo  {journal} {Europhysics Letters}\ }\textbf
  {\bibinfo {volume} {73}},\ \bibinfo {pages} {121} (\bibinfo {year}
  {2006})}\BibitemShut {NoStop}%
\bibitem [{\citenamefont {Forte}\ \emph
  {et~al.}(2008{\natexlab{a}})\citenamefont {Forte}, \citenamefont {Ament},\
  and\ \citenamefont {van~den Brink}}]{PhysRevLett.101.106406}%
  \BibitemOpen
  \bibfield  {author} {\bibinfo {author} {\bibfnamefont {F.}~\bibnamefont
  {Forte}}, \bibinfo {author} {\bibfnamefont {L.~J.~P.}\ \bibnamefont {Ament}},
  \ and\ \bibinfo {author} {\bibfnamefont {J.}~\bibnamefont {van~den Brink}},\
  }\href {\doibase 10.1103/PhysRevLett.101.106406} {\bibfield  {journal}
  {\bibinfo  {journal} {Phys. Rev. Lett.}\ }\textbf {\bibinfo {volume} {101}},\
  \bibinfo {pages} {106406} (\bibinfo {year} {2008}{\natexlab{a}})}\BibitemShut
  {NoStop}%
\bibitem [{\citenamefont {Nagao}\ and\ \citenamefont
  {Igarashi}(2007)}]{PhysRevB.75.214414}%
  \BibitemOpen
  \bibfield  {author} {\bibinfo {author} {\bibfnamefont {T.}~\bibnamefont
  {Nagao}}\ and\ \bibinfo {author} {\bibfnamefont {J.-i.}\ \bibnamefont
  {Igarashi}},\ }\href {\doibase 10.1103/PhysRevB.75.214414} {\bibfield
  {journal} {\bibinfo  {journal} {Phys. Rev. B}\ }\textbf {\bibinfo {volume}
  {75}},\ \bibinfo {pages} {214414} (\bibinfo {year} {2007})}\BibitemShut
  {NoStop}%
\bibitem [{\citenamefont {Forte}\ \emph
  {et~al.}(2008{\natexlab{b}})\citenamefont {Forte}, \citenamefont {Ament},\
  and\ \citenamefont {van~den Brink}}]{PhysRevB.77.134428}%
  \BibitemOpen
  \bibfield  {author} {\bibinfo {author} {\bibfnamefont {F.}~\bibnamefont
  {Forte}}, \bibinfo {author} {\bibfnamefont {L.~J.~P.}\ \bibnamefont {Ament}},
  \ and\ \bibinfo {author} {\bibfnamefont {J.}~\bibnamefont {van~den Brink}},\
  }\href {\doibase 10.1103/PhysRevB.77.134428} {\bibfield  {journal} {\bibinfo
  {journal} {Phys. Rev. B}\ }\textbf {\bibinfo {volume} {77}},\ \bibinfo
  {pages} {134428} (\bibinfo {year} {2008}{\natexlab{b}})}\BibitemShut
  {NoStop}%
\bibitem [{\citenamefont {Luo}\ \emph {et~al.}(2014{\natexlab{a}})\citenamefont
  {Luo}, \citenamefont {Datta},\ and\ \citenamefont
  {Yao}}]{PhysRevB.89.165103}%
  \BibitemOpen
  \bibfield  {author} {\bibinfo {author} {\bibfnamefont {C.}~\bibnamefont
  {Luo}}, \bibinfo {author} {\bibfnamefont {T.}~\bibnamefont {Datta}}, \ and\
  \bibinfo {author} {\bibfnamefont {D.-X.}\ \bibnamefont {Yao}},\ }\href
  {\doibase 10.1103/PhysRevB.89.165103} {\bibfield  {journal} {\bibinfo
  {journal} {Phys. Rev. B}\ }\textbf {\bibinfo {volume} {89}},\ \bibinfo
  {pages} {165103} (\bibinfo {year} {2014}{\natexlab{a}})}\BibitemShut
  {NoStop}%
\bibitem [{\citenamefont {Marra}\ \emph {et~al.}(2012)\citenamefont {Marra},
  \citenamefont {Wohlfeld},\ and\ \citenamefont {van~den
  Brink}}]{PhysRevLett.109.117401}%
  \BibitemOpen
  \bibfield  {author} {\bibinfo {author} {\bibfnamefont {P.}~\bibnamefont
  {Marra}}, \bibinfo {author} {\bibfnamefont {K.}~\bibnamefont {Wohlfeld}}, \
  and\ \bibinfo {author} {\bibfnamefont {J.}~\bibnamefont {van~den Brink}},\
  }\href {\doibase 10.1103/PhysRevLett.109.117401} {\bibfield  {journal}
  {\bibinfo  {journal} {Phys. Rev. Lett.}\ }\textbf {\bibinfo {volume} {109}},\
  \bibinfo {pages} {117401} (\bibinfo {year} {2012})}\BibitemShut {NoStop}%
\bibitem [{\citenamefont {Pakhira}\ \emph {et~al.}(2012)\citenamefont
  {Pakhira}, \citenamefont {Freericks},\ and\ \citenamefont
  {Shvaika}}]{PhysRevB.86.125103}%
  \BibitemOpen
  \bibfield  {author} {\bibinfo {author} {\bibfnamefont {N.}~\bibnamefont
  {Pakhira}}, \bibinfo {author} {\bibfnamefont {J.~K.}\ \bibnamefont
  {Freericks}}, \ and\ \bibinfo {author} {\bibfnamefont {A.~M.}\ \bibnamefont
  {Shvaika}},\ }\href {\doibase 10.1103/PhysRevB.86.125103} {\bibfield
  {journal} {\bibinfo  {journal} {Phys. Rev. B}\ }\textbf {\bibinfo {volume}
  {86}},\ \bibinfo {pages} {125103} (\bibinfo {year} {2012})}\BibitemShut
  {NoStop}%
\bibitem [{\citenamefont {Schlappa}\ \emph {et~al.}(2018)\citenamefont
  {Schlappa}, \citenamefont {Kumar}, \citenamefont {Zhou}, \citenamefont
  {Singh}, \citenamefont {Mourigal}, \citenamefont {Strocov}, \citenamefont
  {Revcolevschi}, \citenamefont {Patthey}, \citenamefont {R{\o}nnow},
  \citenamefont {Johnston},\ and\ \citenamefont {Schmitt}}]{Schlappa2018}%
  \BibitemOpen
  \bibfield  {author} {\bibinfo {author} {\bibfnamefont {J.}~\bibnamefont
  {Schlappa}}, \bibinfo {author} {\bibfnamefont {U.}~\bibnamefont {Kumar}},
  \bibinfo {author} {\bibfnamefont {K.~J.}\ \bibnamefont {Zhou}}, \bibinfo
  {author} {\bibfnamefont {S.}~\bibnamefont {Singh}}, \bibinfo {author}
  {\bibfnamefont {M.}~\bibnamefont {Mourigal}}, \bibinfo {author}
  {\bibfnamefont {V.~N.}\ \bibnamefont {Strocov}}, \bibinfo {author}
  {\bibfnamefont {A.}~\bibnamefont {Revcolevschi}}, \bibinfo {author}
  {\bibfnamefont {L.}~\bibnamefont {Patthey}}, \bibinfo {author} {\bibfnamefont
  {H.~M.}\ \bibnamefont {R{\o}nnow}}, \bibinfo {author} {\bibfnamefont
  {S.}~\bibnamefont {Johnston}}, \ and\ \bibinfo {author} {\bibfnamefont
  {T.}~\bibnamefont {Schmitt}},\ }\href {\doibase 10.1038/s41467-018-07838-y}
  {\bibfield  {journal} {\bibinfo  {journal} {Nature Communications}\ }\textbf
  {\bibinfo {volume} {9}},\ \bibinfo {pages} {5394} (\bibinfo {year}
  {2018})}\BibitemShut {NoStop}%
\bibitem [{\citenamefont {Ellis}\ \emph {et~al.}(2010)\citenamefont {Ellis},
  \citenamefont {Kim}, \citenamefont {Hill}, \citenamefont {Wakimoto},
  \citenamefont {Birgeneau}, \citenamefont {Shvyd'ko}, \citenamefont {Casa},
  \citenamefont {Gog}, \citenamefont {Ishii}, \citenamefont {Ikeuchi},
  \citenamefont {Paramekanti},\ and\ \citenamefont {Kim}}]{PhysRevB.81.085124}%
  \BibitemOpen
  \bibfield  {author} {\bibinfo {author} {\bibfnamefont {D.~S.}\ \bibnamefont
  {Ellis}}, \bibinfo {author} {\bibfnamefont {J.}~\bibnamefont {Kim}}, \bibinfo
  {author} {\bibfnamefont {J.~P.}\ \bibnamefont {Hill}}, \bibinfo {author}
  {\bibfnamefont {S.}~\bibnamefont {Wakimoto}}, \bibinfo {author}
  {\bibfnamefont {R.~J.}\ \bibnamefont {Birgeneau}}, \bibinfo {author}
  {\bibfnamefont {Y.}~\bibnamefont {Shvyd'ko}}, \bibinfo {author}
  {\bibfnamefont {D.}~\bibnamefont {Casa}}, \bibinfo {author} {\bibfnamefont
  {T.}~\bibnamefont {Gog}}, \bibinfo {author} {\bibfnamefont {K.}~\bibnamefont
  {Ishii}}, \bibinfo {author} {\bibfnamefont {K.}~\bibnamefont {Ikeuchi}},
  \bibinfo {author} {\bibfnamefont {A.}~\bibnamefont {Paramekanti}}, \ and\
  \bibinfo {author} {\bibfnamefont {Y.-J.}\ \bibnamefont {Kim}},\ }\href
  {\doibase 10.1103/PhysRevB.81.085124} {\bibfield  {journal} {\bibinfo
  {journal} {Phys. Rev. B}\ }\textbf {\bibinfo {volume} {81}},\ \bibinfo
  {pages} {085124} (\bibinfo {year} {2010})}\BibitemShut {NoStop}%
\bibitem [{\citenamefont {Guarise}\ \emph {et~al.}(2010)\citenamefont
  {Guarise}, \citenamefont {Dalla~Piazza}, \citenamefont {Moretti~Sala},
  \citenamefont {Ghiringhelli}, \citenamefont {Braicovich}, \citenamefont
  {Berger}, \citenamefont {Hancock}, \citenamefont {van~der Marel},
  \citenamefont {Schmitt}, \citenamefont {Strocov}, \citenamefont {Ament},
  \citenamefont {van~den Brink}, \citenamefont {Lin}, \citenamefont {Xu},
  \citenamefont {R\o{}nnow},\ and\ \citenamefont
  {Grioni}}]{PhysRevLett.105.157006}%
  \BibitemOpen
  \bibfield  {author} {\bibinfo {author} {\bibfnamefont {M.}~\bibnamefont
  {Guarise}}, \bibinfo {author} {\bibfnamefont {B.}~\bibnamefont
  {Dalla~Piazza}}, \bibinfo {author} {\bibfnamefont {M.}~\bibnamefont
  {Moretti~Sala}}, \bibinfo {author} {\bibfnamefont {G.}~\bibnamefont
  {Ghiringhelli}}, \bibinfo {author} {\bibfnamefont {L.}~\bibnamefont
  {Braicovich}}, \bibinfo {author} {\bibfnamefont {H.}~\bibnamefont {Berger}},
  \bibinfo {author} {\bibfnamefont {J.~N.}\ \bibnamefont {Hancock}}, \bibinfo
  {author} {\bibfnamefont {D.}~\bibnamefont {van~der Marel}}, \bibinfo {author}
  {\bibfnamefont {T.}~\bibnamefont {Schmitt}}, \bibinfo {author} {\bibfnamefont
  {V.~N.}\ \bibnamefont {Strocov}}, \bibinfo {author} {\bibfnamefont
  {L.~J.~P.}\ \bibnamefont {Ament}}, \bibinfo {author} {\bibfnamefont
  {J.}~\bibnamefont {van~den Brink}}, \bibinfo {author} {\bibfnamefont {P.-H.}\
  \bibnamefont {Lin}}, \bibinfo {author} {\bibfnamefont {P.}~\bibnamefont
  {Xu}}, \bibinfo {author} {\bibfnamefont {H.~M.}\ \bibnamefont {R\o{}nnow}}, \
  and\ \bibinfo {author} {\bibfnamefont {M.}~\bibnamefont {Grioni}},\ }\href
  {\doibase 10.1103/PhysRevLett.105.157006} {\bibfield  {journal} {\bibinfo
  {journal} {Phys. Rev. Lett.}\ }\textbf {\bibinfo {volume} {105}},\ \bibinfo
  {pages} {157006} (\bibinfo {year} {2010})}\BibitemShut {NoStop}%
\bibitem [{\citenamefont {Luo}\ \emph {et~al.}(2014{\natexlab{b}})\citenamefont
  {Luo}, \citenamefont {Datta},\ and\ \citenamefont
  {Yao}}]{LuoPhysRevB.89.165103}%
  \BibitemOpen
  \bibfield  {author} {\bibinfo {author} {\bibfnamefont {C.}~\bibnamefont
  {Luo}}, \bibinfo {author} {\bibfnamefont {T.}~\bibnamefont {Datta}}, \ and\
  \bibinfo {author} {\bibfnamefont {D.-X.}\ \bibnamefont {Yao}},\ }\href
  {\doibase 10.1103/PhysRevB.89.165103} {\bibfield  {journal} {\bibinfo
  {journal} {Phys. Rev. B}\ }\textbf {\bibinfo {volume} {89}},\ \bibinfo
  {pages} {165103} (\bibinfo {year} {2014}{\natexlab{b}})}\BibitemShut
  {NoStop}%
\bibitem [{\citenamefont {Xiong}\ \emph {et~al.}(2017)\citenamefont {Xiong},
  \citenamefont {Datta}, \citenamefont {Stiwinter},\ and\ \citenamefont
  {Yao}}]{PhysRevB.96.144436}%
  \BibitemOpen
  \bibfield  {author} {\bibinfo {author} {\bibfnamefont {Z.}~\bibnamefont
  {Xiong}}, \bibinfo {author} {\bibfnamefont {T.}~\bibnamefont {Datta}},
  \bibinfo {author} {\bibfnamefont {K.}~\bibnamefont {Stiwinter}}, \ and\
  \bibinfo {author} {\bibfnamefont {D.-X.}\ \bibnamefont {Yao}},\ }\href
  {\doibase 10.1103/PhysRevB.96.144436} {\bibfield  {journal} {\bibinfo
  {journal} {Phys. Rev. B}\ }\textbf {\bibinfo {volume} {96}},\ \bibinfo
  {pages} {144436} (\bibinfo {year} {2017})}\BibitemShut {NoStop}%
\bibitem [{\citenamefont {Braicovich}\ \emph {et~al.}(2010)\citenamefont
  {Braicovich}, \citenamefont {Moretti~Sala}, \citenamefont {Ament},
  \citenamefont {Bisogni}, \citenamefont {Minola}, \citenamefont {Balestrino},
  \citenamefont {Di~Castro}, \citenamefont {De~Luca}, \citenamefont {Salluzzo},
  \citenamefont {Ghiringhelli},\ and\ \citenamefont {van~den
  Brink}}]{PhysRevB.81.174533}%
  \BibitemOpen
  \bibfield  {author} {\bibinfo {author} {\bibfnamefont {L.}~\bibnamefont
  {Braicovich}}, \bibinfo {author} {\bibfnamefont {M.}~\bibnamefont
  {Moretti~Sala}}, \bibinfo {author} {\bibfnamefont {L.~J.~P.}\ \bibnamefont
  {Ament}}, \bibinfo {author} {\bibfnamefont {V.}~\bibnamefont {Bisogni}},
  \bibinfo {author} {\bibfnamefont {M.}~\bibnamefont {Minola}}, \bibinfo
  {author} {\bibfnamefont {G.}~\bibnamefont {Balestrino}}, \bibinfo {author}
  {\bibfnamefont {D.}~\bibnamefont {Di~Castro}}, \bibinfo {author}
  {\bibfnamefont {G.~M.}\ \bibnamefont {De~Luca}}, \bibinfo {author}
  {\bibfnamefont {M.}~\bibnamefont {Salluzzo}}, \bibinfo {author}
  {\bibfnamefont {G.}~\bibnamefont {Ghiringhelli}}, \ and\ \bibinfo {author}
  {\bibfnamefont {J.}~\bibnamefont {van~den Brink}},\ }\href {\doibase
  10.1103/PhysRevB.81.174533} {\bibfield  {journal} {\bibinfo  {journal} {Phys.
  Rev. B}\ }\textbf {\bibinfo {volume} {81}},\ \bibinfo {pages} {174533}
  (\bibinfo {year} {2010})}\BibitemShut {NoStop}%
\bibitem [{\citenamefont {Bisogni}\ \emph {et~al.}(2012)\citenamefont
  {Bisogni}, \citenamefont {Simonelli}, \citenamefont {Ament}, \citenamefont
  {Forte}, \citenamefont {Moretti~Sala}, \citenamefont {Minola}, \citenamefont
  {Huotari}, \citenamefont {van~den Brink}, \citenamefont {Ghiringhelli},
  \citenamefont {Brookes},\ and\ \citenamefont
  {Braicovich}}]{PhysRevB.85.214527}%
  \BibitemOpen
  \bibfield  {author} {\bibinfo {author} {\bibfnamefont {V.}~\bibnamefont
  {Bisogni}}, \bibinfo {author} {\bibfnamefont {L.}~\bibnamefont {Simonelli}},
  \bibinfo {author} {\bibfnamefont {L.~J.~P.}\ \bibnamefont {Ament}}, \bibinfo
  {author} {\bibfnamefont {F.}~\bibnamefont {Forte}}, \bibinfo {author}
  {\bibfnamefont {M.}~\bibnamefont {Moretti~Sala}}, \bibinfo {author}
  {\bibfnamefont {M.}~\bibnamefont {Minola}}, \bibinfo {author} {\bibfnamefont
  {S.}~\bibnamefont {Huotari}}, \bibinfo {author} {\bibfnamefont
  {J.}~\bibnamefont {van~den Brink}}, \bibinfo {author} {\bibfnamefont
  {G.}~\bibnamefont {Ghiringhelli}}, \bibinfo {author} {\bibfnamefont {N.~B.}\
  \bibnamefont {Brookes}}, \ and\ \bibinfo {author} {\bibfnamefont
  {L.}~\bibnamefont {Braicovich}},\ }\href {\doibase
  10.1103/PhysRevB.85.214527} {\bibfield  {journal} {\bibinfo  {journal} {Phys.
  Rev. B}\ }\textbf {\bibinfo {volume} {85}},\ \bibinfo {pages} {214527}
  (\bibinfo {year} {2012})}\BibitemShut {NoStop}%
\bibitem [{\citenamefont {Notbohm}\ \emph {et~al.}(2007)\citenamefont
  {Notbohm}, \citenamefont {Ribeiro}, \citenamefont {Lake}, \citenamefont
  {Tennant}, \citenamefont {Schmidt}, \citenamefont {Uhrig}, \citenamefont
  {Hess}, \citenamefont {Klingeler}, \citenamefont {Behr}, \citenamefont
  {B\"uchner}, \citenamefont {Reehuis}, \citenamefont {Bewley}, \citenamefont
  {Frost}, \citenamefont {Manuel},\ and\ \citenamefont
  {Eccleston}}]{PhysRevLett.98.027403}%
  \BibitemOpen
  \bibfield  {author} {\bibinfo {author} {\bibfnamefont {S.}~\bibnamefont
  {Notbohm}}, \bibinfo {author} {\bibfnamefont {P.}~\bibnamefont {Ribeiro}},
  \bibinfo {author} {\bibfnamefont {B.}~\bibnamefont {Lake}}, \bibinfo {author}
  {\bibfnamefont {D.~A.}\ \bibnamefont {Tennant}}, \bibinfo {author}
  {\bibfnamefont {K.~P.}\ \bibnamefont {Schmidt}}, \bibinfo {author}
  {\bibfnamefont {G.~S.}\ \bibnamefont {Uhrig}}, \bibinfo {author}
  {\bibfnamefont {C.}~\bibnamefont {Hess}}, \bibinfo {author} {\bibfnamefont
  {R.}~\bibnamefont {Klingeler}}, \bibinfo {author} {\bibfnamefont
  {G.}~\bibnamefont {Behr}}, \bibinfo {author} {\bibfnamefont {B.}~\bibnamefont
  {B\"uchner}}, \bibinfo {author} {\bibfnamefont {M.}~\bibnamefont {Reehuis}},
  \bibinfo {author} {\bibfnamefont {R.~I.}\ \bibnamefont {Bewley}}, \bibinfo
  {author} {\bibfnamefont {C.~D.}\ \bibnamefont {Frost}}, \bibinfo {author}
  {\bibfnamefont {P.}~\bibnamefont {Manuel}}, \ and\ \bibinfo {author}
  {\bibfnamefont {R.~S.}\ \bibnamefont {Eccleston}},\ }\href {\doibase
  10.1103/PhysRevLett.98.027403} {\bibfield  {journal} {\bibinfo  {journal}
  {Phys. Rev. Lett.}\ }\textbf {\bibinfo {volume} {98}},\ \bibinfo {pages}
  {027403} (\bibinfo {year} {2007})}\BibitemShut {NoStop}%
\bibitem [{\citenamefont {Eccleston}\ \emph {et~al.}(1998)\citenamefont
  {Eccleston}, \citenamefont {Uehara}, \citenamefont {Akimitsu}, \citenamefont
  {Eisaki}, \citenamefont {Motoyama},\ and\ \citenamefont
  {Uchida}}]{PhysRevLett.81.1702}%
  \BibitemOpen
  \bibfield  {author} {\bibinfo {author} {\bibfnamefont {R.~S.}\ \bibnamefont
  {Eccleston}}, \bibinfo {author} {\bibfnamefont {M.}~\bibnamefont {Uehara}},
  \bibinfo {author} {\bibfnamefont {J.}~\bibnamefont {Akimitsu}}, \bibinfo
  {author} {\bibfnamefont {H.}~\bibnamefont {Eisaki}}, \bibinfo {author}
  {\bibfnamefont {N.}~\bibnamefont {Motoyama}}, \ and\ \bibinfo {author}
  {\bibfnamefont {S.-i.}\ \bibnamefont {Uchida}},\ }\href {\doibase
  10.1103/PhysRevLett.81.1702} {\bibfield  {journal} {\bibinfo  {journal}
  {Phys. Rev. Lett.}\ }\textbf {\bibinfo {volume} {81}},\ \bibinfo {pages}
  {1702} (\bibinfo {year} {1998})}\BibitemShut {NoStop}%
\bibitem [{\citenamefont {Matsuda}\ \emph {et~al.}(2000)\citenamefont
  {Matsuda}, \citenamefont {Katsumata}, \citenamefont {Eccleston},
  \citenamefont {Brehmer},\ and\ \citenamefont {Mikeska}}]{PhysRevB.62.8903}%
  \BibitemOpen
  \bibfield  {author} {\bibinfo {author} {\bibfnamefont {M.}~\bibnamefont
  {Matsuda}}, \bibinfo {author} {\bibfnamefont {K.}~\bibnamefont {Katsumata}},
  \bibinfo {author} {\bibfnamefont {R.~S.}\ \bibnamefont {Eccleston}}, \bibinfo
  {author} {\bibfnamefont {S.}~\bibnamefont {Brehmer}}, \ and\ \bibinfo
  {author} {\bibfnamefont {H.-J.}\ \bibnamefont {Mikeska}},\ }\href {\doibase
  10.1103/PhysRevB.62.8903} {\bibfield  {journal} {\bibinfo  {journal} {Phys.
  Rev. B}\ }\textbf {\bibinfo {volume} {62}},\ \bibinfo {pages} {8903}
  (\bibinfo {year} {2000})}\BibitemShut {NoStop}%
\bibitem [{\citenamefont {R\"uegg}\ \emph {et~al.}(2007)\citenamefont
  {R\"uegg}, \citenamefont {McMorrow}, \citenamefont {Normand}, \citenamefont
  {R\o{}nnow}, \citenamefont {Sebastian}, \citenamefont {Fisher}, \citenamefont
  {Batista}, \citenamefont {Gvasaliya}, \citenamefont {Niedermayer},\ and\
  \citenamefont {Stahn}}]{PhysRevLett.98.017202}%
  \BibitemOpen
  \bibfield  {author} {\bibinfo {author} {\bibfnamefont {C.}~\bibnamefont
  {R\"uegg}}, \bibinfo {author} {\bibfnamefont {D.~F.}\ \bibnamefont
  {McMorrow}}, \bibinfo {author} {\bibfnamefont {B.}~\bibnamefont {Normand}},
  \bibinfo {author} {\bibfnamefont {H.~M.}\ \bibnamefont {R\o{}nnow}}, \bibinfo
  {author} {\bibfnamefont {S.~E.}\ \bibnamefont {Sebastian}}, \bibinfo {author}
  {\bibfnamefont {I.~R.}\ \bibnamefont {Fisher}}, \bibinfo {author}
  {\bibfnamefont {C.~D.}\ \bibnamefont {Batista}}, \bibinfo {author}
  {\bibfnamefont {S.~N.}\ \bibnamefont {Gvasaliya}}, \bibinfo {author}
  {\bibfnamefont {C.}~\bibnamefont {Niedermayer}}, \ and\ \bibinfo {author}
  {\bibfnamefont {J.}~\bibnamefont {Stahn}},\ }\href {\doibase
  10.1103/PhysRevLett.98.017202} {\bibfield  {journal} {\bibinfo  {journal}
  {Phys. Rev. Lett.}\ }\textbf {\bibinfo {volume} {98}},\ \bibinfo {pages}
  {017202} (\bibinfo {year} {2007})}\BibitemShut {NoStop}%
\bibitem [{\citenamefont {R\"uegg}\ \emph {et~al.}(2008)\citenamefont
  {R\"uegg}, \citenamefont {Normand}, \citenamefont {Matsumoto}, \citenamefont
  {Furrer}, \citenamefont {McMorrow}, \citenamefont {Kr\"amer}, \citenamefont
  {G\"udel}, \citenamefont {Gvasaliya}, \citenamefont {Mutka},\ and\
  \citenamefont {Boehm}}]{PhysRevLett.100.205701}%
  \BibitemOpen
  \bibfield  {author} {\bibinfo {author} {\bibfnamefont {C.}~\bibnamefont
  {R\"uegg}}, \bibinfo {author} {\bibfnamefont {B.}~\bibnamefont {Normand}},
  \bibinfo {author} {\bibfnamefont {M.}~\bibnamefont {Matsumoto}}, \bibinfo
  {author} {\bibfnamefont {A.}~\bibnamefont {Furrer}}, \bibinfo {author}
  {\bibfnamefont {D.~F.}\ \bibnamefont {McMorrow}}, \bibinfo {author}
  {\bibfnamefont {K.~W.}\ \bibnamefont {Kr\"amer}}, \bibinfo {author}
  {\bibfnamefont {H.~U.}\ \bibnamefont {G\"udel}}, \bibinfo {author}
  {\bibfnamefont {S.~N.}\ \bibnamefont {Gvasaliya}}, \bibinfo {author}
  {\bibfnamefont {H.}~\bibnamefont {Mutka}}, \ and\ \bibinfo {author}
  {\bibfnamefont {M.}~\bibnamefont {Boehm}},\ }\href {\doibase
  10.1103/PhysRevLett.100.205701} {\bibfield  {journal} {\bibinfo  {journal}
  {Phys. Rev. Lett.}\ }\textbf {\bibinfo {volume} {100}},\ \bibinfo {pages}
  {205701} (\bibinfo {year} {2008})}\BibitemShut {NoStop}%
\bibitem [{\citenamefont {Savici}\ \emph {et~al.}(2009)\citenamefont {Savici},
  \citenamefont {Granroth}, \citenamefont {Broholm}, \citenamefont
  {Pajerowski}, \citenamefont {Brown}, \citenamefont {Talham}, \citenamefont
  {Meisel}, \citenamefont {Schmidt}, \citenamefont {Uhrig},\ and\ \citenamefont
  {Nagler}}]{PhysRevB.80.094411}%
  \BibitemOpen
  \bibfield  {author} {\bibinfo {author} {\bibfnamefont {A.~T.}\ \bibnamefont
  {Savici}}, \bibinfo {author} {\bibfnamefont {G.~E.}\ \bibnamefont
  {Granroth}}, \bibinfo {author} {\bibfnamefont {C.~L.}\ \bibnamefont
  {Broholm}}, \bibinfo {author} {\bibfnamefont {D.~M.}\ \bibnamefont
  {Pajerowski}}, \bibinfo {author} {\bibfnamefont {C.~M.}\ \bibnamefont
  {Brown}}, \bibinfo {author} {\bibfnamefont {D.~R.}\ \bibnamefont {Talham}},
  \bibinfo {author} {\bibfnamefont {M.~W.}\ \bibnamefont {Meisel}}, \bibinfo
  {author} {\bibfnamefont {K.~P.}\ \bibnamefont {Schmidt}}, \bibinfo {author}
  {\bibfnamefont {G.~S.}\ \bibnamefont {Uhrig}}, \ and\ \bibinfo {author}
  {\bibfnamefont {S.~E.}\ \bibnamefont {Nagler}},\ }\href {\doibase
  10.1103/PhysRevB.80.094411} {\bibfield  {journal} {\bibinfo  {journal} {Phys.
  Rev. B}\ }\textbf {\bibinfo {volume} {80}},\ \bibinfo {pages} {094411}
  (\bibinfo {year} {2009})}\BibitemShut {NoStop}%
\bibitem [{\citenamefont {Lorenzo}\ \emph {et~al.}(2010)\citenamefont
  {Lorenzo}, \citenamefont {Regnault}, \citenamefont {Boullier}, \citenamefont
  {Martin}, \citenamefont {Moudden}, \citenamefont {Vanishri}, \citenamefont
  {Marin},\ and\ \citenamefont {Revcolevschi}}]{PhysRevLett.105.097202}%
  \BibitemOpen
  \bibfield  {author} {\bibinfo {author} {\bibfnamefont {J.~E.}\ \bibnamefont
  {Lorenzo}}, \bibinfo {author} {\bibfnamefont {L.~P.}\ \bibnamefont
  {Regnault}}, \bibinfo {author} {\bibfnamefont {C.}~\bibnamefont {Boullier}},
  \bibinfo {author} {\bibfnamefont {N.}~\bibnamefont {Martin}}, \bibinfo
  {author} {\bibfnamefont {A.~H.}\ \bibnamefont {Moudden}}, \bibinfo {author}
  {\bibfnamefont {S.}~\bibnamefont {Vanishri}}, \bibinfo {author}
  {\bibfnamefont {C.}~\bibnamefont {Marin}}, \ and\ \bibinfo {author}
  {\bibfnamefont {A.}~\bibnamefont {Revcolevschi}},\ }\href {\doibase
  10.1103/PhysRevLett.105.097202} {\bibfield  {journal} {\bibinfo  {journal}
  {Phys. Rev. Lett.}\ }\textbf {\bibinfo {volume} {105}},\ \bibinfo {pages}
  {097202} (\bibinfo {year} {2010})}\BibitemShut {NoStop}%
\bibitem [{\citenamefont {Lorenzo}\ \emph {et~al.}(2011)\citenamefont
  {Lorenzo}, \citenamefont {Regnault}, \citenamefont {Boullier}, \citenamefont
  {Martin}, \citenamefont {Vanishri},\ and\ \citenamefont
  {Marin}}]{PhysRevB.83.140413}%
  \BibitemOpen
  \bibfield  {author} {\bibinfo {author} {\bibfnamefont {J.~E.}\ \bibnamefont
  {Lorenzo}}, \bibinfo {author} {\bibfnamefont {L.~P.}\ \bibnamefont
  {Regnault}}, \bibinfo {author} {\bibfnamefont {C.}~\bibnamefont {Boullier}},
  \bibinfo {author} {\bibfnamefont {N.}~\bibnamefont {Martin}}, \bibinfo
  {author} {\bibfnamefont {S.}~\bibnamefont {Vanishri}}, \ and\ \bibinfo
  {author} {\bibfnamefont {C.}~\bibnamefont {Marin}},\ }\href {\doibase
  10.1103/PhysRevB.83.140413} {\bibfield  {journal} {\bibinfo  {journal} {Phys.
  Rev. B}\ }\textbf {\bibinfo {volume} {83}},\ \bibinfo {pages} {140413}
  (\bibinfo {year} {2011})}\BibitemShut {NoStop}%
\bibitem [{\citenamefont {Schmidiger}\ \emph {et~al.}(2012)\citenamefont
  {Schmidiger}, \citenamefont {Bouillot}, \citenamefont {M\"uhlbauer},
  \citenamefont {Gvasaliya}, \citenamefont {Kollath}, \citenamefont
  {Giamarchi},\ and\ \citenamefont {Zheludev}}]{PhysRevLett.108.167201}%
  \BibitemOpen
  \bibfield  {author} {\bibinfo {author} {\bibfnamefont {D.}~\bibnamefont
  {Schmidiger}}, \bibinfo {author} {\bibfnamefont {P.}~\bibnamefont
  {Bouillot}}, \bibinfo {author} {\bibfnamefont {S.}~\bibnamefont
  {M\"uhlbauer}}, \bibinfo {author} {\bibfnamefont {S.}~\bibnamefont
  {Gvasaliya}}, \bibinfo {author} {\bibfnamefont {C.}~\bibnamefont {Kollath}},
  \bibinfo {author} {\bibfnamefont {T.}~\bibnamefont {Giamarchi}}, \ and\
  \bibinfo {author} {\bibfnamefont {A.}~\bibnamefont {Zheludev}},\ }\href
  {\doibase 10.1103/PhysRevLett.108.167201} {\bibfield  {journal} {\bibinfo
  {journal} {Phys. Rev. Lett.}\ }\textbf {\bibinfo {volume} {108}},\ \bibinfo
  {pages} {167201} (\bibinfo {year} {2012})}\BibitemShut {NoStop}%
\bibitem [{\citenamefont {Deng}\ \emph {et~al.}(2013)\citenamefont {Deng},
  \citenamefont {Tsyrulin}, \citenamefont {Bourges}, \citenamefont {Lamago},
  \citenamefont {Ronnow}, \citenamefont {Kenzelmann}, \citenamefont {Danilkin},
  \citenamefont {Pomjakushina},\ and\ \citenamefont
  {Conder}}]{PhysRevB.88.014504}%
  \BibitemOpen
  \bibfield  {author} {\bibinfo {author} {\bibfnamefont {G.}~\bibnamefont
  {Deng}}, \bibinfo {author} {\bibfnamefont {N.}~\bibnamefont {Tsyrulin}},
  \bibinfo {author} {\bibfnamefont {P.}~\bibnamefont {Bourges}}, \bibinfo
  {author} {\bibfnamefont {D.}~\bibnamefont {Lamago}}, \bibinfo {author}
  {\bibfnamefont {H.}~\bibnamefont {Ronnow}}, \bibinfo {author} {\bibfnamefont
  {M.}~\bibnamefont {Kenzelmann}}, \bibinfo {author} {\bibfnamefont
  {S.}~\bibnamefont {Danilkin}}, \bibinfo {author} {\bibfnamefont
  {E.}~\bibnamefont {Pomjakushina}}, \ and\ \bibinfo {author} {\bibfnamefont
  {K.}~\bibnamefont {Conder}},\ }\href {\doibase 10.1103/PhysRevB.88.014504}
  {\bibfield  {journal} {\bibinfo  {journal} {Phys. Rev. B}\ }\textbf {\bibinfo
  {volume} {88}},\ \bibinfo {pages} {014504} (\bibinfo {year}
  {2013})}\BibitemShut {NoStop}%
\bibitem [{\citenamefont {Schmidt}\ \emph {et~al.}(2001)\citenamefont
  {Schmidt}, \citenamefont {Knetter},\ and\ \citenamefont
  {Uhrig}}]{Schmidt_2001}%
  \BibitemOpen
  \bibfield  {author} {\bibinfo {author} {\bibfnamefont {K.~P.}\ \bibnamefont
  {Schmidt}}, \bibinfo {author} {\bibfnamefont {C.}~\bibnamefont {Knetter}}, \
  and\ \bibinfo {author} {\bibfnamefont {G.~S.}\ \bibnamefont {Uhrig}},\ }\href
  {\doibase 10.1209/epl/i2001-00601-y} {\bibfield  {journal} {\bibinfo
  {journal} {Europhysics Letters ({EPL})}\ }\textbf {\bibinfo {volume} {56}},\
  \bibinfo {pages} {877} (\bibinfo {year} {2001})}\BibitemShut {NoStop}%
\bibitem [{\citenamefont {Schmidt}\ \emph {et~al.}(2005)\citenamefont
  {Schmidt}, \citenamefont {G\"ossling}, \citenamefont {Kuhlmann},
  \citenamefont {Thomsen}, \citenamefont {L\"offert}, \citenamefont {Gross},\
  and\ \citenamefont {Assmus}}]{PhysRevB.72.094419}%
  \BibitemOpen
  \bibfield  {author} {\bibinfo {author} {\bibfnamefont {K.~P.}\ \bibnamefont
  {Schmidt}}, \bibinfo {author} {\bibfnamefont {A.}~\bibnamefont {G\"ossling}},
  \bibinfo {author} {\bibfnamefont {U.}~\bibnamefont {Kuhlmann}}, \bibinfo
  {author} {\bibfnamefont {C.}~\bibnamefont {Thomsen}}, \bibinfo {author}
  {\bibfnamefont {A.}~\bibnamefont {L\"offert}}, \bibinfo {author}
  {\bibfnamefont {C.}~\bibnamefont {Gross}}, \ and\ \bibinfo {author}
  {\bibfnamefont {W.}~\bibnamefont {Assmus}},\ }\href {\doibase
  10.1103/PhysRevB.72.094419} {\bibfield  {journal} {\bibinfo  {journal} {Phys.
  Rev. B}\ }\textbf {\bibinfo {volume} {72}},\ \bibinfo {pages} {094419}
  (\bibinfo {year} {2005})}\BibitemShut {NoStop}%
\bibitem [{\citenamefont {Schmidt}\ \emph {et~al.}(2003)\citenamefont
  {Schmidt}, \citenamefont {Knetter}, \citenamefont {Gr\"uninger},\ and\
  \citenamefont {Uhrig}}]{PhysRevLett.90.167201}%
  \BibitemOpen
  \bibfield  {author} {\bibinfo {author} {\bibfnamefont {K.~P.}\ \bibnamefont
  {Schmidt}}, \bibinfo {author} {\bibfnamefont {C.}~\bibnamefont {Knetter}},
  \bibinfo {author} {\bibfnamefont {M.}~\bibnamefont {Gr\"uninger}}, \ and\
  \bibinfo {author} {\bibfnamefont {G.~S.}\ \bibnamefont {Uhrig}},\ }\href
  {\doibase 10.1103/PhysRevLett.90.167201} {\bibfield  {journal} {\bibinfo
  {journal} {Phys. Rev. Lett.}\ }\textbf {\bibinfo {volume} {90}},\ \bibinfo
  {pages} {167201} (\bibinfo {year} {2003})}\BibitemShut {NoStop}%
\bibitem [{\citenamefont {Schmidiger}\ \emph {et~al.}(2013)\citenamefont
  {Schmidiger}, \citenamefont {M\"uhlbauer}, \citenamefont {Zheludev},
  \citenamefont {Bouillot}, \citenamefont {Giamarchi}, \citenamefont {Kollath},
  \citenamefont {Ehlers},\ and\ \citenamefont {Tsvelik}}]{PhysRevB.88.094411}%
  \BibitemOpen
  \bibfield  {author} {\bibinfo {author} {\bibfnamefont {D.}~\bibnamefont
  {Schmidiger}}, \bibinfo {author} {\bibfnamefont {S.}~\bibnamefont
  {M\"uhlbauer}}, \bibinfo {author} {\bibfnamefont {A.}~\bibnamefont
  {Zheludev}}, \bibinfo {author} {\bibfnamefont {P.}~\bibnamefont {Bouillot}},
  \bibinfo {author} {\bibfnamefont {T.}~\bibnamefont {Giamarchi}}, \bibinfo
  {author} {\bibfnamefont {C.}~\bibnamefont {Kollath}}, \bibinfo {author}
  {\bibfnamefont {G.}~\bibnamefont {Ehlers}}, \ and\ \bibinfo {author}
  {\bibfnamefont {A.~M.}\ \bibnamefont {Tsvelik}},\ }\href {\doibase
  10.1103/PhysRevB.88.094411} {\bibfield  {journal} {\bibinfo  {journal} {Phys.
  Rev. B}\ }\textbf {\bibinfo {volume} {88}},\ \bibinfo {pages} {094411}
  (\bibinfo {year} {2013})}\BibitemShut {NoStop}%
\bibitem [{\citenamefont {Windt}\ \emph {et~al.}(2001)\citenamefont {Windt},
  \citenamefont {Gr\"uninger}, \citenamefont {Nunner}, \citenamefont {Knetter},
  \citenamefont {Schmidt}, \citenamefont {Uhrig}, \citenamefont {Kopp},
  \citenamefont {Freimuth}, \citenamefont {Ammerahl}, \citenamefont
  {B\"uchner},\ and\ \citenamefont {Revcolevschi}}]{PhysRevLett.87.127002}%
  \BibitemOpen
  \bibfield  {author} {\bibinfo {author} {\bibfnamefont {M.}~\bibnamefont
  {Windt}}, \bibinfo {author} {\bibfnamefont {M.}~\bibnamefont {Gr\"uninger}},
  \bibinfo {author} {\bibfnamefont {T.}~\bibnamefont {Nunner}}, \bibinfo
  {author} {\bibfnamefont {C.}~\bibnamefont {Knetter}}, \bibinfo {author}
  {\bibfnamefont {K.~P.}\ \bibnamefont {Schmidt}}, \bibinfo {author}
  {\bibfnamefont {G.~S.}\ \bibnamefont {Uhrig}}, \bibinfo {author}
  {\bibfnamefont {T.}~\bibnamefont {Kopp}}, \bibinfo {author} {\bibfnamefont
  {A.}~\bibnamefont {Freimuth}}, \bibinfo {author} {\bibfnamefont
  {U.}~\bibnamefont {Ammerahl}}, \bibinfo {author} {\bibfnamefont
  {B.}~\bibnamefont {B\"uchner}}, \ and\ \bibinfo {author} {\bibfnamefont
  {A.}~\bibnamefont {Revcolevschi}},\ }\href {\doibase
  10.1103/PhysRevLett.87.127002} {\bibfield  {journal} {\bibinfo  {journal}
  {Phys. Rev. Lett.}\ }\textbf {\bibinfo {volume} {87}},\ \bibinfo {pages}
  {127002} (\bibinfo {year} {2001})}\BibitemShut {NoStop}%
\bibitem [{\citenamefont {Schlappa}\ \emph {et~al.}(2009)\citenamefont
  {Schlappa}, \citenamefont {Schmitt}, \citenamefont {Vernay}, \citenamefont
  {Strocov}, \citenamefont {Ilakovac}, \citenamefont {Thielemann},
  \citenamefont {R\o{}nnow}, \citenamefont {Vanishri}, \citenamefont
  {Piazzalunga}, \citenamefont {Wang}, \citenamefont {Braicovich},
  \citenamefont {Ghiringhelli}, \citenamefont {Marin}, \citenamefont {Mesot},
  \citenamefont {Delley},\ and\ \citenamefont
  {Patthey}}]{PhysRevLett.103.047401}%
  \BibitemOpen
  \bibfield  {author} {\bibinfo {author} {\bibfnamefont {J.}~\bibnamefont
  {Schlappa}}, \bibinfo {author} {\bibfnamefont {T.}~\bibnamefont {Schmitt}},
  \bibinfo {author} {\bibfnamefont {F.}~\bibnamefont {Vernay}}, \bibinfo
  {author} {\bibfnamefont {V.~N.}\ \bibnamefont {Strocov}}, \bibinfo {author}
  {\bibfnamefont {V.}~\bibnamefont {Ilakovac}}, \bibinfo {author}
  {\bibfnamefont {B.}~\bibnamefont {Thielemann}}, \bibinfo {author}
  {\bibfnamefont {H.~M.}\ \bibnamefont {R\o{}nnow}}, \bibinfo {author}
  {\bibfnamefont {S.}~\bibnamefont {Vanishri}}, \bibinfo {author}
  {\bibfnamefont {A.}~\bibnamefont {Piazzalunga}}, \bibinfo {author}
  {\bibfnamefont {X.}~\bibnamefont {Wang}}, \bibinfo {author} {\bibfnamefont
  {L.}~\bibnamefont {Braicovich}}, \bibinfo {author} {\bibfnamefont
  {G.}~\bibnamefont {Ghiringhelli}}, \bibinfo {author} {\bibfnamefont
  {C.}~\bibnamefont {Marin}}, \bibinfo {author} {\bibfnamefont
  {J.}~\bibnamefont {Mesot}}, \bibinfo {author} {\bibfnamefont
  {B.}~\bibnamefont {Delley}}, \ and\ \bibinfo {author} {\bibfnamefont
  {L.}~\bibnamefont {Patthey}},\ }\href {\doibase
  10.1103/PhysRevLett.103.047401} {\bibfield  {journal} {\bibinfo  {journal}
  {Phys. Rev. Lett.}\ }\textbf {\bibinfo {volume} {103}},\ \bibinfo {pages}
  {047401} (\bibinfo {year} {2009})}\BibitemShut {NoStop}%
\bibitem [{\citenamefont {Nagao}\ and\ \citenamefont
  {Igarashi}(2012)}]{PhysRevB.85.224436}%
  \BibitemOpen
  \bibfield  {author} {\bibinfo {author} {\bibfnamefont {T.}~\bibnamefont
  {Nagao}}\ and\ \bibinfo {author} {\bibfnamefont {J.-i.}\ \bibnamefont
  {Igarashi}},\ }\href {\doibase 10.1103/PhysRevB.85.224436} {\bibfield
  {journal} {\bibinfo  {journal} {Phys. Rev. B}\ }\textbf {\bibinfo {volume}
  {85}},\ \bibinfo {pages} {224436} (\bibinfo {year} {2012})}\BibitemShut
  {NoStop}%
\bibitem [{\citenamefont {Kumar}\ \emph {et~al.}(2019)\citenamefont {Kumar},
  \citenamefont {Nocera}, \citenamefont {Dagotto},\ and\ \citenamefont
  {Johnston}}]{kumarPhysRevB.99.205130}%
  \BibitemOpen
  \bibfield  {author} {\bibinfo {author} {\bibfnamefont {U.}~\bibnamefont
  {Kumar}}, \bibinfo {author} {\bibfnamefont {A.}~\bibnamefont {Nocera}},
  \bibinfo {author} {\bibfnamefont {E.}~\bibnamefont {Dagotto}}, \ and\
  \bibinfo {author} {\bibfnamefont {S.}~\bibnamefont {Johnston}},\ }\href
  {\doibase 10.1103/PhysRevB.99.205130} {\bibfield  {journal} {\bibinfo
  {journal} {Phys. Rev. B}\ }\textbf {\bibinfo {volume} {99}},\ \bibinfo
  {pages} {205130} (\bibinfo {year} {2019})}\BibitemShut {NoStop}%
\bibitem [{\citenamefont {Rokhsar}\ and\ \citenamefont
  {Kivelson}(1988)}]{PhysRevLett.61.2376}%
  \BibitemOpen
  \bibfield  {author} {\bibinfo {author} {\bibfnamefont {D.~S.}\ \bibnamefont
  {Rokhsar}}\ and\ \bibinfo {author} {\bibfnamefont {S.~A.}\ \bibnamefont
  {Kivelson}},\ }\href {\doibase 10.1103/PhysRevLett.61.2376} {\bibfield
  {journal} {\bibinfo  {journal} {Phys. Rev. Lett.}\ }\textbf {\bibinfo
  {volume} {61}},\ \bibinfo {pages} {2376} (\bibinfo {year}
  {1988})}\BibitemShut {NoStop}%
\bibitem [{\citenamefont {Qin}\ \emph {et~al.}(2015)\citenamefont {Qin},
  \citenamefont {Normand}, \citenamefont {Sandvik},\ and\ \citenamefont
  {Meng}}]{PhysRevB.92.214401}%
  \BibitemOpen
  \bibfield  {author} {\bibinfo {author} {\bibfnamefont {Y.~Q.}\ \bibnamefont
  {Qin}}, \bibinfo {author} {\bibfnamefont {B.}~\bibnamefont {Normand}},
  \bibinfo {author} {\bibfnamefont {A.~W.}\ \bibnamefont {Sandvik}}, \ and\
  \bibinfo {author} {\bibfnamefont {Z.~Y.}\ \bibnamefont {Meng}},\ }\href
  {\doibase 10.1103/PhysRevB.92.214401} {\bibfield  {journal} {\bibinfo
  {journal} {Phys. Rev. B}\ }\textbf {\bibinfo {volume} {92}},\ \bibinfo
  {pages} {214401} (\bibinfo {year} {2015})}\BibitemShut {NoStop}%
\bibitem [{\citenamefont {Jaime}\ \emph {et~al.}(2004)\citenamefont {Jaime},
  \citenamefont {Correa}, \citenamefont {Harrison}, \citenamefont {Batista},
  \citenamefont {Kawashima}, \citenamefont {Kazuma}, \citenamefont {Jorge},
  \citenamefont {Stern}, \citenamefont {Heinmaa}, \citenamefont {Zvyagin},
  \citenamefont {Sasago},\ and\ \citenamefont
  {Uchinokura}}]{PhysRevLett.93.087203}%
  \BibitemOpen
  \bibfield  {author} {\bibinfo {author} {\bibfnamefont {M.}~\bibnamefont
  {Jaime}}, \bibinfo {author} {\bibfnamefont {V.~F.}\ \bibnamefont {Correa}},
  \bibinfo {author} {\bibfnamefont {N.}~\bibnamefont {Harrison}}, \bibinfo
  {author} {\bibfnamefont {C.~D.}\ \bibnamefont {Batista}}, \bibinfo {author}
  {\bibfnamefont {N.}~\bibnamefont {Kawashima}}, \bibinfo {author}
  {\bibfnamefont {Y.}~\bibnamefont {Kazuma}}, \bibinfo {author} {\bibfnamefont
  {G.~A.}\ \bibnamefont {Jorge}}, \bibinfo {author} {\bibfnamefont
  {R.}~\bibnamefont {Stern}}, \bibinfo {author} {\bibfnamefont
  {I.}~\bibnamefont {Heinmaa}}, \bibinfo {author} {\bibfnamefont {S.~A.}\
  \bibnamefont {Zvyagin}}, \bibinfo {author} {\bibfnamefont {Y.}~\bibnamefont
  {Sasago}}, \ and\ \bibinfo {author} {\bibfnamefont {K.}~\bibnamefont
  {Uchinokura}},\ }\href {\doibase 10.1103/PhysRevLett.93.087203} {\bibfield
  {journal} {\bibinfo  {journal} {Phys. Rev. Lett.}\ }\textbf {\bibinfo
  {volume} {93}},\ \bibinfo {pages} {087203} (\bibinfo {year}
  {2004})}\BibitemShut {NoStop}%
\bibitem [{\citenamefont {Sundar}\ \emph {et~al.}(2019)\citenamefont {Sundar},
  \citenamefont {Rutkowski}, \citenamefont {Mueller},\ and\ \citenamefont
  {Lawler}}]{PhysRevA.99.043623}%
  \BibitemOpen
  \bibfield  {author} {\bibinfo {author} {\bibfnamefont {B.}~\bibnamefont
  {Sundar}}, \bibinfo {author} {\bibfnamefont {T.~C.}\ \bibnamefont
  {Rutkowski}}, \bibinfo {author} {\bibfnamefont {E.~J.}\ \bibnamefont
  {Mueller}}, \ and\ \bibinfo {author} {\bibfnamefont {M.~J.}\ \bibnamefont
  {Lawler}},\ }\href {\doibase 10.1103/PhysRevA.99.043623} {\bibfield
  {journal} {\bibinfo  {journal} {Phys. Rev. A}\ }\textbf {\bibinfo {volume}
  {99}},\ \bibinfo {pages} {043623} (\bibinfo {year} {2019})}\BibitemShut
  {NoStop}%
\bibitem [{\citenamefont {Singh}\ \emph {et~al.}(1988)\citenamefont {Singh},
  \citenamefont {Gelfand},\ and\ \citenamefont {Huse}}]{PhysRevLett.61.2484}%
  \BibitemOpen
  \bibfield  {author} {\bibinfo {author} {\bibfnamefont {R.~R.~P.}\
  \bibnamefont {Singh}}, \bibinfo {author} {\bibfnamefont {M.~P.}\ \bibnamefont
  {Gelfand}}, \ and\ \bibinfo {author} {\bibfnamefont {D.~A.}\ \bibnamefont
  {Huse}},\ }\href {\doibase 10.1103/PhysRevLett.61.2484} {\bibfield  {journal}
  {\bibinfo  {journal} {Phys. Rev. Lett.}\ }\textbf {\bibinfo {volume} {61}},\
  \bibinfo {pages} {2484} (\bibinfo {year} {1988})}\BibitemShut {NoStop}%
\bibitem [{\citenamefont {Chubukov}\ \emph {et~al.}(1994)\citenamefont
  {Chubukov}, \citenamefont {Sachdev},\ and\ \citenamefont
  {Ye}}]{PhysRevB.49.11919}%
  \BibitemOpen
  \bibfield  {author} {\bibinfo {author} {\bibfnamefont {A.~V.}\ \bibnamefont
  {Chubukov}}, \bibinfo {author} {\bibfnamefont {S.}~\bibnamefont {Sachdev}}, \
  and\ \bibinfo {author} {\bibfnamefont {J.}~\bibnamefont {Ye}},\ }\href
  {\doibase 10.1103/PhysRevB.49.11919} {\bibfield  {journal} {\bibinfo
  {journal} {Phys. Rev. B}\ }\textbf {\bibinfo {volume} {49}},\ \bibinfo
  {pages} {11919} (\bibinfo {year} {1994})}\BibitemShut {NoStop}%
\bibitem [{\citenamefont {Senthil}\ \emph {et~al.}(2004)\citenamefont
  {Senthil}, \citenamefont {Vishwanath}, \citenamefont {Balents}, \citenamefont
  {Sachdev},\ and\ \citenamefont {Fisher}}]{Senthil1490}%
  \BibitemOpen
  \bibfield  {author} {\bibinfo {author} {\bibfnamefont {T.}~\bibnamefont
  {Senthil}}, \bibinfo {author} {\bibfnamefont {A.}~\bibnamefont {Vishwanath}},
  \bibinfo {author} {\bibfnamefont {L.}~\bibnamefont {Balents}}, \bibinfo
  {author} {\bibfnamefont {S.}~\bibnamefont {Sachdev}}, \ and\ \bibinfo
  {author} {\bibfnamefont {M.~P.~A.}\ \bibnamefont {Fisher}},\ }\href {\doibase
  10.1126/science.1091806} {\bibfield  {journal} {\bibinfo  {journal}
  {Science}\ }\textbf {\bibinfo {volume} {303}},\ \bibinfo {pages} {1490}
  (\bibinfo {year} {2004})}\BibitemShut {NoStop}%
\bibitem [{\citenamefont {Kuklov}\ \emph {et~al.}(2008)\citenamefont {Kuklov},
  \citenamefont {Matsumoto}, \citenamefont {Prokof'ev}, \citenamefont
  {Svistunov},\ and\ \citenamefont {Troyer}}]{PhysRevLett.101.050405}%
  \BibitemOpen
  \bibfield  {author} {\bibinfo {author} {\bibfnamefont {A.~B.}\ \bibnamefont
  {Kuklov}}, \bibinfo {author} {\bibfnamefont {M.}~\bibnamefont {Matsumoto}},
  \bibinfo {author} {\bibfnamefont {N.~V.}\ \bibnamefont {Prokof'ev}}, \bibinfo
  {author} {\bibfnamefont {B.~V.}\ \bibnamefont {Svistunov}}, \ and\ \bibinfo
  {author} {\bibfnamefont {M.}~\bibnamefont {Troyer}},\ }\href {\doibase
  10.1103/PhysRevLett.101.050405} {\bibfield  {journal} {\bibinfo  {journal}
  {Phys. Rev. Lett.}\ }\textbf {\bibinfo {volume} {101}},\ \bibinfo {pages}
  {050405} (\bibinfo {year} {2008})}\BibitemShut {NoStop}%
\bibitem [{\citenamefont {Sandvik}(2007)}]{PhysRevLett.98.227202}%
  \BibitemOpen
  \bibfield  {author} {\bibinfo {author} {\bibfnamefont {A.~W.}\ \bibnamefont
  {Sandvik}},\ }\href {\doibase 10.1103/PhysRevLett.98.227202} {\bibfield
  {journal} {\bibinfo  {journal} {Phys. Rev. Lett.}\ }\textbf {\bibinfo
  {volume} {98}},\ \bibinfo {pages} {227202} (\bibinfo {year}
  {2007})}\BibitemShut {NoStop}%
\bibitem [{\citenamefont {Kotov}\ \emph {et~al.}(2009)\citenamefont {Kotov},
  \citenamefont {Yao}, \citenamefont {Castro~Neto},\ and\ \citenamefont
  {Campbell}}]{PhysRevB.80.174403}%
  \BibitemOpen
  \bibfield  {author} {\bibinfo {author} {\bibfnamefont {V.~N.}\ \bibnamefont
  {Kotov}}, \bibinfo {author} {\bibfnamefont {D.-X.}\ \bibnamefont {Yao}},
  \bibinfo {author} {\bibfnamefont {A.~H.}\ \bibnamefont {Castro~Neto}}, \ and\
  \bibinfo {author} {\bibfnamefont {D.~K.}\ \bibnamefont {Campbell}},\ }\href
  {\doibase 10.1103/PhysRevB.80.174403} {\bibfield  {journal} {\bibinfo
  {journal} {Phys. Rev. B}\ }\textbf {\bibinfo {volume} {80}},\ \bibinfo
  {pages} {174403} (\bibinfo {year} {2009})}\BibitemShut {NoStop}%
\bibitem [{\citenamefont {Banerjee}\ \emph {et~al.}(2014)\citenamefont
  {Banerjee}, \citenamefont {B\"ogli}, \citenamefont {Hofmann}, \citenamefont
  {Jiang}, \citenamefont {Widmer},\ and\ \citenamefont
  {Wiese}}]{PhysRevB.90.245143}%
  \BibitemOpen
  \bibfield  {author} {\bibinfo {author} {\bibfnamefont {D.}~\bibnamefont
  {Banerjee}}, \bibinfo {author} {\bibfnamefont {M.}~\bibnamefont {B\"ogli}},
  \bibinfo {author} {\bibfnamefont {C.~P.}\ \bibnamefont {Hofmann}}, \bibinfo
  {author} {\bibfnamefont {F.-J.}\ \bibnamefont {Jiang}}, \bibinfo {author}
  {\bibfnamefont {P.}~\bibnamefont {Widmer}}, \ and\ \bibinfo {author}
  {\bibfnamefont {U.-J.}\ \bibnamefont {Wiese}},\ }\href {\doibase
  10.1103/PhysRevB.90.245143} {\bibfield  {journal} {\bibinfo  {journal} {Phys.
  Rev. B}\ }\textbf {\bibinfo {volume} {90}},\ \bibinfo {pages} {245143}
  (\bibinfo {year} {2014})}\BibitemShut {NoStop}%
\bibitem [{\citenamefont {Banerjee}\ \emph {et~al.}(2016)\citenamefont
  {Banerjee}, \citenamefont {B\"ogli}, \citenamefont {Hofmann}, \citenamefont
  {Jiang}, \citenamefont {Widmer},\ and\ \citenamefont
  {Wiese}}]{PhysRevB.94.115120}%
  \BibitemOpen
  \bibfield  {author} {\bibinfo {author} {\bibfnamefont {D.}~\bibnamefont
  {Banerjee}}, \bibinfo {author} {\bibfnamefont {M.}~\bibnamefont {B\"ogli}},
  \bibinfo {author} {\bibfnamefont {C.~P.}\ \bibnamefont {Hofmann}}, \bibinfo
  {author} {\bibfnamefont {F.-J.}\ \bibnamefont {Jiang}}, \bibinfo {author}
  {\bibfnamefont {P.}~\bibnamefont {Widmer}}, \ and\ \bibinfo {author}
  {\bibfnamefont {U.-J.}\ \bibnamefont {Wiese}},\ }\href {\doibase
  10.1103/PhysRevB.94.115120} {\bibfield  {journal} {\bibinfo  {journal} {Phys.
  Rev. B}\ }\textbf {\bibinfo {volume} {94}},\ \bibinfo {pages} {115120}
  (\bibinfo {year} {2016})}\BibitemShut {NoStop}%
\bibitem [{\citenamefont {Kaul}\ \emph {et~al.}(2007)\citenamefont {Kaul},
  \citenamefont {Kolezhuk}, \citenamefont {Levin}, \citenamefont {Sachdev},\
  and\ \citenamefont {Senthil}}]{PhysRevB.75.235122}%
  \BibitemOpen
  \bibfield  {author} {\bibinfo {author} {\bibfnamefont {R.~K.}\ \bibnamefont
  {Kaul}}, \bibinfo {author} {\bibfnamefont {A.}~\bibnamefont {Kolezhuk}},
  \bibinfo {author} {\bibfnamefont {M.}~\bibnamefont {Levin}}, \bibinfo
  {author} {\bibfnamefont {S.}~\bibnamefont {Sachdev}}, \ and\ \bibinfo
  {author} {\bibfnamefont {T.}~\bibnamefont {Senthil}},\ }\href {\doibase
  10.1103/PhysRevB.75.235122} {\bibfield  {journal} {\bibinfo  {journal} {Phys.
  Rev. B}\ }\textbf {\bibinfo {volume} {75}},\ \bibinfo {pages} {235122}
  (\bibinfo {year} {2007})}\BibitemShut {NoStop}%
\bibitem [{\citenamefont {Matsumoto}\ \emph {et~al.}(2001)\citenamefont
  {Matsumoto}, \citenamefont {Yasuda}, \citenamefont {Todo},\ and\
  \citenamefont {Takayama}}]{PhysRevB.65.014407}%
  \BibitemOpen
  \bibfield  {author} {\bibinfo {author} {\bibfnamefont {M.}~\bibnamefont
  {Matsumoto}}, \bibinfo {author} {\bibfnamefont {C.}~\bibnamefont {Yasuda}},
  \bibinfo {author} {\bibfnamefont {S.}~\bibnamefont {Todo}}, \ and\ \bibinfo
  {author} {\bibfnamefont {H.}~\bibnamefont {Takayama}},\ }\href {\doibase
  10.1103/PhysRevB.65.014407} {\bibfield  {journal} {\bibinfo  {journal} {Phys.
  Rev. B}\ }\textbf {\bibinfo {volume} {65}},\ \bibinfo {pages} {014407}
  (\bibinfo {year} {2001})}\BibitemShut {NoStop}%
\bibitem [{\citenamefont {Wenzel}\ and\ \citenamefont
  {Janke}(2009)}]{PhysRevB.79.014410}%
  \BibitemOpen
  \bibfield  {author} {\bibinfo {author} {\bibfnamefont {S.}~\bibnamefont
  {Wenzel}}\ and\ \bibinfo {author} {\bibfnamefont {W.}~\bibnamefont {Janke}},\
  }\href {\doibase 10.1103/PhysRevB.79.014410} {\bibfield  {journal} {\bibinfo
  {journal} {Phys. Rev. B}\ }\textbf {\bibinfo {volume} {79}},\ \bibinfo
  {pages} {014410} (\bibinfo {year} {2009})}\BibitemShut {NoStop}%
\bibitem [{\citenamefont {Wenzel}\ \emph {et~al.}(2008)\citenamefont {Wenzel},
  \citenamefont {Bogacz},\ and\ \citenamefont
  {Janke}}]{PhysRevLett.101.127202}%
  \BibitemOpen
  \bibfield  {author} {\bibinfo {author} {\bibfnamefont {S.}~\bibnamefont
  {Wenzel}}, \bibinfo {author} {\bibfnamefont {L.}~\bibnamefont {Bogacz}}, \
  and\ \bibinfo {author} {\bibfnamefont {W.}~\bibnamefont {Janke}},\ }\href
  {\doibase 10.1103/PhysRevLett.101.127202} {\bibfield  {journal} {\bibinfo
  {journal} {Phys. Rev. Lett.}\ }\textbf {\bibinfo {volume} {101}},\ \bibinfo
  {pages} {127202} (\bibinfo {year} {2008})}\BibitemShut {NoStop}%
\bibitem [{\citenamefont {Gelfand}\ \emph {et~al.}(1989)\citenamefont
  {Gelfand}, \citenamefont {Singh},\ and\ \citenamefont
  {Huse}}]{PhysRevB.40.10801}%
  \BibitemOpen
  \bibfield  {author} {\bibinfo {author} {\bibfnamefont {M.~P.}\ \bibnamefont
  {Gelfand}}, \bibinfo {author} {\bibfnamefont {R.~R.~P.}\ \bibnamefont
  {Singh}}, \ and\ \bibinfo {author} {\bibfnamefont {D.~A.}\ \bibnamefont
  {Huse}},\ }\href {\doibase 10.1103/PhysRevB.40.10801} {\bibfield  {journal}
  {\bibinfo  {journal} {Phys. Rev. B}\ }\textbf {\bibinfo {volume} {40}},\
  \bibinfo {pages} {10801} (\bibinfo {year} {1989})}\BibitemShut {NoStop}%
\bibitem [{\citenamefont {Sachdev}\ and\ \citenamefont
  {Bhatt}(1990)}]{PhysRevB.41.9323}%
  \BibitemOpen
  \bibfield  {author} {\bibinfo {author} {\bibfnamefont {S.}~\bibnamefont
  {Sachdev}}\ and\ \bibinfo {author} {\bibfnamefont {R.~N.}\ \bibnamefont
  {Bhatt}},\ }\href {\doibase 10.1103/PhysRevB.41.9323} {\bibfield  {journal}
  {\bibinfo  {journal} {Phys. Rev. B}\ }\textbf {\bibinfo {volume} {41}},\
  \bibinfo {pages} {9323} (\bibinfo {year} {1990})}\BibitemShut {NoStop}%
\bibitem [{\citenamefont {Sushkov}\ \emph {et~al.}(2001)\citenamefont
  {Sushkov}, \citenamefont {Oitmaa},\ and\ \citenamefont
  {Weihong}}]{PhysRevB.63.104420}%
  \BibitemOpen
  \bibfield  {author} {\bibinfo {author} {\bibfnamefont {O.~P.}\ \bibnamefont
  {Sushkov}}, \bibinfo {author} {\bibfnamefont {J.}~\bibnamefont {Oitmaa}}, \
  and\ \bibinfo {author} {\bibfnamefont {Z.}~\bibnamefont {Weihong}},\ }\href
  {\doibase 10.1103/PhysRevB.63.104420} {\bibfield  {journal} {\bibinfo
  {journal} {Phys. Rev. B}\ }\textbf {\bibinfo {volume} {63}},\ \bibinfo
  {pages} {104420} (\bibinfo {year} {2001})}\BibitemShut {NoStop}%
\bibitem [{\citenamefont {Chubukov}\ and\ \citenamefont
  {Jolicoeur}(1991)}]{PhysRevB.44.12050}%
  \BibitemOpen
  \bibfield  {author} {\bibinfo {author} {\bibfnamefont {A.~V.}\ \bibnamefont
  {Chubukov}}\ and\ \bibinfo {author} {\bibfnamefont {T.}~\bibnamefont
  {Jolicoeur}},\ }\href {\doibase 10.1103/PhysRevB.44.12050} {\bibfield
  {journal} {\bibinfo  {journal} {Phys. Rev. B}\ }\textbf {\bibinfo {volume}
  {44}},\ \bibinfo {pages} {12050} (\bibinfo {year} {1991})}\BibitemShut
  {NoStop}%
\bibitem [{\citenamefont {Ma}\ \emph {et~al.}(2018)\citenamefont {Ma},
  \citenamefont {Weinberg}, \citenamefont {Shao}, \citenamefont {Guo},
  \citenamefont {Yao},\ and\ \citenamefont {Sandvik}}]{PhysRevLett.121.117202}%
  \BibitemOpen
  \bibfield  {author} {\bibinfo {author} {\bibfnamefont {N.}~\bibnamefont
  {Ma}}, \bibinfo {author} {\bibfnamefont {P.}~\bibnamefont {Weinberg}},
  \bibinfo {author} {\bibfnamefont {H.}~\bibnamefont {Shao}}, \bibinfo {author}
  {\bibfnamefont {W.}~\bibnamefont {Guo}}, \bibinfo {author} {\bibfnamefont
  {D.-X.}\ \bibnamefont {Yao}}, \ and\ \bibinfo {author} {\bibfnamefont
  {A.~W.}\ \bibnamefont {Sandvik}},\ }\href {\doibase
  10.1103/PhysRevLett.121.117202} {\bibfield  {journal} {\bibinfo  {journal}
  {Phys. Rev. Lett.}\ }\textbf {\bibinfo {volume} {121}},\ \bibinfo {pages}
  {117202} (\bibinfo {year} {2018})}\BibitemShut {NoStop}%
\bibitem [{\citenamefont {Sandvik}(2010)}]{doi:10.1063/1.3518900}%
  \BibitemOpen
  \bibfield  {author} {\bibinfo {author} {\bibfnamefont {A.~W.}\ \bibnamefont
  {Sandvik}},\ }\href {\doibase 10.1063/1.3518900} {\bibfield  {journal}
  {\bibinfo  {journal} {AIP Conference Proceedings}\ }\textbf {\bibinfo
  {volume} {1297}},\ \bibinfo {pages} {135} (\bibinfo {year} {2010})},\ \Eprint
  {http://arxiv.org/abs/https://aip.scitation.org/doi/pdf/10.1063/1.3518900}
  {https://aip.scitation.org/doi/pdf/10.1063/1.3518900} \BibitemShut {NoStop}%
\bibitem [{\citenamefont {Yasuda}\ and\ \citenamefont
  {Todo}(2013)}]{PhysRevE.88.061301}%
  \BibitemOpen
  \bibfield  {author} {\bibinfo {author} {\bibfnamefont {S.}~\bibnamefont
  {Yasuda}}\ and\ \bibinfo {author} {\bibfnamefont {S.}~\bibnamefont {Todo}},\
  }\href {\doibase 10.1103/PhysRevE.88.061301} {\bibfield  {journal} {\bibinfo
  {journal} {Phys. Rev. E}\ }\textbf {\bibinfo {volume} {88}},\ \bibinfo
  {pages} {061301} (\bibinfo {year} {2013})}\BibitemShut {NoStop}%
\bibitem [{\citenamefont {Luo}\ \emph {et~al.}(2015)\citenamefont {Luo},
  \citenamefont {Datta}, \citenamefont {Huang},\ and\ \citenamefont
  {Yao}}]{LuoPhysRevB.92.035109}%
  \BibitemOpen
  \bibfield  {author} {\bibinfo {author} {\bibfnamefont {C.}~\bibnamefont
  {Luo}}, \bibinfo {author} {\bibfnamefont {T.}~\bibnamefont {Datta}}, \bibinfo
  {author} {\bibfnamefont {Z.}~\bibnamefont {Huang}}, \ and\ \bibinfo {author}
  {\bibfnamefont {D.-X.}\ \bibnamefont {Yao}},\ }\href {\doibase
  10.1103/PhysRevB.92.035109} {\bibfield  {journal} {\bibinfo  {journal} {Phys.
  Rev. B}\ }\textbf {\bibinfo {volume} {92}},\ \bibinfo {pages} {035109}
  (\bibinfo {year} {2015})}\BibitemShut {NoStop}%
\bibitem [{\citenamefont {Doretto}(2014)}]{PhysRevB.89.104415}%
  \BibitemOpen
  \bibfield  {author} {\bibinfo {author} {\bibfnamefont {R.~L.}\ \bibnamefont
  {Doretto}},\ }\href {\doibase 10.1103/PhysRevB.89.104415} {\bibfield
  {journal} {\bibinfo  {journal} {Phys. Rev. B}\ }\textbf {\bibinfo {volume}
  {89}},\ \bibinfo {pages} {104415} (\bibinfo {year} {2014})}\BibitemShut
  {NoStop}%
\bibitem [{\citenamefont {Chabot-Couture}\ \emph {et~al.}(2010)\citenamefont
  {Chabot-Couture}, \citenamefont {Hancock}, \citenamefont {Mang},
  \citenamefont {Casa}, \citenamefont {Gog},\ and\ \citenamefont
  {Greven}}]{PhysRevB.82.035113}%
  \BibitemOpen
  \bibfield  {author} {\bibinfo {author} {\bibfnamefont {G.}~\bibnamefont
  {Chabot-Couture}}, \bibinfo {author} {\bibfnamefont {J.~N.}\ \bibnamefont
  {Hancock}}, \bibinfo {author} {\bibfnamefont {P.~K.}\ \bibnamefont {Mang}},
  \bibinfo {author} {\bibfnamefont {D.~M.}\ \bibnamefont {Casa}}, \bibinfo
  {author} {\bibfnamefont {T.}~\bibnamefont {Gog}}, \ and\ \bibinfo {author}
  {\bibfnamefont {M.}~\bibnamefont {Greven}},\ }\href {\doibase
  10.1103/PhysRevB.82.035113} {\bibfield  {journal} {\bibinfo  {journal} {Phys.
  Rev. B}\ }\textbf {\bibinfo {volume} {82}},\ \bibinfo {pages} {035113}
  (\bibinfo {year} {2010})}\BibitemShut {NoStop}%
\bibitem [{\citenamefont {Coldea}\ \emph {et~al.}(2001)\citenamefont {Coldea},
  \citenamefont {Hayden}, \citenamefont {Aeppli}, \citenamefont {Perring},
  \citenamefont {Frost}, \citenamefont {Mason}, \citenamefont {Cheong},\ and\
  \citenamefont {Fisk}}]{PhysRevLett.86.5377}%
  \BibitemOpen
  \bibfield  {author} {\bibinfo {author} {\bibfnamefont {R.}~\bibnamefont
  {Coldea}}, \bibinfo {author} {\bibfnamefont {S.~M.}\ \bibnamefont {Hayden}},
  \bibinfo {author} {\bibfnamefont {G.}~\bibnamefont {Aeppli}}, \bibinfo
  {author} {\bibfnamefont {T.~G.}\ \bibnamefont {Perring}}, \bibinfo {author}
  {\bibfnamefont {C.~D.}\ \bibnamefont {Frost}}, \bibinfo {author}
  {\bibfnamefont {T.~E.}\ \bibnamefont {Mason}}, \bibinfo {author}
  {\bibfnamefont {S.-W.}\ \bibnamefont {Cheong}}, \ and\ \bibinfo {author}
  {\bibfnamefont {Z.}~\bibnamefont {Fisk}},\ }\href {\doibase
  10.1103/PhysRevLett.86.5377} {\bibfield  {journal} {\bibinfo  {journal}
  {Phys. Rev. Lett.}\ }\textbf {\bibinfo {volume} {86}},\ \bibinfo {pages}
  {5377} (\bibinfo {year} {2001})}\BibitemShut {NoStop}%
\bibitem [{\citenamefont {Braicovich}\ \emph {et~al.}(2009)\citenamefont
  {Braicovich}, \citenamefont {Ament}, \citenamefont {Bisogni}, \citenamefont
  {Forte}, \citenamefont {Aruta}, \citenamefont {Balestrino}, \citenamefont
  {Brookes}, \citenamefont {De~Luca}, \citenamefont {Medaglia}, \citenamefont
  {Granozio}, \citenamefont {Radovic}, \citenamefont {Salluzzo}, \citenamefont
  {van~den Brink},\ and\ \citenamefont
  {Ghiringhelli}}]{PhysRevLett.102.167401}%
  \BibitemOpen
  \bibfield  {author} {\bibinfo {author} {\bibfnamefont {L.}~\bibnamefont
  {Braicovich}}, \bibinfo {author} {\bibfnamefont {L.~J.~P.}\ \bibnamefont
  {Ament}}, \bibinfo {author} {\bibfnamefont {V.}~\bibnamefont {Bisogni}},
  \bibinfo {author} {\bibfnamefont {F.}~\bibnamefont {Forte}}, \bibinfo
  {author} {\bibfnamefont {C.}~\bibnamefont {Aruta}}, \bibinfo {author}
  {\bibfnamefont {G.}~\bibnamefont {Balestrino}}, \bibinfo {author}
  {\bibfnamefont {N.~B.}\ \bibnamefont {Brookes}}, \bibinfo {author}
  {\bibfnamefont {G.~M.}\ \bibnamefont {De~Luca}}, \bibinfo {author}
  {\bibfnamefont {P.~G.}\ \bibnamefont {Medaglia}}, \bibinfo {author}
  {\bibfnamefont {F.~M.}\ \bibnamefont {Granozio}}, \bibinfo {author}
  {\bibfnamefont {M.}~\bibnamefont {Radovic}}, \bibinfo {author} {\bibfnamefont
  {M.}~\bibnamefont {Salluzzo}}, \bibinfo {author} {\bibfnamefont
  {J.}~\bibnamefont {van~den Brink}}, \ and\ \bibinfo {author} {\bibfnamefont
  {G.}~\bibnamefont {Ghiringhelli}},\ }\href {\doibase
  10.1103/PhysRevLett.102.167401} {\bibfield  {journal} {\bibinfo  {journal}
  {Phys. Rev. Lett.}\ }\textbf {\bibinfo {volume} {102}},\ \bibinfo {pages}
  {167401} (\bibinfo {year} {2009})}\BibitemShut {NoStop}%
\bibitem [{\citenamefont {Ament}\ \emph {et~al.}(2009)\citenamefont {Ament},
  \citenamefont {Ghiringhelli}, \citenamefont {Sala}, \citenamefont
  {Braicovich},\ and\ \citenamefont {van~den Brink}}]{PhysRevLett.103.117003}%
  \BibitemOpen
  \bibfield  {author} {\bibinfo {author} {\bibfnamefont {L.~J.~P.}\
  \bibnamefont {Ament}}, \bibinfo {author} {\bibfnamefont {G.}~\bibnamefont
  {Ghiringhelli}}, \bibinfo {author} {\bibfnamefont {M.~M.}\ \bibnamefont
  {Sala}}, \bibinfo {author} {\bibfnamefont {L.}~\bibnamefont {Braicovich}}, \
  and\ \bibinfo {author} {\bibfnamefont {J.}~\bibnamefont {van~den Brink}},\
  }\href {\doibase 10.1103/PhysRevLett.103.117003} {\bibfield  {journal}
  {\bibinfo  {journal} {Phys. Rev. Lett.}\ }\textbf {\bibinfo {volume} {103}},\
  \bibinfo {pages} {117003} (\bibinfo {year} {2009})}\BibitemShut {NoStop}%
\bibitem [{\citenamefont {Sala}\ \emph {et~al.}(2011)\citenamefont {Sala},
  \citenamefont {Bisogni}, \citenamefont {Aruta}, \citenamefont {Balestrino},
  \citenamefont {Berger}, \citenamefont {Brookes}, \citenamefont {de~Luca},
  \citenamefont {Castro}, \citenamefont {Grioni}, \citenamefont {Guarise},
  \citenamefont {Medaglia}, \citenamefont {Granozio}, \citenamefont {Minola},
  \citenamefont {Perna}, \citenamefont {Radovic}, \citenamefont {Salluzzo},
  \citenamefont {Schmitt}, \citenamefont {Zhou}, \citenamefont {Braicovich},\
  and\ \citenamefont {Ghiringhelli}}]{1367-2630-13-4-043026}%
  \BibitemOpen
  \bibfield  {author} {\bibinfo {author} {\bibfnamefont {M.~M.}\ \bibnamefont
  {Sala}}, \bibinfo {author} {\bibfnamefont {V.}~\bibnamefont {Bisogni}},
  \bibinfo {author} {\bibfnamefont {C.}~\bibnamefont {Aruta}}, \bibinfo
  {author} {\bibfnamefont {G.}~\bibnamefont {Balestrino}}, \bibinfo {author}
  {\bibfnamefont {H.}~\bibnamefont {Berger}}, \bibinfo {author} {\bibfnamefont
  {N.~B.}\ \bibnamefont {Brookes}}, \bibinfo {author} {\bibfnamefont {G.~M.}\
  \bibnamefont {de~Luca}}, \bibinfo {author} {\bibfnamefont {D.~D.}\
  \bibnamefont {Castro}}, \bibinfo {author} {\bibfnamefont {M.}~\bibnamefont
  {Grioni}}, \bibinfo {author} {\bibfnamefont {M.}~\bibnamefont {Guarise}},
  \bibinfo {author} {\bibfnamefont {P.~G.}\ \bibnamefont {Medaglia}}, \bibinfo
  {author} {\bibfnamefont {F.~M.}\ \bibnamefont {Granozio}}, \bibinfo {author}
  {\bibfnamefont {M.}~\bibnamefont {Minola}}, \bibinfo {author} {\bibfnamefont
  {P.}~\bibnamefont {Perna}}, \bibinfo {author} {\bibfnamefont
  {M.}~\bibnamefont {Radovic}}, \bibinfo {author} {\bibfnamefont
  {M.}~\bibnamefont {Salluzzo}}, \bibinfo {author} {\bibfnamefont
  {T.}~\bibnamefont {Schmitt}}, \bibinfo {author} {\bibfnamefont {K.~J.}\
  \bibnamefont {Zhou}}, \bibinfo {author} {\bibfnamefont {L.}~\bibnamefont
  {Braicovich}}, \ and\ \bibinfo {author} {\bibfnamefont {G.}~\bibnamefont
  {Ghiringhelli}},\ }\href {http://stacks.iop.org/1367-2630/13/i=4/a=043026}
  {\bibfield  {journal} {\bibinfo  {journal} {New Journal of Physics}\ }\textbf
  {\bibinfo {volume} {13}},\ \bibinfo {pages} {043026} (\bibinfo {year}
  {2011})}\BibitemShut {NoStop}%
\bibitem [{\citenamefont {Nomura}(2017)}]{NomuraPhysRevB.96.165128}%
  \BibitemOpen
  \bibfield  {author} {\bibinfo {author} {\bibfnamefont {T.}~\bibnamefont
  {Nomura}},\ }\href {\doibase 10.1103/PhysRevB.96.165128} {\bibfield
  {journal} {\bibinfo  {journal} {Phys. Rev. B}\ }\textbf {\bibinfo {volume}
  {96}},\ \bibinfo {pages} {165128} (\bibinfo {year} {2017})}\BibitemShut
  {NoStop}%
\bibitem [{\citenamefont {Jia}\ \emph {et~al.}(2016)\citenamefont {Jia},
  \citenamefont {Wohlfeld}, \citenamefont {Wang}, \citenamefont {Moritz},\ and\
  \citenamefont {Devereaux}}]{PhysRevX.6.021020}%
  \BibitemOpen
  \bibfield  {author} {\bibinfo {author} {\bibfnamefont {C.}~\bibnamefont
  {Jia}}, \bibinfo {author} {\bibfnamefont {K.}~\bibnamefont {Wohlfeld}},
  \bibinfo {author} {\bibfnamefont {Y.}~\bibnamefont {Wang}}, \bibinfo {author}
  {\bibfnamefont {B.}~\bibnamefont {Moritz}}, \ and\ \bibinfo {author}
  {\bibfnamefont {T.~P.}\ \bibnamefont {Devereaux}},\ }\href {\doibase
  10.1103/PhysRevX.6.021020} {\bibfield  {journal} {\bibinfo  {journal} {Phys.
  Rev. X}\ }\textbf {\bibinfo {volume} {6}},\ \bibinfo {pages} {021020}
  (\bibinfo {year} {2016})}\BibitemShut {NoStop}%
\bibitem [{\citenamefont {Vaknin}\ \emph {et~al.}(1990)\citenamefont {Vaknin},
  \citenamefont {Sinha}, \citenamefont {Stassis}, \citenamefont {Miller},\ and\
  \citenamefont {Johnston}}]{PhysRevB.41.1926}%
  \BibitemOpen
  \bibfield  {author} {\bibinfo {author} {\bibfnamefont {D.}~\bibnamefont
  {Vaknin}}, \bibinfo {author} {\bibfnamefont {S.~K.}\ \bibnamefont {Sinha}},
  \bibinfo {author} {\bibfnamefont {C.}~\bibnamefont {Stassis}}, \bibinfo
  {author} {\bibfnamefont {L.~L.}\ \bibnamefont {Miller}}, \ and\ \bibinfo
  {author} {\bibfnamefont {D.~C.}\ \bibnamefont {Johnston}},\ }\href {\doibase
  10.1103/PhysRevB.41.1926} {\bibfield  {journal} {\bibinfo  {journal} {Phys.
  Rev. B}\ }\textbf {\bibinfo {volume} {41}},\ \bibinfo {pages} {1926}
  (\bibinfo {year} {1990})}\BibitemShut {NoStop}%
\bibitem [{\citenamefont {Gopalan}\ \emph {et~al.}(1994)\citenamefont
  {Gopalan}, \citenamefont {Rice},\ and\ \citenamefont
  {Sigrist}}]{PhysRevB.49.8901}%
  \BibitemOpen
  \bibfield  {author} {\bibinfo {author} {\bibfnamefont {S.}~\bibnamefont
  {Gopalan}}, \bibinfo {author} {\bibfnamefont {T.~M.}\ \bibnamefont {Rice}}, \
  and\ \bibinfo {author} {\bibfnamefont {M.}~\bibnamefont {Sigrist}},\ }\href
  {\doibase 10.1103/PhysRevB.49.8901} {\bibfield  {journal} {\bibinfo
  {journal} {Phys. Rev. B}\ }\textbf {\bibinfo {volume} {49}},\ \bibinfo
  {pages} {8901} (\bibinfo {year} {1994})}\BibitemShut {NoStop}%
\end{thebibliography}%
\end{document}